\documentclass[a4paper,11pt]{article}
\pdfoutput=1 

\usepackage{jheppub} 

\usepackage[T1]{fontenc} 

\usepackage[usenames,table]{xcolor}
\usepackage{dsfont}
\usepackage{mathrsfs}
\usepackage{tensor}
\usepackage{verbatim}
\usepackage[vcentermath]{youngtab}
\usepackage{multirow}
\usepackage{multicol}
\usepackage{tikz-cd}
\usepackage{soul}

\def\cA{{\cal A}}
\def\cB{{\cal B}}
\def\cC{{\cal C}}

\def\cG{{\cal G}}

\def\cI{{\cal I}}
\def\cJ{{\cal J}}
\def\cK{{\cal K}}
\def\cL{{\cal L}}
\def\cM{{\cal M}}

\def\cO{{\cal O}}
\def\cP{{\cal P}}
\def\cQ{{\cal Q}}
\def\cR{{\cal R}}
\def\cS{{\cal S}}

\def\cU{{\cal U}}

\def\cZ{{\cal Z}}

\def\a{\alpha} 
  
\def\g{\gamma} 
\def\G{\Gamma}
\def\d{\delta} 

\def\e{\epsilon} 
\def\ve{\varepsilon}

\def\h{\eta}
 
\def\Th{\Theta}
\def\k{\kappa}
\def\l{\lambda}
\def\L{\Lambda}
\def\m{\mu}
\def\n{\nu}

\def\r{\rho}
\def\s{\sigma}

\def\t{\tau}
\def\f{\phi}

\def\vf{\varphi}

\def\be{\begin{equation}}
\def\ee{\end{equation}}

\def\dim{D}

\def\pr{\partial}

\def\nn{\nonumber}

\newcommand{\mfk}[1]{\mathfrak{#1}}
\newcommand{\hs}{\mfk{hs}}
\newcommand{\ihs}{\mfk{ihs}}
\newcommand{\gchs}{\mfk{gchs}}
\renewcommand{\sl}{\mfk{sl}}
\renewcommand{\so}{\mfk{so}} 
\newcommand{\iso}{\mfk{iso}}
\newcommand{\gca}{\mfk{gca}}
\newcommand{\galanyD}[4]{\left(\,\scriptsize{#1}\,, \mathbf{#2}\right)^{(#3, #4)}}
\newcommand{\galfiveD}[4]{\left(\mathbf{#1}, \mathbf{#2}\right)^{(#3, #4)}}
\newcommand{\pol}[3]{\text P^{#1 #2}_{#3}}
\def\Th{\hat T}
\def\Tt{\tilde T}
\def\Lb{\bar L}
\def\Wb{\bar W}
\def\tb{\bar \t}
\def\sh{\hat \s}
\def\sb{\bar \s}
\def\rb{\bar \r}
\def\rt{\tilde \r}
\def\pib{\bar \pi}
\def\pit{\tilde \pi}
\def\Bb{\bar B}
\def\Xb{\bar X}
\def\Yb{\bar Y}
\def\qb{\bar q}

\newcommand{\nocontentsline}[3]{}
\newcommand{\tocless}[2]{\bgroup\let\addcontentsline=\nocontentsline#1{#2}\egroup}
\newcommand{\toclesslab}[3]{\bgroup\let\addcontentsline=\nocontentsline#1{#2\label{#3}}\egroup}

\title{Carrollian and Galilean conformal higher-spin algebras in any dimensions}

\author[1]{Andrea~Campoleoni\note{Research Associate of the Fund for Scientific Research -- FNRS, Belgium.}}
\author[2]{and Simon Pekar\note{FRIA grantee of the Fund for Scientific Research -- FNRS, Belgium.}}
\affiliation{Service de Physique de l'Univers, Champs et Gravitation\\ Universit\'e de Mons -- UMONS\\ 20 place du Parc, 7000 Mons, 
Belgium}

\emailAdd{andrea.campoleoni@umons.ac.be}
\emailAdd{simon.pekar@umons.ac.be}

\abstract{We present higher-spin algebras containing a Poincar\'e subalgebra and with the same set of generators as the Lie algebras that are relevant to Vasiliev's equations in any space-time dimension $D \geq 3$. Given these properties, they can be considered either as candidate rigid symmetries for higher-spin gauge theories in Minkowski space or as Carrollian conformal higher-spin symmetries in one less dimension. We build these Lie algebras as quotients of the universal enveloping algebra of $\iso(1,D-1)$ and we show how to recover them as In\"on\"u-Wigner contractions of the rigid symmetries of higher-spin gauge theories in Anti de Sitter space or, equivalently, of relativistic conformal higher-spin symmetries. We use the same techniques to also define higher-spin algebras with the same set of generators and containing a Galilean conformal subalgebra, to be interpreted as non-relativistic limits of the conformal symmetries of a free scalar field. We begin by showing that the known flat-space higher-spin algebras in three dimensions can be obtained as quotients of the universal enveloping algebra of $\iso(1,2)$ and then we extend the analysis along the same lines to a generic number of space-time dimensions. We also discuss the peculiarities that emerge for $D=5$.}

\begin{document} 
\maketitle
\flushbottom

\section{Introduction}

The Poincar\'e and (Anti) de Sitter algebras describe the isometries of the vacuum in gravitational theories. In the Cartan formulation they are also instrumental in identifying the basic geometric quantities, connections and curvatures, upon which the field theory is built. So, on the one hand, these Lie algebras emerge as the vacuum-preserving part of the diffeomorphism symmetry and, on the other hand, Einstein gravity can be recovered from their gauging. A similar pattern is expected to apply to massless (or partially-massless) fields of any spin. Any Lorentz-covariant description of their dynamics involves gauge fields, and gauge transformations preserving the vacuum should give rise to a Lie algebra, including the Poincar\'e or (Anti) de Sitter isometries as a subalgebra. Interacting field theories may then result from the gauging of these rigid, often called global, higher-spin symmetries. 

This scheme has been successfully implemented in Anti de Sitter (AdS) space: a candidate Lie algebra for the role of global higher-spin symmetry in four dimensions was proposed in \cite{Fradkin:1986ka} and Vasiliev's equations bring about its gauging \cite{Vasiliev:1990en}. The same strategy has then been applied to build other examples of higher-spin theories. Extended supersymmetry was included in four dimensions shortly thereafter \cite{Vasiliev:1986qx, Konstein:1989ij}, while extensions to five and seven dimensions were studied in \cite{Sezgin:2001zs, Vasiliev:2001wa, Sezgin:2001ij}. Higher-spin algebras for massless fields in (A)dS spaces of any dimensions were eventually presented in \cite{Eastwood:2002su, Vasiliev:2003ev, Vasiliev:2004qz, Vasiliev:2004cm} together with the corresponding interacting equations of motion. More recently, Lie algebras appropriate for the role of global symmetries for mixed-symmetry and partially-massless fields have been identified \cite{Boulanger:2011se, Joung:2015jza} and interacting equations of motion involving partially-massless fields have been proposed \cite{Alkalaev:2014nsa, Brust:2016zns}. Global symmetry algebras play an important role in three dimensions as well: they determine the interactions of massless (and partially-massless) gauge fields via Chern-Simons actions \cite{Blencowe:1988gj, Bergshoeff:1989ns, Campoleoni:2010zq, Grigoriev:2020lzu} and rule their interactions with matter \`a la Vasiliev \cite{Prokushkin:1998bq}.

The quest for global\footnote{Following a common notation in the literature, we shall often denote as \emph{global} the gauge symmetries preserving the \emph{vacuum}. We recall however that, in gauge theories, bona fide global symmetries correspond to asymptotic ones, which preserve the space of allowed asymptotic field configurations. In AdS$_\dim$ spaces with $\dim \geq 4$ the two concepts should lead to the same algebras even when considering higher-spin fields \cite{Henneaux:1985ey, Campoleoni:2016uwr, Campoleoni:2017vds} (at least within the standard reflective boundary conditions), while in $\dim=3$ asymptotic symmetries greatly enhance the rigid symmetries of the vacuum \cite{Brown:1986nw, Henneaux:2010xg, Campoleoni:2010zq, Gaberdiel:2011wb, Campoleoni:2011hg}. Similar enhancements are expected on Minkowski backgrounds of any dimensions \cite{Campoleoni:2017mbt, Campoleoni:2020ejn}.} higher-spin symmetries went side by side with studies on the structure of these Lie algebras, that are typically infinite dimensional because their closure requires infinitely many gauge fields. A key observation is that bosonic higher-spin algebras in AdS$_\dim$ are isomorphic to the conformal symmetries of a free scalar in \mbox{$\dim-1$} dimensions \cite{Mikhailov:2002bp, Eastwood:2002su}. This result led to characterise those algebras as quotients of the universal enveloping algebra of the $\so(2,\dim-1)$ symmetry of the vacuum \cite{Eastwood:2002su, Bekaert:2007mi, Iazeolla:2008ix, Bekaert:2008sa, Boulanger:2011se, Joung:2014qya}.\footnote{Actually, higher-spin algebras in three dimensions have been presented as quotients of the universal enveloping algebra of the isometries of the vacuum even before \cite{Feigin:1988, Bergshoeff:1989ns, Bordemann:1989zi, Fradkin:1990ir}.} The Lie algebras obtained in this way suffice to describe the vacuum-preserving part of all known non-linear deformations to the free gauge symmetry when $\dim = 4$ and $\dim \geq 7$ (up to Chan-Paton factors) \cite{Boulanger:2013zza}. The same conclusion can be inferred, via the AdS/CFT correspondence, from the analysis of the correlation functions of conformal field theories admitting at least one conserved higher-spin current \cite{Maldacena:2011jn, Stanev:2013qra, Alba:2013yda, Alba:2015upa}. Besides proving rather universal, the coset construction of higher-spin algebras has also been fruitful: for instance, it is at the basis of the generalisations including mixed-symmetry and partially-massless fields \cite{Boulanger:2011se, Joung:2015jza}.

The study of higher-spin extensions of the Poincar\'e algebra is definitely less developed. Although this issue was raised soon after the first ``yes-go'' results in flat space \cite{Bengtsson:1986bz} and revisited lately in \cite{Bekaert:2006us, Bekaert:2008sa, Joung:2013nma, Sleight:2016xqq, Ponomarev:2017nrr}, a systematic study of possible flat-space counterparts of the bosonic algebras relevant to Vasiliev's equations is still missing in $\dim \geq 4$.
The situation is instead radically different in three-dimensional Minkowski space, where global higher-spin symmetries can be naturally obtained as In\"on\"u-Wigner contractions of the AdS ones \cite{Blencowe:1988gj, Campoleoni:2011tn, Afshar:2013vka, Gonzalez:2013oaa, Ammon:2017vwt}. In this context, one can then easily build interacting gauge theories using Chern-Simons actions as in AdS$_3$, while specific interactions with matter were introduced following Vasiliev's approach \cite{Ammon:2020fxs}. Another notable exception is given by higher-spin fermionic extensions of the Poincar\'e algebra, where one can consistently add a single generator of spin $s+1/2$ \cite{Fuentealba:2015jma, Fuentealba:2019bgb}.

In this paper, we start filling the gap by presenting bosonic higher-spin extensions of the Poincar\'e algebra in any space-time dimensions. We reach this goal by classifying the quotients of the universal enveloping algebra of $\iso(1,\dim-1)$ that have the same set of generators as the Lie algebras entering Vasiliev's equations. Our approach is thus rather pragmatic: we aim at providing examples of flat-space higher-spin algebras, while leaving to future investigations important questions like the study of their representations and their possible realisation in interacting gauge theories. The underlying idea is that ---~in spite of the ``no-go'' results that strongly constrain higher-spin interactions in flat space as, e.g.,  \cite{Weinberg:1964ew, Bekaert:2010hp, Bekaert:2010hw, Porrati:2012rd, Taronna:2017wbx}~--- interacting gauge theories involving fields with spin greater than two may still exist in some, possibly exotic, form. This is suggested, at least, by the long-held conjecture that string theory may describe the broken phase of a higher-spin gauge theory (see, e.g., \cite{Sagnotti:2013bha} for a review). A classification of cubic vertices for massless higher-spin particles in flat space is also known since the eighties \cite{Bengtsson:1983pd, Bengtsson:1986kh, Fradkin:1991iy, Metsaev:1993ap, Metsaev:2005ar, Metsaev:2007rn, Manvelyan:2010jr, Sagnotti:2010at, Francia:2016weg} and concrete proposals for interacting, although ``unconventional'', higher-spin gauge theories in Minkowski space are by now available \cite{Ponomarev:2016lrm, Skvortsov:2018jea, Bonora:2018ggh, Bonora:2021pcj, Krasnov:2021nsq}.\footnote{Roughly speaking, the no-go results spot a tension between unitarity, locality and a spin greater than two. Dropping one of these requirements allows one to build manageable toy models with higher spins, displaying simple spectra compared to string theory. Another remarkable example of interacting, but non-unitary, gauge theory breaking the spin-two barrier is provided by conformal higher-spin gravity \cite{Fradkin:1985am, Segal:2002gd}.} Our results may help in clarifying the structure of the latter and foster the development of new models. Moreover, in Minkowski space free fields of any spin admit infinite-dimensional asymptotic symmetries \cite{Campoleoni:2017mbt, Campoleoni:2020ejn}. In gravity the same mechanism eventually leads to the Bondi-Metzner-Sachs (BMS) symmetry of asymptotically-flat spaces and thus suggests the existence of BMS-like higher-spin algebras. These should contain our ``rigid'' symmetries as subalgebras (or wedge algebras).\footnote{While this paper was in preparation, we became aware of a proposal to define directly BMS-like higher-spin algebras as quotients of the universal enveloping algebra of $\mfk{bms}_D$  \cite{talk_quarks}.} Similar ``higher-spin supertranslations'' also emerged in the analysis of the near-horizon symmetries of black holes \cite{Grumiller:2019fmp}.

Aside from classifying the quotients of the universal enveloping algebra of \mbox{$\iso(1,D-1)$} with the correct set of generators, we also show how the resulting algebras can be recovered as In\"on\"u-Wigner contractions of AdS$_D$ higher-spin algebras. Recalling that $\so(2,D-1)$ describes both the isometries of AdS$_D$ and the algebra of conformal transformations in $D-1$ dimensions, these contractions can also be interpreted as ultra-relativistic, namely Carrollian, limits \cite{Duval:2014uva} of the symmetries of conformal higher-spin gravity (see, e.g., \cite{Fradkin:1985am, Segal:2002gd} and the more recent \cite{Basile:2018eac, Grigoriev:2019xmp}).

We begin our analysis by revisiting the construction of flat-space higher-spin algebras in three dimensions. In this case, the Poincar\'e algebra is also isomorphic to the global conformal Galilean algebra in two dimensions, named $\gca_{2}$ \cite{Bagchi:2009my, Bagchi:2009pe, Bagchi:2010zz}. In a holographic setup, three-dimensional flat-space higher-spin algebras can thus be interpreted either as ultra-relativistic limits of conformal higher-spin algebras in two dimensions or as non-relativistic limits thereof.
Given that our extensions of the Poincar\'e algebra for $D \geq 4$ can be interpreted as Carrollian contractions of conformal higher-spin symmetries, it is natural to ask whether interesting Galilean contractions exist too. With this motivation in mind, we also classify the quotients of the universal enveloping algebra of $\gca_{D-1}$ that are defined on the same vector space as the Lie algebras discussed before.

Another peculiarity of three dimensions is the existence of one-parameter families of higher-spin algebras that admit consistent finite-dimensional truncations. 
The same property holds in five dimensions and, for this reason, we shall discuss separately this case. Our analysis of three-dimensional global symmetries also discloses a further way of realising higher-spin algebras as a subspace of the Killing tensors of the vacuum equipped with a suitable generalisation of the Lie bracket. Similar Lie algebras exist in any space-time dimensions and we close the paper by discussing how they could be interpreted as the global symmetries of certain gauge theories with a wide and, alas, non-unitary spectrum.

The paper is organised as follows. In section~\ref{sec:relativistic} we review the coset construction of higher-spin algebras in AdS backgrounds of any dimensions. Section~\ref{sec:3D} focuses on the known higher-spin algebras in three-dimensional Minkowski space, that we use as guiding examples to develop the strategy to tackle the case of arbitrary $D \geq 4$. In particular, we show how these Lie algebras can be obtained as quotients of the universal enveloping algebra of $\iso(1,2)$. Section~\ref{sec:carrollian} then presents flat-space or, equivalently, Carrollian-conformal higher-spin algebras in any dimensions along the lines of section~\ref{sec:3D-flat}, with a separate survey of the five-dimensional case. Section~\ref{sec:galilean} mirrors the previous analysis for the Galilean limit of conformal higher-spin algebras. Finally, section \ref{sec:other-carroll} generalises section~\ref{sec:3D-other}: it exhibits additional higher-spin extensions of the Poincar\'e algebra defined on the space of Killing tensors of Minkowski space and proposes to interpret them as the global symmetries of certain non-unitary gauge theories.\\

\section{AdS$_D$ higher-spin algebras} \label{sec:relativistic}

In this section we review the construction of higher-spin algebras in AdS$_\dim$ as quotients of the universal enveloping algebra of $\so(2,\dim-1)$. We first recall how one can identify the right set of generators from the study of vacuum-preserving gauge symmetries. We then introduce the Lie algebras, mainly following \cite{Boulanger:2011se, Joung:2014qya, Joung:2015jza}. Even if in this paper we focus on massless fields, we include a discussion of the global symmetries of partially-massless fields because they display structures that we shall also encounter in flat space.

\subsection{Higher-spin ``isometries'' of the vacuum}\label{sec:isometries}

A free massless particle of spin $s$ propagating on a constant-curvature spacetime of any dimension can be described, e.g., using a symmetric tensor of rank $s$ admitting gauge transformations of the form
\be \label{free-gauge-fronsdal}
\d \vf_{\m_1 \cdots \m_s} = s\, \bar{\nabla}_{\!(\m_1} \e_{\m_2 \cdots \m_s)} \, ,
\quad \textrm{with} \
\e_{\m_1 \cdots \m_{s-3}\l}{}^\l = 0 \, .
\ee
This is the field content of Fronsdal's formulation of the dynamics \cite{Fronsdal:1978rb, Fronsdal:1978vb}, where $\bar{\nabla}$ denotes the background covariant derivative and indices enclosed between parentheses are symmetrised with weight one (i.e.\ dividing by the number of terms used in the symmetrisation is understood). To write an action principle, one also has to impose that the field be doubly traceless, but this does not bring any new condition on the gauge parameter. Indeed, \eqref{free-gauge-fronsdal} is doubly traceless thanks to the trace constraint on $\e$.

Interactions typically bring deformations $\cO(\vf)$ of the free gauge transformation \eqref{free-gauge-fronsdal}. Still, preserving the vacuum solution $\vf_{\m_1 \cdots \m_s} = 0$ only requires that the gauge parameters be Killing tensors, satisfying
\be \label{killing}
\bar{\nabla}_{\!(\m_1} \e_{\m_2 \cdots \m_s)} = 0 \, .
\ee
Gauge transformations generated by traceless Killing tensors can thus be interpreted as global symmetries for particles of any spin. In Minkowski space, the general solution of the Killing equation \eqref{killing} for traceless tensors takes the simple form
\be \label{sol-killing}
\e_{\m_1 \cdots \m_{s-1}} = \sum_{k=0}^{s-1} \cM_{\m_1 \cdots \m_{s-1} | \n_1 \cdots \n_{k}} x^{\n_1} \cdots x^{\n_k} \, ,
\ee
where the $\cM$'s are constant $\so(1,D-1)$-irreducible tensors  \cite{Bekaert:2005ka}. Indices in both sets are manifestly symmetrised, while irreducibility means that these tensors are fully traceless and satisfy 
\be
\cM_{(\m_1 \cdots \m_{s-1} | \m_s)\n_1 \cdots \n_{k}} = 0 \, , \quad 0 \leq k \leq s-2 \, .
\ee
In (A)dS the number of independent solutions of the Killing equation is the same \cite{Thompson:1986} and one can characterise them using the same set of tensors as in flat space \cite{Barnich:2005bn}. As it is manifest if one solves \eqref{killing} using ambient-space techniques, the $\cM$'s can be collected into a single two-row irreducible $\so(2,D-1)$ tensor $M_{A_1 \cdots A_{s-1} | B_1 \cdots B_{s-1}}$ \cite{McLenaghan:2004, Barnich:2005bn}.\footnote{Alternatively, one can describe a particle of spin $s$ using a set of one-forms that can be collected in a field $W^{A_1 \cdots A_{s-1}|B_{1} \cdots B_{s-1}}$ transforming irreducibly under $\so(2,D-1)$ (or $\so(1,D)$) and admitting gauge transformations 
$\d W^{A_1 \cdots A_{s-1}|B_{1} \cdots B_{s-1}} = d \e^{A_1 \cdots A_{s-1}|B_{1} \cdots B_{s-1}} + \cdots$ \cite{Vasiliev:1986td}. In this approach, global symmetries are manifestly spanned by two-row irreducible tensors in $D+1$ dimensions. This is an example of a general fact: global symmetries should not depend on the particular formulation of the dynamics one chooses.} Its indices $A_n$ and $B_n$ are symmetrised and take values in the range $\left\{ 0, \ldots, D \right\}$, while the tensor is traceless and satisfies
\be
M_{(A_1 \cdots A_{s-1}|B_1)B_2 \cdots B_{s-1}} = 0 \, .
\ee

The traceless solutions of the Killing equation form a vector space and one can define a Lie bracket on it, for instance, by looking at the field-independent part of the commutator of two gauge transformations generated by traceless Killing tensors (see, e.g., \cite{Bekaert:2006us, Boulanger:2013zza, Joung:2013nma}). In general relativity one can perform this task directly at the full non-linear level by computing the commutator of two infinitesimal diffeomorphisms generated by Killing vectors. Equivalently, one can isolate a background metric with the split $g_{\m\n} = \bar{g}_{\m\n} + h_{\m\n}$ and consider the commutator of two transformations
\be
\d_\e h_{\m\n} = 2\, \bar{\nabla}_{\!(\m}\,\e_{\n)} + 2\, \bar{\G}^\a{}_{\m\n}(h) \e_\a + \cO(h^2) \, ,
\ee
where $\bar{\G}^\a{}_{\m\n}$ denotes the linearised Christoffel symbols. This suffices to recover the algebra of isometries of the vacuum from
\be
[ \d_{\e_1} , \d_{\e_2} ] h_{\m\n} = 2\, \bar{\nabla}_{\!(\m}\,[\e_1 , \e_2 ]_{\n)} + \cO(h) \, ,
\ee
where $[\e_1 , \e_2 ]$ is the Lie bracket of the two vectors. In a similar fashion, in higher-spin theories one can approach the problem perturbatively in a weak-field expansion. In both (A)dS and Minkowski spaces one soon realises the need to introduce an infinite number of fields with increasing spin in order to close the algebra and in AdS one recovers in this way the higher-spin algebras entering Vasiliev's equations \cite{Boulanger:2013zza, Joung:2013nma}. In flat space, on the contrary, using the deformations of the free gauge transformations induced by the covariant and local cubic vertices classified in \cite{Manvelyan:2010jr, Sagnotti:2010at} one cannot find any Lie algebra at all \cite{Bekaert:2010hp, Joung:2013nma}. However, these negative results do not exclude completely the option of defining a Lie bracket for traceless Killing tensors: they only imply that the resulting global symmetries may be realised in an interacting theory not admitting either a covariant description or a perturbatively-local expansion.\footnote{The no-go results of \cite{Bekaert:2010hp, Joung:2013nma} depend on the choice of the spectrum: covariant and perturbatively-local interacting theories in flat space may still be obtained, for instance, including mixed-symmetry fields \cite{Karapetyan:2021wdc}.} Indeed, in section~\ref{sec:carrollian} we shall provide examples of such Lie algebras and discuss their properties.

In both Fronsdal's and Vasiliev's descriptions of the free dynamics, a single particle of spin $s$ is considered at the time \cite{Fronsdal:1978vb, Vasiliev:1986td}. On the other hand, there exist other free gauge theories describing more involved spectra. A prototypical example is the string-inspired triplet system \cite{Bengtsson:1986ys, Henneaux:1988, Francia:2002pt, Sagnotti:2003qa}, that accounts for a multiplet of particles of spin $s$, $s-2$ and so on, up to spin $0$ or $1$. In its ``Maxwell-like'' formulation \cite{Campoleoni:2012th, Francia:2016weg}, this multiplet is described by a single, symmetric and traceful tensor of rank $s$ admitting gauge transformations
\be \label{free-triplet}
\d \vf_{\m_1 \cdots \m_s} = s\, \bar{\nabla}_{\!(\m_1} \e_{\m_2 \cdots \m_s)} + \cO(\vf)
\quad \textrm{with} \quad
\bar{\nabla}\cdot \e_{\m_1 \cdots \m_{s-2}} = \cO(\vf) \, .
\ee
Therefore, in this context global symmetries are still described by particular solutions of the Killing equation \eqref{killing}, now involving traceful but divergenceless tensors (traceless solutions of the Killing equation must also be divergenceless, while the contrary is not true). In Minkowski space, the global symmetries of the triplet thus still take the form \eqref{sol-killing}, but only traces of the $\cM$'s involving both the first and second set of indices or only the second vanish \cite{Barnich:2005bn}. Indeed, in flat space the divergence constraint also implies $\Box\,\e_{\m_1 \cdots \m_{s-1}} = 0$ and
\begin{subequations}
\begin{align}
\pr\cdot \e_{\m_1 \cdots \m_{s-2}} & = \sum_{k=0}^{s-1} k\, \cM_{\m_1 \cdots \m_{s-2}\l | \n_1 \cdots \n_{k-1}}{}^\l\, x^{\n_1} \cdots x^{\n_{k-1}} \, , \label{divergence} \\
\Box\,\e_{\m_1 \cdots \m_{s-1}} & = \sum_{k=0}^{s-1} \binom{k}{2} \cM_{\m_1 \cdots \m_{s-1} | \n_1 \cdots \n_{k-2}\l}{}^\l\, x^{\n_1} \cdots x^{\n_{k-2}} \, . \label{trace}
\end{align}
\end{subequations}
One can eventually rearrange the independent components in the tensors $M_{A_1 \cdots A_{s-1} | B_1 \cdots B_{s-1}}$, $M_{A_1 \cdots A_{s-3} | B_1 \cdots B_{s-3}}$, etc.,\footnote{The branching of the sum of the $\so(D+1)$ tensors $M_{A_1 \cdots A_{n} | B_1 \cdots B_{n}}$ into $\so(D)$ components tallies with the branching of the sum of the tensors $\cM_{\m_1 \cdots \m_{s-1}|\n_1 \cdots \n_k}$ into their fully-traceless, $\so(D)$, components. Notice also that, thanks to the conditions \eqref{divergence} and \eqref{trace}, the problem manifestly coincides with the analysis of the global symmetries of a Fierz system involving traceful fields.} so that the vector space of global symmetries corresponds to the direct sum of those associated with the individual particles contained in the reducible spectrum. This is not always the case though: in section \ref{sec:other-carroll} we shall discuss other examples with reducible, but non-unitary, spectra that lead to global symmetries with a different structure even in flat space. Partially-massless fields, discussed in section~\ref{sec:PM}, actually provide the simplest example of global symmetries that cannot be interpreted in terms of those of massless fields with given helicity.

\subsection{Global symmetries for massless fields}\label{sec:global}

We now focus on the global symmetries of gauge theories involving only Fronsdal's fields in AdS$_\dim$ and on their construction as quotients of the universal enveloping algebra (UEA) of $\so(2,D-1)$. For arbitrary values of the space-time dimension, there exists a unique quotient of the UEA of the isometries of the vacuum that gives a vector space appropriate to describe the global symmetries of massless fields. When $\dim = 3$ and $\dim = 5$ one can instead obtain one-parameter families of non-isomorphic higher-spin algebras. We discuss these two cases in detail and, in particular, we bridge the gap between the customary presentation of three-dimensional higher-spin algebras and that applying to any dimension.

\subsubsection{Spacetime of generic dimension: the $\hs_D$ algebra}\label{sec:global_anyD}

The isometries of the AdS$_\dim$ background are given by the $\so(2,D-1)$ algebra
\be \label{g-commutators}
[J_{AB} \,,\, J_{CD}] = \tilde \h_{AC}\, J_{BD} - \tilde \h_{BC}\, J_{AD} - \tilde \h_{AD}\, J_{BC} + \tilde \h_{BD}\, J_{AC}\,,
\ee
where the $J_{AB}$ are antisymmetric tensors with $A, B \in \left\{ 0, \ldots, \dim \right\}$, while $\tilde{\h}_{AB}$ denotes the matrix $ \text{diag}(-,+,\ldots,+,-)$ that we shall also employ to raise and lower indices. For simplicity, we shall often use the shorthand $\mfk g \equiv \so(2,D-1)$.

The Eastwood-Vasiliev higher-spin algebra $\hs_D$ is a coset algebra obtained by quotienting the UEA $\cU(\mfk g)$ by the two-sided ideal
\be
\langle \cI \rangle \equiv \cU(\mfk g) \,\star\, \cI \,\star\, \cU(\mfk g) \, ,
\ee
where $\star$ denotes the associative product on the UEA (that we shall omit in the following), while $\cI$ will be specified below. We recall that $\cU(\mfk g)$ is obtained by considering tensor products of the $J_{AB}$ modulo the relation \eqref{g-commutators} and, thanks to the Poincar\'e-Birkhoff-Witt theorem, a basis of representatives is given by symmetrised products of the generators (for more details and further references see, e.g., \cite{Boulanger:2011se, Joung:2014qya}). Symmetrised products of the $J_{AB}$ are typically reducible under permutations of their indices. For instance, one can decompose a generic quadratic symmetrised product 
\be
J_{AB} \odot J_{CD} \equiv \frac12 \left\{ J_{AB} , J_{CD} \right\} = \frac12 \left( J_{AB} J_{CD} + J_{CD} J_{AB} \right) \,,
\ee
into\footnote{In general, the product of two antisymmetric tensors would contain also a ``hook'' Young diagram with a single box in the second column. This is however absent in the commutative product of two identical tensors as a result of the symmetry under exchanges of the factors.}
\be \label{products}
J_{A(B} \odot J_{C)D} \simeq\, {\footnotesize \yng(2,2)} \ , \qquad
J_{[AB} \odot J_{CD]} \simeq\, {\footnotesize \yng(1,1,1,1)} \ , 
\ee
where Young diagrams characterise here $\mfk{gl}(D+1)$ irreducible components, while curly (resp.\ square) brackets indicate a symmetrisation (resp.\ antisymmetrisation) over the indices with the same conventions as in \eqref{free-gauge-fronsdal}.

According to the discussion in section~\ref{sec:isometries}, the traceless part of the first combination in \eqref{products} has the right structure to describe the global symmetries of a massless spin-three field, while its trace and the fully antisymmetric product do not fit into the vector space of global symmetries of any Fronsdal's field. As a result, one has to eliminate them from the set of generators and this is achieved by building the ideal out of a  $\cI = \cI_{AB} \oplus \cI_{ABCD}$ comprising the following two parts:
\be \label{rel_anyD:ideal}
\cI_{AB} \equiv J_{C(A} \odot {J_{B)}}^C - \frac{2}{D+1}\, \tilde\h_{AB}\, C_2 \simeq\, {\footnotesize \yng(2)} \ , \qquad \cI_{ABCD} \equiv J_{[AB} \odot J_{CD]} \simeq\, {\footnotesize \yng(1,1,1,1)} \ ,
\ee
where, differently from \eqref{products}, here and in the rest of this section Young diagrams identify $\mfk{o}(D+1)$ irreducible components, e.g.\ $\h^{AB} \cI_{AB} = 0$, while $C_2$ is the quadratic Casimir
\be \label{C2}
C_2 \equiv \frac{1}{2}\, J_{AB} \odot J^{BA} \, .
\ee
Notice that one could substitute the $\odot$ in eqs.~\eqref{rel_anyD:ideal} and \eqref{C2} with a simple tensor product because of the projections and contractions entering these expressions. When this is the case, we shall often omit the symmetrised product altogether. The first part of the ideal $\langle\cI\rangle$ then factors out all traces from products of $\so(2,D-1)$ generators, while the second part factors out products associated with Young diagrams containing more than two rows. 

In order to define the higher-spin algebra as the quotient 
\be
\hs_D = \cU(\so(2,D-1))/ \langle \cI \rangle \,,
\ee
the quadratic Casimir, as well as the higher-order Casimir operators, must also be fixed \cite{Boulanger:2011se}. For instance, 
\be \label{I-C2}
0 \sim \frac{3}{2}\, \cI_{ABCD} J^{CD} - \cI_{C[A} \,J_{B]}{}^C = \frac{1-D}{D+1} \left( C_2 + \frac{(D+1)(D-3)}{4}\,id\right) J_{AB}  \, ,
\ee
where $\sim$ means that the identification is taking place in the quotient. We wish that the combination between parentheses on the right-hand side vanish to avoid that $\langle \cI \rangle$ coincide with the whole $\cU(\mfk g)$ and this fixes the value of the quadratic Casimir.
Assuming that $C_2$ is proportional to the identity, the same result can also be inferred from
\begin{align}
0 & \sim \frac12\, \cI_{AB} \cI^{AB} = C_4 - \frac{2}{D+1}\, (C_2)^2 - \frac{(D-1)^2}{4}\, C_2 \,,\label{I_AB^2} \\
0 & \sim -\frac34\, \cI_{ABCD} \cI^{ABCD} = C_4 - (C_2)^2 - \frac{(D-2)(D-1)}{2}\, C_2 \,, \label{I_ABCD^2}
\end{align}
with
\be \label{C4}
C_4 \equiv \frac12\, J_{AB} J^{BC} J_{CD} J^{DA} \,.
\ee
Factoring out the ideal \eqref{rel_anyD:ideal} thus implies two polynomial equations linking the quartic Casimir $C_4$ to the quadratic one $C_2$. In their turn, consistently with \eqref{I-C2}, they imply
\be \label{rel_anyD:Casimir}
C_2 \sim - \frac{(D+1)(D-3)}{4}\,id \,,
\ee
which corresponds to the value of the quadratic Casimir in the scalar singleton representation.
Higher-order polynomial equations then fix the higher-order Casimir operators in a similar way \cite{Boulanger:2011se}, while still being consistent with the value of $C_2$. As noticed in \cite{Vasiliev:1999ba, Iazeolla:2008ix}, the ideal $\cI$ indeed corresponds to the annihilator of the scalar singleton representation, and one can think at the Eastwood-Vasiliev algebra as the UEA of $\so(2,D-1)$ evaluated on the scalar singleton module. The scalar singleton is the minimal representation of the isometry algebra, so that higher-spin algebras can also be viewed as the quotient of $\cU(\mfk g)$ by its Joseph ideal \cite{Joung:2014qya}.

In any $D \geq 3$ the resulting algebra decomposes under the adjoint action of $\so(2,D-1)$ into irreducible representations associated with the following two-row (traceless) Young diagrams:\footnote{The case $D=2$ is degenerate, but its relation with the generic-$D$ construction was studied in \cite{Alkalaev:2014qpa}.}
\be \label{EV-spectrum}
\bullet \,\oplus\, \yng(1,1) \,\oplus\, \yng(2,2) \,\oplus\, \yng(3,3) \,\oplus\, \cdots \,.
\ee
The Young diagram with no rows at all (denoted by $\bullet$) represents the identity element in $\cU(\mfk g)$ and corresponds to the $U(1)$ generator of gauge transformations for a field of spin one.
The second element corresponds to the ``usual'' isometries of AdS$_D$ (Lorentz transformations and translations) and we shall loosely refer to it in the following as the spin-two sector. The other Young diagrams correspond to the global symmetries of Fronsdal's fields of spin three, four, etc. Colour indices can also be included in higher-spin algebras \cite{Konstein:1989ij}, but in this work we focus on the minimal spectrum resulting from the UEA construction.

\subsubsection{Three dimensions: the $\hs_3[\l]$ family} \label{sec:rel_3D}

The case of three space-time dimensions is quite peculiar because of the isomorphism $\so(2,2) \simeq \so(1,2) \oplus \so(1,2) \simeq \sl(2,\mathbb R) \oplus \sl(2,\mathbb R)$ meaning that the isometries of the vacuum are not described anymore by a simple Lie algebra. To make this manifest, one can begin by decomposing the generators of $\so(2,2)$ as
\be
m_{ab} \equiv J_{ab} \,, \quad p_a \equiv J_{a3} \,,
\ee
where $a,b \in \{0,1,2\}$ and $m_{ab}$ can be dualised into
\be \label{dual-j}
j^a \equiv \frac{1}{2}\, \ve^{abc} m_{bc} \quad \Leftrightarrow \quad m_{ab} = - \ve_{abc} j^c
\ee
using the Levi-Civita tensor. For the latter we adopt the convention $\ve^{012} = 1$ and we use the metric $\h = (-,+,+)$ to raise or lower indices. In this basis the commutation relations read
\be
[j_a, j_b] = \ve_{abc} j^c \,, \quad
[j_a, p_b] = \ve_{abc} p^c \,, \quad
[p_a, p_b] = \ve_{abc} j^c \,. 
\ee
One can then rearrange the components of $p_a$ and $j_a$ into the generators
\be \label{p->P}
P_m \equiv \left(p_0 - p_1,\, p_2,\, p_0 + p_1 \right) \,, \quad L_m \equiv \left(j_0 - j_1,\, j_2,\, j_0 + j_1 \right) ,
\ee
where $m \in \{-1,0,1\}$. This gives the algebra
\be \label{comm:lorentz}
[L_m,L_n] = (m-n)L_{m+n} \,, \quad
[L_m,P_n] = (m-n)P_{m+n} \,, \quad
[P_m,P_n] = (m-n)L_{m+n} \,.
\ee
One can eventually identify the two orthogonal copies of $\sl(2,\mathbb R)$  by introducing the linear combinations
\be
\cL_m = \frac{1}{2} \left(L_m + P_m \right) , \quad \bar \cL_m = \frac{1}{2} \left(L_m - P_m \right) ,
\ee
verifying
\be \label{sl2}
[\cL_m , \cL_n ] = (m-n) \cL_{m+n} \,, \quad
[\bar \cL_m , \bar \cL_n ] = (m-n) \bar \cL_{m+n} \,, \quad
[\cL_m , \bar \cL_n ] = 0 \,.
\ee
Notice that the change of basis leading from \eqref{comm:lorentz} to \eqref{sl2} is not unique because of the automorphisms $\cL_m \to -\cL_{-m}$ and $\bar\cL_m \to -\bar\cL_{-m}$. The definition we gave is usually employed in the study of non-relativistic limits of $\so(2,2)$, while using the automorphisms one can recover that employed in the study of ultra-relativistic limits (see section~\ref{sec:3D}). 
The quadratic Casimir operator of $\so(2,2)$ introduced in \eqref{C2} is then expressed in terms of the Casimir operators of the two $\sl(2,\mathbb R)$ algebras as
\be \label{C2_3D}
C_2 = L^2 + P^2 = 2 \left(\cL^2 + \bar \cL^2 \right) ,
\ee
where we used the inverse of the $\sl(2,\mathbb R)$ Killing metric $\g_{mn}$ to contract indices ---~e.g., $L^2 = \g^{mn} L_m L_n$ and $\cL^2 = \g^{mn} \cL_m \cL_n$~--- with the convention
\be \label{sl2-killing}
\g_{mn} =
\begin{pmatrix}
0 & 0 & -2 \\
0 & 1 & 0 \\
-2 & 0 & 0 
\end{pmatrix} .
\ee

We now move to the construction of higher-spin algebras, following the approach of section~\ref{sec:global_anyD}. The element $\cI_{AB}$ defined in eq.~\eqref{rel_anyD:ideal} must be factored out from the UEA even when $D=3$ in order to eliminate the traces of all two-row Young diagrams appearing in the decomposition of products of $\so(2,2)$ generators. It carries nine independent components, that can be conveniently presented as follows
\begin{subequations}
\begin{align}
\cI_{ab} - \h_{ab}\, \cI_{33} & = p_{(a} p_{b)} - j_{(a} j_{b)} \sim 0 \,, \\[3pt]
- 2\, \cI^a{}_3 & = \ve^{abc}\left(p_b j_c + j_c p_b\right) \sim 0 \,,
\end{align}
\end{subequations}
where $a,b \in \{ 0,1,2 \}$ and we recall that $\sim$ means that the identification is taking place in the quotient. Being traceful, the first expression contains six independent components, while the second gives the remaining three. After performing the change of basis \eqref{p->P}, we can express those relations as
\begin{subequations}
\begin{align}
\label{rel_3D:ideal_1} \{P_m, P_n\} - \{L_m, L_n\} &\sim 0 \,, \\[3pt]
\label{rel_3D:ideal_2} \{P_m, L_n\} - \{L_m, P_n\} &\sim 0 \,,
\end{align}
\end{subequations}
where we can take $m \geq n$ in the first expression and we can consider the last expression, e.g., only for $m > n$. When written in terms of the $\cL$ and $\bar \cL$ generators, they turn into
\begin{subequations}\label{3D_ideal_LbarL}
\begin{align}
\{\cL_m, \bar \cL_n\} + \{\cL_n, \bar \cL_m\} &\sim 0 \,, \\[3pt]
\{\cL_m, \bar \cL_n\} - \{\cL_n, \bar \cL_m\} &\sim 0 \,,
\end{align}
\end{subequations}
where the first expression is again considered for $m \geq n$ and the last expression for $m > n$. By combining the above relations, one obtains
\be \label{rel_3D:LLbar}
\cI_{AB} \sim 0 \quad \Rightarrow \quad \cL_m \bar \cL_n \sim 0 \,,
\ee
with no restrictions on $m,n$ and where we omitted the anticommutator since the two copies of $\sl(2,\mathbb R)$ commute. This explains why, in the construction of higher-spin algebras in three dimensions, it is enough to consider the direct sum of two copies of the UEA of $\sl(2,\mathbb R)$ (see, e.g., \cite{Blencowe:1988gj} and the review \cite{Gaberdiel:2012uj}): factoring out $\cI_{AB}$ sets all mixed products to zero.\footnote{An alternative construction, in which mixed products of the $\cL$ and $\bar{\cL}$ generators are allowed, leads to an extended algebra dubbed ``large AdS higher-spin algebra'' \cite{Ammon:2020fxs}. Similar extended algebras also appear in the description of partially-massless fields in three dimensions \cite{Grigoriev:2020lzu}.}

On the other hand, contrary to the generic case, in three dimensions we do not need to factor out the element $\cI_{ABCD}$. One can dualise it into a singlet, that fits in the vector space of global symmetries of massless particles (as an additional $U(1)$ factor besides the identity). Actually, its dualisation gives the other independent quadratic Casimir of $\so(2,2)$:
\be \label{W_3D}
W \equiv \frac{1}{8}
\,\ve^{ABCD} \cI_{ABCD} = j^a p_a = \gamma^{mn} L_m P_n = \cL^2 - \bar \cL^2 \,.
\ee
Factoring out $\cI_{AB}$ then implies a constraint on $W$ via eq.~\eqref{I_AB^2}, that fixes $C_4$ as a function of $C_2$ only. Substituting the result in the expansion of $W^2 = 4!\, \cI_{ABCD}\cI^{ABCD}$ one obtains
\be \label{rel_3D:W}
W^2 \sim \frac{1}{4}\, (C_2)^2 \, .
\ee
Imposing $\cI_{ABCD} = \frac13\, \ve_{ABCD} W \sim 0$ as in section~\ref{sec:global_anyD} thus implies $C_2 \sim 0$ consistently with the $D$-dimensional result \eqref{rel_anyD:Casimir}. However, when $D=3$, we have the option to work only with the weaker condition \eqref{rel_3D:W} that leaves the quadratic Casimir $C_2$ free. For the latter, one can eventually require
\be \label{rel_3D:C2}
C_2 \sim \frac{\l^2-1}{2}\, id \,,
\ee
in order to remove multiplicities from the spectrum, that is to identify, e.g., the elements $J_{AB}$ and $C_2 J_{AB}$ in the UEA. This condition also guarantees that products of $W$ with other elements in the UEA do not introduce new generators since the relation \eqref{W_3D} implies
\be
W \cL_m \sim \frac{1}{2}\,C_2\,\cL_m \, , \quad
W \bar{\cL}_m \sim -\,\frac{1}{2}\,C_2\,\bar{\cL}_m \, .
\ee
Eventually, eq.~\eqref{rel_3D:C2} leads to the one-parameter family of higher-spin algebras that has been considered in the literature on massless fields in three dimensions \cite{Bergshoeff:1989ns, Bordemann:1989zi, Fradkin:1990ir, Gaberdiel:2012uj}. 
The conditions \eqref{rel_3D:LLbar} and \eqref{rel_3D:C2} imply that these algebras are obtained by evaluating the UEA of $\so(2,2)$ on a reducible module built upon two $\sl(2,\mathbb R)$ irreps with highest-weight vectors
\be
\left| h,\bar{h} \right\rangle_{\so(2,2)} = \left| h \right\rangle_{\sl(2,\mathbb R)} \, \otimes \, \left| \bar h \right\rangle_{\sl(2,\mathbb R)}
\ee
with conformal dimensions $h = \bar h = h_{\pm} = \frac{1}{2}\left( 1 \pm \l \right)$ (or from their conjugate representations) \cite{Gaberdiel:2011wb}. According to the decompositions \eqref{C2_3D} and \eqref{W_3D}, on a generic vector in this module the Casimir operators $C_2 = 2\,(\cL^2 + \bar \cL^2) = \frac{\l^2-1}{2}\,(\mathds{1} \otimes \bar 0 + 0 \otimes \bar{\mathds{1}})$ and $W = \cL^2 - \bar \cL^2 = \frac{\l^2-1}{4}\,(\mathds{1} \otimes \bar 0 - 0 \otimes \bar{\mathds{1}})$ act as
\begin{subequations}
\begin{align}
\frac12\, C_2 \left| h,\bar{h} \right\rangle &= \left(\cL^2 \left| h \right\rangle \right) \, \otimes \, \left(\bar \cL^2 \left| \bar h \right\rangle \right) = \frac{\l^2-1}{4} \left| h,\bar h \right \rangle ,\\
W \left| h,\bar{h} \right\rangle &= \left(\cL^2 \left| h \right\rangle\right) \, \otimes \, \left(-\bar \cL^2 \left| \bar h \right\rangle \right) = \frac{\l^2-1}{4}\, \h\, \left|h,\bar h\right\rangle ,
\end{align}
\end{subequations}
where we introduced the twist operator $\h$ flipping the sign of one copy of $\sl(2,\mathbb R)$ while leaving the other untouched. Of course, $\h^2 = id$ which respects \eqref{rel_3D:W}. 
This leads to the presentation of the one-parameter family of higher-spin algebras as
\be
\hs_3[\l] = id \,\oplus\, \h \,\oplus\, \hs[\l] \oplus \hs[\l] \,.
\ee
The algebra $\hs[\l]$ is defined as
\be \label{hs[lambda]_def}
\mathds{1} \oplus \hs[\l] = \frac{\cU(\sl(2,\mathbb R))}{\left\langle \cC_2 - \frac{\l^2-1}{4}\, \mathds{1} \right\rangle} \, ,
\ee
where $\cC_2$ denotes the $\sl(2,\mathbb R)$ Casimir operator (say $\cL^2$ or $\bar\cL^2$). When $\l = N \in \mathbb{N}$ its eigenvalue corresponds to that of a finite-dimensional irreducible representation and a further infinite-dimensional ideal appears. Factoring it out leads to the $\sl(N,\mathbb R)$ algebra, that can thus be interpreted as a higher-spin algebra involving a finite number of fields with spin $2 \leq s \leq N$.

It is worth revisiting the previous construction in the finite-dimensional case, where it becomes particularly neat. The absence of mixed products of $\cL$ and $\bar{\cL}$ means that one has to consider $\so(2,2)$ representations of the form 
\be
\cL_m = \begin{pmatrix} l_m & 0 \\ 0 & 0 \end{pmatrix} , \quad \bar \cL_m = \begin{pmatrix} 0 & 0 \\ 0 & \bar{l}_m \end{pmatrix} ,
\ee
where, a priori, $l_m$ and $\bar{l}_m$ might be two different finite-dimensional representations of the $\sl(2,\mathbb R)$ algebra. If this is the case, however, $C_2$ is not a multiple of the identity (which is possible because the module is not irreducible). To fulfil the condition \eqref{rel_3D:C2} one thus has to work with the same $\sl(2,\mathbb R)$ representation in both blocks, $l_m = \bar{l}_m$. Choosing that of dimension $N$, the Casimir operators take the form
\be
C_2 = \frac{N^2-1}{2} \begin{pmatrix} \mathds{1} & 0 \\ 0 & \mathds{1} \end{pmatrix} , \quad W = \frac{N^2-1}{4} \begin{pmatrix} \mathds{1} & 0 \\ 0 & - \mathds{1} \end{pmatrix} ,
\ee
where $W$ manifestly satisfies the relation \eqref{rel_3D:W} imposed by $\cI_{AB} \sim 0$.

\subsubsection{Five dimensions: the $\hs_5[\l]$ family}\label{sec:5D_rel}

The five-dimensional case is also special compared to the generic one. Indeed, it is possible to evaluate the UEA on a module different from the scalar singleton, while keeping the same set of generators: this leads to a one-parameter family of higher-spin algebras \cite{Boulanger:2011se, Joung:2014qya} (see also \cite{Fernando:2009fq, Boulanger:2013zza}). This is most conveniently seen at the level of the complex algebra, where one can take the quotient of the UEA of $D_3$ by a one-parameter ideal $\langle\cI^\l\rangle$:
\be
\hs_5[\l] = \cU(D_3) / \langle \cI^\l \rangle \,,
\ee
where $\cI^\l = \cI_{AB} \oplus \cI_{ABCD}^\l$ with
\begin{align}
\cI_{AB} &\equiv J_{C(A} {J_{B)}}^C - \frac{1}{3}\,\tilde\h_{AB} \, C_2 \,, \label{5D-ideal_1}\\
\cI_{ABCD}^\l &\equiv J_{[AB} J_{CD]} - i\, \frac{\l}{6}\, \ve_{ABCDEF} J^{EF} \,. \label{5D-ideal_2}
\end{align}
As in three dimensions, one still has to factorise $\cI_{AB}$ in order to eliminate the traces of products of the $J_{AB}$ associated with two-row Young diagrams. The second part of the ideal involves instead the dualisation of four-row Young diagrams into two-row Young diagrams. 
These steps fix the value of the higher-order Casimir operators as functions of the quadratic one, as is the case in generic dimensions \cite{Boulanger:2011se}. On the other hand, the value of the quadratic Casimir is not fixed in the same way as \eqref{rel_anyD:Casimir} since now
\be
0 \sim - \frac{3}{4}\, \cI^\l_{ABCD}\, \cI^{\l\,ABCD} = C_4 - (C_2)^2 + 2( \l^ 2 -3 )\, C_2 \,,
\ee
and this relation combined with \eqref{I_AB^2} implies that
\be \label{C2_5D}
C_2 \sim 3(\l^2-1)\,id \,.
\ee

When $\l = \frac{M}{2}+1$ with $M \geq 1$ the higher-spin algebra develops an infinite-dimensional ideal which, upon further factorisation, gives a finite-dimensional truncation describing only fields of spin $1 \leq s \leq M+1$.\footnote{The case $M=2$ was studied in detail in \cite{Manvelyan:2013oua} and used to couple fields of spin two and three via a five-dimensional Chern-Simons action. Finite-dimensional higher-spin algebras can thus lead to interesting topological systems, while they are not expected to emerge in gauge theories describing the unitary propagation of higher-spin particles, since they do not satisfy the so-called admissibility condition \cite{Boulanger:2013zza}.} The existence of a one-parameter family of higher-spin algebras is also explained by the isomorphism $\so(2,4) \simeq \mfk{su}(2,2)$ ($\simeq \sl(4,\mathbb R)$ up to signature). All $\sl(N,\mathbb R)$ algebras admit indeed a one-parameter family of minimal representations, as explained in Appendix~\ref{app:finite-dim_irreps_5D}. Quotienting their UEA by the associated Joseph ideal gives ``higher-spin algebras'' labelled by a free real parameter $\l$, with finite-dimensional truncations occurring for $N(\l\pm 1) \in -2\, \mathbb N$ \cite{Joung:2014qya}. The cases $N=2$ and $N=4$ are isomorphic to $\so(d)$ algebras with, respectively, $d=3$ and $d=6$. Recalling that the construction of section~\ref{sec:global_anyD} can be interpreted as the quotient of the UEA of $\so(2,D-1)$ by its Joseph ideal, this confirms the existence of one-parameter families of higher-spin algebras in two, three (where one considers a double copy of the $N=2$ case) and five dimensions.

\subsection{Global symmetries for partially-massless fields: the $\cA_D[\m]$ family}\label{sec:PM}

The simplest example of global symmetries whose generators cannot be presented as a sum of two-row rectangular Young diagrams in $D+1$ dimensions is given by partially-massless fields, that are defined on backgrounds with a non-vanishing cosmological constant \cite{Deser:1983mm, Higuchi:1986wu, Deser:2001us}. They describe irreducible short representations of the isometry groups $SO(1,D)$ or $SO(2,D-1)$ (which are unitary in the first case and non-unitary in the second) by means of the Fierz system
\begin{subequations} \label{Fierz-PM}
\begin{align}
& \left[\Box + \frac{(D+s-2) - (s-t-1)(D+s-t-4)}{L^2}\right] \vf_{\m_1 \cdots \m_s} = 0 \, , \\
& \bar{\nabla}\cdot \vf_{\m_1 \cdots \m_{s-1}} = 0 \, , \\[7pt]
& \vf_{\m_1 \cdots \m_{s-2}\l}{}^\l = 0 \, .
\end{align}
\end{subequations}
These equations of motion are left invariant by gauge transformations of the form
\be \label{gauge-PM}
\d \vf_{\m_1 \cdots \m_s} = \frac{s!}{(s-t-1)!}\, \bar{\nabla}_{\!(\m_1} \cdots \bar{\nabla}_{\!\m_{t+1}} \e_{\m_{t+2} \cdots \m_s)} + \cO(L^{-2}) \, ,
\ee
with
\begin{subequations} 
\begin{align}
& \left[ \Box + \frac{(D-t-3)-s(D+s-4)}{L^2} \right] \e_{\m_1 \cdots \m_{s-t-1}} = 0 \, , \\
& \bar{\nabla}\cdot \e_{\m_1 \cdots \m_{s-t-2}} = 0 \, , \\[7pt]
& \e_{\m_1 \cdots \m_{s-t-3}\l}{}^\l = 0 \, .
\end{align}
\end{subequations}
The parameter $t$ is called depth and belongs to the range $0 \leq t \leq s-1$ ($t=0$ describes the massless case), while $L$ is the AdS radius, related to the cosmological constant by $\L = - \frac{(D-1)(D-2)}{2L^2}$. Even if in the flat limit a partially-massless field of depth $t$ decomposes into a collection of massless fields of spin $s, s-1,\ldots, s-t$ \cite{Zinoviev:2001dt}, the solutions of the higher-order Killing equation $\d \vf_{\m_1 \cdots \m_s} = 0$ are characterised by $\mfk{o}(D+1)$ irreducible tensors associated with Young diagrams with two rows of length $s-1$ and $s-t-1$ \cite{Skvortsov:2006at, Joung:2015jza}. One can grasp the structure of these higher-order Killing tensors by looking at the flat-space limit of the Killing equation, that is
\be \label{killing-PM}
\pr_{(\m_1} \cdots \pr_{\m_{t+1}} \e_{\m_{t+2} \cdots \m_s)} = 0 \, ,
\ee
with
\be \label{constraints-PM}
\Box\,\e_{\m_1 \cdots \m_{s-t-1}} = 0 \, , \qquad
\partial\cdot \e_{\m_1 \cdots \m_{s-t-2}} = 0 \, , \qquad
\e_{\m_1 \cdots \m_{s-t-3}\l}{}^\l = 0 \, .
\ee
One can indeed check that the general solution takes the form
\be \label{sol-killing-PM}
\e_{\m_1 \cdots \m_{s-t-1}} = \sum_{k=0}^{s-1} \cM_{\m_1 \cdots \m_{s-t-1} | \n_1 \cdots \n_{k}} x^{\n_1} \cdots x^{\n_k} \, ,
\ee
where the tensors $\cM_{\m_1 \cdots \m_{s-t-1} | \n_1 \cdots \n_{k}}$ are fully traceless in order to fulfil the constraints \eqref{constraints-PM} but, in this case, they are in general not irreducible. They indeed have to satisfy only the constraint
\be
\cM_{(\m_1 \cdots \m_{s-t-1} | \m_{s-t} \cdots \m_s)\n_1 \cdots \n_{k-t-1}} = 0 \qquad \textrm{for}\ k > t \, .
\ee
If one further decomposes the resulting tensors in irreducible components, one reconstructs the branching of a representation of $\mfk{o}(D+1)$ associated with a Young diagram with two rows of length $s-1$ and $s-t-1$ in terms of irreducible representations of $\mfk{o}(D)$. One thus recovers the solution obtained with ambient-space techniques in \cite{Joung:2015jza}.

Having recalled the structure of the symmetry generators associated with a partially-massless field, we can now review the proposal for a global symmetry algebra including some of them (together with a $\so(2,D-1)$ subalgebra) \cite{Joung:2015jza}. This construction is based, once again, on a suitable quotient of the UEA of $\so(2,D-1)$:
\be \label{PM-algebra}
\cA_D[\m] \equiv \cU\!\left(\so(2,D-1)\right)/\left\langle \cI_{ABCD} \oplus \left( C_2 - \n[\m]\,id  \right) \right\rangle \,,
\ee
where $\cI_{ABCD}$ is the antisymmetric product of two $\so(2,D-1)$ generators introduced in eq.~\eqref{rel_anyD:ideal} and where we parameterise the eigenvalue of the quadratic Casimir as
\be \label{eigenvalue_PM}
\n[\m] = - \frac{(D-1+2\m)(D-1-2\m)}{4} \,.
\ee
Compared to the massless case, here one does not factorise the symmetric traceless diagram $\cI_{AB} \sim \scriptsize \yng(2)$. Decomposing the algebra resulting from the factorisation of $\cI_{ABCD}$ under the adjoint action of its $\so(2,D-1)$ subalgebra, one still finds the two-row traceless Young diagrams of the massless case, together with traces. The former can be identified with the global symmetries of a tower of massless fields, while the latter can be divided into two classes. Those obtained by taking traces only in the second row are identified with the global symmetries of a tower of partially massless fields with depth $t = 2, 4, \ldots$, while the others (i.e.\ those involving at least a trace in both the first and the second row) contain the Casimir operators since $\h^{A[B} \h^{C]D} J_{AB} J_{CD} = C_2$. They are thus identified with generators already included in the spectrum on account of the identification $C_2 \sim \n[\m]\,id$. Looking at the decomposition of the first few products of the generators one obtains, for instance,
\be \label{spectrum-PM}
\cA_D[\m] \simeq \bullet \,\oplus\ \yng(1,1) \,\oplus \left(\, \yng(2,2) \oplus \yng(2) \,\right) \oplus \left(\, \yng(3,3) \oplus \yng(3,1) \,\right) \oplus\, \cdots \,.
\ee
The terms displayed explicitly correspond to massless fields of spin $1$, $2$, $3$ and $4$, together with two partially-massless field of depth $2$, one of spin $3$ and one of spin $4$.

As was noticed in \cite{AIF_2014__64_4_1581_0, Joung:2015jza}, for any integer value of $\m$ the algebra $\cA_D[\m]$ develops an infinite-dimensional ideal comprising fields whose depth is greater or equal to $2\,\m$. Quotienting it out  gives the algebra of higher symmetries of the higher-order singleton \mbox{$\Box^\m\,\phi = 0$} \cite{Bekaert:2013zya, gover2012higher, Basile:2014wua}. In this setup, the Eastwood-Vasiliev algebra can be recovered by setting the parameter to the ``singleton point'' $\m = 1$, which leads to an algebra containing only diagrams of zero depth, thus corresponding to massless fields. Furthermore, it should be noted that $\cA_D[\m]$ admits finite-dimensional truncations for $\m = \frac{D-1}{2} + k$ with $k \geq 0$ an integer, whose spectrum involves only diagrams with at most $2\,k$ boxes \cite{Joung:2015jza}.\\

\section{Three space-time dimensions} \label{sec:3D}

In this section we show how the one-parameter family of flat-space higher-spin algebras $\ihs_3[\l]$ defined in \cite{Ammon:2017vwt} can be obtained as a quotient of the UEA of the isometries of Minkowski space (see also \cite{Ammon:2020fxs} for similar considerations). We also show that one can recover the algebra obtained in the limit $\l \to \infty$ by equipping the vector space of traceless Killing tensors of Minkowski spacetime with a suitable generalisation of the Lie bracket.

\subsection{Carrollian and Galilean contractions} \label{sec:3D-flat}

The algebras $\ihs_3[\l]$ are obtained from the In\"on\"u-Wigner contraction of the AdS$_3$ higher-spin algebras $\hs[\l] \,\oplus\, \hs[\l] $ that we introduced in section~\ref{sec:rel_3D}. We denote the generators of the two copies of $\hs[\l]$ by  $\cL^{(s)}_m$ and $\bar \cL^{(s)}_m$, with $s \geq 2$ and $-s+1 \leq m \leq s-1$. Generators of a given spin $s$ can be realised as products of $s-1$ generators $\cL_m$ (or $\bar{\cL}_m$) of $\sl(2,\mathbb R)$ (with the conventions \eqref{sl2}) and correspond to the independent vectors in an irreducible representation of $\sl(2,\mathbb R)$ of dimension $2s-1$. To build them, one can start from
\be \label{highest-weight}
\cL^{(s)}_{\pm (s-1)} \equiv \left(\cL_{\pm 1}\right)^{s-1}
\ee
and obtain all other components of $\cL^{(s)}_m$ by acting with the operators $\cL_{1}$ or $\cL_{-1}$ as
\be \label{L^(s)}
\cL^{(s)}_{m \mp 1} \equiv \frac{\mp 1}{s \pm m -1}\, [\, \cL_{\mp 1} , \cL^{(s)}_{m} \,] \,,
\ee
while taking into account the constraint on the $\sl(2,\mathbb R)$ quadratic Casimir introduced in \eqref{hs[lambda]_def}:
\be \label{C2_hs[lambda]}
\cC_2 \equiv \cL_0 \cL_0 - \frac12 \left(\cL_1 \cL_{-1} + \cL_{-1} \cL_1\right) \sim \frac{\l^2-1}{4}\,\mathds{1} \,.
\ee
The resulting $\hs[\l]$ algebra reads \cite{Pope:1989sr}
\be \label{hs[lambda]}
\left[ \cL_m^{(s)} \,,\, \cL_n^{(t)} \right] = \underset{s+t+u \text{ even}}{\sum_{u=|s-t|+2}^{s+t-2}} g^{st}_{s+t-u}(m,n;\l) \cL_{m+n}^{(u)} \,,
\ee
and similarly for the barred sector, with
\be \label{hs[lambda]_structure-consts}
\begin{split}
g^{st}_{u}(m,n;\l) & = \frac{q^{u-2}}{2(u-1)!}\, \f^{st}_{u}(\l) N^{st}_{u}(m,n) \\
& = \frac{N^{st}_u(n,m)}{2(u-1)!} \frac{\left(\frac{1-u}{2}\right)_r  (q\l)^{u-2}}{\left(\frac32-s\right)_r \left(\frac32-t\right)_r \left(\frac12+s+t-u\right)_r} +
\cO(\l^{u-4}) \,,
\end{split}
\ee
where $(a)_n = a(a+1) \ldots (a+n-1)$ denotes the raising Pochhammer symbol and $q = \frac14$ for any finite $\l$, while the polynomials $N_u^{st}(m,n)$ are defined in Appendix~\ref{app:hs[lambda]}.
The structure constants are polynomials in $\l$ and we showed explicitly here only their leading term, corresponding to the structure constants of the algebra of area-preserving diffeomorphisms of the two-dimensional hyperboloid \cite{Bergshoeff:1989ns}. The complete expression for $\f^{st}_u(\l)$ can be found in Appendix~\ref{app:hs[lambda]}.
These algebras satisfy a series of properties that are a general feature of the quotient construction in AdS$_D$ and that correspond to desirable physical requirements: in particular, they allow minimal coupling to gravity in the sense that $[2,s] \propto s$ and they imply a non-zero contribution of higher spins to the gravitational energy-momentum tensor, that is $[s,s] \propto 2 + \cdots$.

The In\"on\"u-Wigner contraction  leading to $\ihs_3[\l]$ can be performed by defining
\be \label{galilean_contraction_3D}
P_m^{(s)} \equiv \e \left(\cL_m^{(s)} - \bar \cL_m^{(s)} \right) , \quad L_m^{(s)} \equiv \cL_m^{(s)} + \bar \cL_m^{(s)}
\ee
and taking the limit $\e \to 0$.
In the basis \eqref{galilean_contraction_3D}, most of the structure constants obtained in the limit are the same as in the original AdS$_3$ higher-spin algebras, the main difference being that now all $P_m^{(s)}$ commute:
\begin{subequations}\label{ihs3}
\begin{align}
\left[P_m^{(s)}, P_n^{(t)}\right] & = 0 \,, \label{ihs3_trabslations}\\[5pt]
\left[L_m^{(s)}, P_n^{(t)}\right] & =  \underset{s+t+u \text{ even}}{\sum_{u=|s-t|+2}^{s+t-2}} g^{st}_{s+t-u}(m,n;\l) P_{m+n}^{(u)} \,, \label{ihs3_mixed} \\
\left[L_m^{(s)}, L_n^{(t)}\right] & =  \underset{s+t+u \text{ even}}{\sum_{u=|s-t|+2}^{s+t-2}} g^{st}_{s+t-u}(m,n;\l) L_{m+n}^{(u)} \,, \label{ihs3_lorentz}
\end{align}
\end{subequations}
where the structure constants $g^{st}_u$ are the same as in eq.~\eqref{hs[lambda]_structure-consts}.
The contracted algebra contains a Poincar\'e subalgebra generated by the $P^{(2)}_m$ and the $L^{(2)}_m$. Its Lorentz subalgebra fits into a $\hs[\l]$ subalgebra and $\ihs_3[\l]$ inherits from the AdS$_3$ higher-spin algebra several properties (spin addition rules, minimal coupling to gravity, non-degenerate bilinear form, finite-dimensional truncability).

Following the pattern that applies to any dimension $D \geq 3$, the Poincar\'e algebra $\iso(1,2)$ is isomorphic to the global two-dimensional Carrollian conformal algebra $\mfk{cca}_2$ (for more details see Appendix~\ref{app:diff}). However, in this particular case it is also isomorphic to the global two-dimensional Galilean conformal algebra $\gca_2$ \cite{Bagchi:2010zz}. Looking at a differential realisation of the generators $P_m$ and $L_m$ \cite{Bagchi:2009pe}, it is natural to interpret the limit \eqref{galilean_contraction_3D} as a non-relativistic, Galilean contraction of the conformal algebra $\so(2,2)$. A ultra-relativistic, Carrollian contraction of the conformal algebra $\so(2,2)$ is instead naturally defined by introducing \cite{Bagchi:2012cy, Campoleoni:2016vsh}
\be \label{carrollian_contraction_3D}
P_m^{(s)} \equiv \frac{1}{L} \left(\cL_m^{(s)} + \bar \cL_{-m}^{(s)} \right) \,, \quad L_m^{(s)} \equiv \cL_m^{(s)} - \bar \cL_{-m}^{(s)}
\ee
and taking the limit $L \to \infty$. The two limits are actually isomorphic thanks to the $\hs[\l]$ automorphism $\bar\cL^{(s)}_{m\phantom{-}} \!\to -\bar\cL^{(s)}_{-m}$, which boils down to the option of exchanging the time and space coordinates in a two-dimensional space. Since the contractions \eqref{galilean_contraction_3D} and \eqref{carrollian_contraction_3D} are equivalent, in the rest of this subsection we shall deal only with the former. We recall, however, that the two setups lead to different asymptotic symmetries due to the different behaviour of non-linear algebras in the two limits \cite{Campoleoni:2016vsh}.

\subsubsection{The contracted algebras as quotients of $\cU(\iso(1,2))$}\label{sec:3D_limit-from-quotient}

We are now going to reconstruct the algebra $id \oplus W \oplus \ihs_3[\l]$ ---~obtained by adding the identity and a central element $W$ satisfying $W^2 = 0$ to the algebra \eqref{ihs3}~--- as a quotient of the UEA of the three-dimensional Poincar\'e algebra $\iso(1,2)$. To this end, in analogy with eq.~\eqref{comm:lorentz}, we denote by $P_m$ and $L_m$ the generators of the latter:
\be \label{poincare}
[L_m\,,\,L_n] = (m-n)\,L_{m+n} \,, \quad [L_m\,,\,P_n] = (m-n)\,P_{m+n} \,, \quad [P_m\,,\,P_n] = 0\,,
\ee
where $m,n \in \{-1,0,1\}$. The algebras we wish to reproduce contain only two classes of higher-spin generators: the $L_m^{(s)}$ and $P_m^{(s)}$, corresponding to generalisations of Lorentz transformations and translations respectively.  We shall denote the two sets as the ``higher-Lorentz'' and ``higher-translation'' sectors and we shall analyse them separately.

\paragraph{Higher-Lorentz sector}

The higher-Lorentz generators $L^{(s)}_m$ span a $\hs[\l]$ subalgebra of $\ihs_3[\l]$ (see eq.~\eqref{ihs3_lorentz}). It is therefore natural to consider an Ansatz in which they are realised as products of Lorentz generators as in eqs.~\eqref{highest-weight} and \eqref{L^(s)}, that is to consider
\be \label{3D-flat:Ansatz_L}
L_{\pm(s-1)}^{(s)} \equiv \left(L_{\pm 1} \right)^{s-1} \,, \quad L_{m\mp 1}^{(s)} \equiv \frac{\mp 1}{s\pm m-1} \left[ L_{\mp 1} \,,\, L_m^{(s)} \right] .
\ee
For instance, for $s=3$ we find
\be \label{spin-3_L}
L_{\pm 2}^{(3)} = L_{\pm 1} L_{\pm 1} \,, \qquad
L_{\pm 1}^{(3)} = L_0 L_{\pm 1} \pm \frac{1}{2}\, L_{\pm 1} \,, \qquad
L_0^{(3)} = L_0 L_0 - \frac13\, L^2 \,.
\ee
To recover the $\hs[\l]$ algebra we also have to impose that the quadratic Casimir $L^2$ of the Lorentz subalgebra is proportional to the identity with the following $\l$ dependence:
\be \label{3D-flat:Casimir}
L^2 \equiv \g^{mn} L_m L_n = L_0 L_0 - \frac12 (L_1 L_{-1} + L_{-1} L_1) \sim \frac{\l^2 -1}{4}\,id \,,
\ee
where $\g^{mn}$ is the inverse Killing metric of $\so(1,2) \simeq \sl(2,\mathbb R)$ with the conventions of eq.~\eqref{sl2-killing}. Notice that this is a rather strong constraint since in general $L^2$ does not commute with translations: we shall check its consistency later.

\paragraph{Higher-translation sector}

The other class of generators, denoted by $P^{(s)}_m$, can be recovered from the $L^{(s)}_m$ via the adjoint action of the Poincar\'e subalgebra. Indeed, from \eqref{ihs3_mixed}, we have that
\be
\left[P_m\,,\,L_n^{(s)}\right] = \left((s-1)\,m-n\right) P_{m+n}^{(s)} \, ,
\ee
and we can use this relation to define all $P_m^{(s)}$, that will thus be linear in $P_m$. For instance, for $s=3$ we have
\be \label{spin-3_P}
P_{\pm 2}^{(3)} = L_{\pm 1} P_{\pm 1} \,, \qquad
P_{\pm 1}^{(3)} = L_0 P_{\pm 1} \pm \frac{1}{2}\, P_{\pm 1} \,, \qquad
P_0^{(3)} = L_0 P_0 - \frac{1}{3}\, W \,,
\ee
where
\be \label{Pauli-Lubanski_3D}
W \equiv \g^{mn} L_m P_n = L_0 P_0 - \frac{1}{2}\, L_1 P_{-1} - \frac{1}{2}\, L_{-1} P_1 \,,
\ee
is the three-dimensional analogue of the Pauli-Lubanski vector, which in this case is a Casimir operator.

Eqs.~\eqref{spin-3_P} can be readily generalised to arbitrary values of the spin by noticing that the adjoint action of $P_m$ is consistent with the definitions
\be \label{3D-flat:Ansatz_P}
P_{\pm(s-1)}^{(s)} \equiv \left(L_{\pm 1} \right)^{s-2}P_{\pm 1} \,, \quad P_{m\mp 1}^{(s)} \equiv \frac{\mp 1}{s\pm m-1} \left[ L_{\mp 1} \,,\, P_m^{(s)} \right] .
\ee
The expression for $P^{(s)}_{\pm(s-1)}$ follows from the adjoint action of $P_0$ on $L^{(s)}_{\pm(s-1)}$ and the position of the operator $P_{\pm 1}$ is irrelevant since $[L_{\pm 1},P_{\pm 1}] = 0$. The other components are then fixed by the known action of Lorentz transformations on the $P^{(s)}_m$. 
Still, the consistency of the whole set of relations \eqref{3D-flat:Ansatz_L}, \eqref{3D-flat:Casimir} and \eqref{3D-flat:Ansatz_P} requires some additional constraints that eventually specify the ideal one has to factor out from the UEA of $\iso(1,2)$.

\paragraph{Consistency conditions}

Let us observe from \eqref{ihs3} that
\be \label{first-consistency}
\left[P_m\,,\,L_n^{(s)}\right] = \left((s-1)\,m - n\right) P_{m+n}^{(s)} = \left[L_m\,,\,P_n^{(s)}\right]  .
\ee
These relations give rise to a first set of consistency conditions, since the two commutators must agree. We can obtain them from the analysis of the case $s=3$:
\begin{subequations}
\begin{align}
\left[P_\mp, L_{\pm 2}^{(3)}\right] &= \left[L_\mp, P_{\pm 2}^{(3)}\right] 
\quad \Rightarrow \quad
L_\pm P_0 \sim P_\pm L_0 \,,\\
\left[P_\mp, L_{\pm 1}^{(3)}\right] &= \left[L_\mp, P_{\pm 1}^{(3)}\right] 
\quad \Rightarrow \quad
L_\pm P_\mp \sim P_\pm L_\mp \,,
\end{align}
\end{subequations}
which can be summed up in
\be \label{3D-flat:consistency_1}
L_m P_n \sim P_m L_n \,.
\ee
One can verify that the remaining relations in \eqref{first-consistency} are identically satisfied: the cases with $s \geq 4$ give rise to the same conditions, multiplied on the left or the right by some elements of $\cU(\iso(1,2))$.  
Using eq.~\eqref{3D-flat:consistency_1}, one can also check that
\be \label{[L^2,P]}
\left[ L^2, P_m \right] \sim 0 \,.
\ee
Therefore, in this setup $L^2$ commutes with all elements of the Poincar\'e algebra and this confirms the consistency of the relation \eqref{3D-flat:Casimir}, in which we imposed $L^2 \propto id$.

We now have to check that the higher-translation generators previously defined form an Abelian ideal and satisfy the commutation relations \eqref{ihs3_mixed}. 
When developing $[P^{(s)}_m, P^{(t)}_n]$ any factors of $P$ can be pushed to the right thanks to \eqref{3D-flat:consistency_1} and, for $s=2$ and $t=3$, one obtains the following set of consistency conditions:
\begin{subequations}
\begin{align}
\left[P_0\,,\, P^{(3)}_{\pm 2}\right] = 0 & 
\quad \Rightarrow \quad
P_{\pm 1} P_{\pm 1} \sim 0 \,, \\
\left[P_{\mp 1}\,,\, P^{(3)}_{\pm 2}\right] = 0 &
\quad \Rightarrow \quad
P_{\pm 1} P_0 \sim 0 \,, \\
\left[P_{\mp 1}\,,\, P^{(3)}_{\pm 1}\right] = 0 &
\quad \Rightarrow \quad
P_{\pm 1} P_{\mp 1} \sim 0 \,.
\end{align}
\end{subequations}
Taking advantage of the relation \eqref{3D-flat:consistency_1} to get $P^{(3)}_{\pm 1} \sim P_0 L_{\pm 1} \pm \frac{1}{2}\,P_{\pm 1}$ from \eqref{spin-3_P}, the last commutator also gives
\begin{align}
\left[P_{\mp 1}\,,\, P^{(3)}_{\pm 1}\right] = 0 
\quad \Rightarrow \quad
P_0 P_0 \sim 0 \,.
\end{align}
In conclusion, the product of any two translation generators must vanish:
\be \label{3D-flat:consistency_2}
P_m P_n \sim 0\,.
\ee
In particular, this implies that the quadratic Casimir $P^2$ must vanish.

From the consistency conditions \eqref{3D-flat:consistency_1} and \eqref{3D-flat:consistency_2} one can also fix the action of the second quadratic Casimir $W$ (that we defined in eq.~\eqref{Pauli-Lubanski_3D}) on the generators of $\iso(1,2)$:
\begin{subequations} \label{[W,Poincare]}
\begin{align}
W L_k &= \g^{mn} L_m P_n L_k \sim \g^{mn} L_m L_n P_k = L^2 P_k \sim \frac{\l^2-1}{4}\, P_k \,, \\
W P_k &= \g^{mn} L_m P_n P_k \sim 0 \,.
\end{align}
\end{subequations}
Notice also that the relations \eqref{3D-flat:consistency_1} and \eqref{3D-flat:consistency_2} imply
\be
W^2 \sim L^2 P^2 \sim 0 \,.
\ee
The whole set of consistency conditions,
\begin{subequations} \label{ideal_3D}
\begin{align}
\cP_{mn} \equiv P_m P_n &\sim 0 \,, \label{ideal-3D_1}\\[6pt]
\cI_{m|n} \equiv L_m P_n - P_m L_n &\sim 0 \,, \label{ideal-3D_2} \\
L^2 - \frac{\l^2-1}{4}\, id &\sim 0 \,, \label{ideal-3D_3}
\end{align}
\end{subequations}
defines an ideal because
\begin{subequations}
\begin{align}
[L_k, \cI_{m|n}] &= (k-m)\, \cI_{m+k|n} + (k-n)\, \cI_{m|n+k}  \,, \\[5pt]
[L_k, \cP_{mn}] &= (k-m)\, \cP_{(m+k)n} + (k-n)\, \cP_{m(n+k)} \,,
\end{align}
\end{subequations}
which show that $\cP_{mn}$ and $\cI_{m|n}$ transform as Lorentz tensors, and
\begin{subequations}
\begin{align}
[P_k, \cI_{m|n} ] &= (k-m)\, \cP_{(m+k)n} - (k-n)\, \cP_{m(n+k)} \,, \\[5pt]
[P_k, \cP_{mn} ] &= 0 \,.
\end{align}
\end{subequations}
Furthermore, we already showed that $L^2$ is central in the quotient and in section~\ref{sec:contraction-ideal_3D} we shall see how one can recover the ideal \eqref{ideal_3D} from the flat limit of the ideal introduced in the quotient construction of higher-spin algebra in AdS$_3$. In conclusion, the one-parameter family of flat-space higher-spin algebras can be obtained as
\be
id \oplus W \oplus \ihs_3[\l] = \cU(\iso(1,2)) / \left\langle \cI \right\rangle ,
\ee
where the ideal $\cI$ is defined by the relations \eqref{ideal_3D}.

\paragraph{Commutation relations and finite-dimensional truncations}

In fixing the form of the ideal \eqref{ideal_3D}, we already checked the commutators of all higher-spin generators with the Poincar\'e subalgebra. Moreover, we already know that the $P^{(s)}_m$ form an Abelian ideal and that the $L^{(s)}_m$ form a $\hs[\l]$ subalgebra. This information suffices to conclude that the commutators $[L^{(s)}_m,P^{(t)}_n]$ take the form \eqref{ihs3_mixed} via the Jacobi identities. First of all, the commutators under scrutiny belong to the higher-translation sector because all $P^{(s)}_m$ are linear in $P_m$. The form \eqref{3D-flat:Ansatz_P} of the $P^{(s)}_m$ also makes manifest that they must take the form
\be \label{hs[lambda]_2}
\left[ L_m^{(s)} \,,\, P_n^{(t)} \right] = \underset{s+t+u \text{ even}}{\sum_{u=|s-t|+2}^{s+t-2}} f^{st}_{s+t-u}(m,n;\l) P_{m+n}^{(u)} \,,
\ee
in the sense that both the extrema of the sum and the parity rule follow from our Ansatz.
The identities
\begin{subequations}
\begin{align}
[L_k,[L^{(s)}_m,P^{(t)}_n]] + [L^{(s)}_m,[P^{(t)}_n,L_k]] + [P^{(t)}_n,[L_k,L^{(s)}_m]] & = 0 \, , \\
[L_k,[L^{(s)}_m,L^{(t)}_n]] + [L^{(s)}_m,[L^{(t)}_n,L_k]] + [L^{(t)}_n,[L_k,L^{(s)}_m]] & = 0 \, , \label{[L,L,L]}
\end{align}
\end{subequations}
then give rise to the same constraints on the functions $f$ and on the $\hs[\l]$ structure constants. The latter are the unique solution of \eqref{[L,L,L]} when all $s \geq 2$ are included \cite{Bergshoeff:1989ns, Bordemann:1989zi} and this allows one to conclude that the $f$'s must correspond to the $\hs[\l]$ structure constants too. The structure constants in the commutators $[L^{(s)}_m,L^{(t)}_n]$ and $[L^{(s)}_m,P^{(t)}_n]$ being the same, one can also conclude that $\ihs_3[\l]$ admits finite-dimensional truncations, which can also be recovered directly from those of the AdS$_3$ algebras via the contraction \eqref{galilean_contraction_3D}.

One can also check explicitly a set of commutators that suffice to fix the form of the whole $\ihs_3[\l]$ algebra (again via the Jacobi identities, see \cite{Gaberdiel:2012ku}). For instance, let us introduce
\be \label{3D-flat:P4}
P_{1}^{(4)} = L_0 L_0 P_1 + L_0 P_1 - \frac{\l^2 - 9}{20}\, P_1
\ee
and the corresponding $L_{1}^{(4)}$ obtained by replacing everywhere $P_m$ with $L_m$ (we prove in Appendix~\ref{app:hs[lambda]} that this rule holds for any couple $P^{(s)}_m$, $L^{(s)}_m$ of generators with the same quantum numbers). Combining these expressions with those for the spin-three generators given in eqs.~\eqref{spin-3_L} and \eqref{spin-3_P}, we obtain the $\hs[\l]$ commutator
\be \label{comm1}
\left[L_{2}^{(3)}\,,\,L_{-1}^{(3)}\right] = 6\, L_{1}^{(4)} - \frac{\l^2-4}{5}\,L_{1}^{(2)} \,,
\ee
together with
\be \label{comm2}
\left[P_{2}^{(3)}\,,\,L_{-1}^{(3)}\right] = \left[L_{2}^{(3)}\,,\,P_{-1}^{(3)}\right] = 6\, P_{1}^{(4)} - \frac{\l^2 - 4}{5}\, P_{1}^{(2)} \,,
\ee
which show precisely the same structure constants. Similarly, introducing
\begin{subequations}
\begin{align}
\label{3D-flat:P4_3} P_{3}^{(4)} & = L_1 L_1 P_1 \,, \\
\label{3D-flat:P5} P_{2}^{(5)} & = L_0 L_0 L_1 P_1 + 2\, L_0 L_1 P_1 - \frac{\l^2 - 37}{28}\, L_1 P_1 \,,
\end{align}
\end{subequations}
and the generators $L^{(4)}_3$ and $L^{(5)}_2$ obtained by substituting all $P_m$ with the corresponding $L_m$, one obtains
\be
\left[P_{-1}^{(3)}\,,\,L_{3}^{(4)}\right] = \left[L_{-1}^{(3)}\,,\,P_{3}^{(4)}\right] = -9\, P_{2}^{(5)} + \frac{3(\l^2-9)}{7}\, P_{2}^{(3)} \,, \label{comm3}
\ee
and $[L_{-1}^{(3)}\,,\,L_{3}^{(4)}]$ takes the same form with all $P$'s substituted by the corresponding $L$'s. 
As we anticipated, the remaining structure constants can be fixed from these inputs via the Jacobi identities \cite{Gaberdiel:2012ku}.\footnote{For simplicity we displayed only some commutators with fixed values of the axial quantum numbers $m$ and $n$, since our main goal was to fix the $\l$ dependence of the structure constants. The full $m,n$ dependence can be reconstructed via the adjoint action of the Poincar\'e subalgebra.} 

\paragraph{Finite-dimensional matrix representation} \label{3D-flat:matrix}

For $\l = N \in \mathbb N$, the family of flat-space algebras admits finite-dimensional truncations of the form $\ihs_3[N] \simeq \mfk{isl}(N,\mathbb R) \simeq \sl(N,\mathbb R) \ \mathbin{\rotatebox[origin=c]{-90}{$\uplus$}} \ \sl(N,\mathbb R)_\text{Ab}$. In this particular case, one can recover the previous construction by evaluating the UEA of $\iso(1,2)$ on the following finite-dimensional representation of the Poincar\'e algebra:
\be \label{irrep_3D}
L_m =
\begin{pmatrix}
l_m & 0 \\
0 & l_m
\end{pmatrix}
\,, \quad
P_m =
\begin{pmatrix}
0 & l_m \\
0 & 0
\end{pmatrix}
\,,
\ee
where the $l_m$ are $N \times N$ matrices giving an irreducible representation of $\so(1,2) \simeq \sl(2,\mathbb R)$. Thanks to the upper-triangular form of the generators $P_m$, the conditions \eqref{ideal-3D_1} and \eqref{ideal-3D_2} in the definition of the ideal are clearly satisfied. Moreover,
\be
L^2 = \begin{pmatrix}
l^2 & 0 \\
0 & l^2
\end{pmatrix}
= \frac{N^2-1}{4} \begin{pmatrix}
\mathds{1} & 0 \\
0 & \mathds{1}
\end{pmatrix}
, \quad
W = \begin{pmatrix}
0 & l^2 \\
0 & 0
\end{pmatrix}
= \frac{N^2-1}{4} \begin{pmatrix}
0 & \mathds{1} \\
0 & 0
\end{pmatrix}
,
\ee
so that the conditions \eqref{[W,Poincare]} and \eqref{ideal-3D_3} are satisfied too. Notice that $W$ is manifestly a central element, but it is not proportional to the identity: this is consistent with the structure of the representation \eqref{irrep_3D}, which is indecomposable but not irreducible. 

The semi-direct structure is realised by simple block $2 \times 2$ matrix multiplication, so that the whole set of generators takes the form
\be
L^{(s)}_m =
\begin{pmatrix}
l^{(s)}_m & 0 \\
0 & l^{(s)}_m
\end{pmatrix}
, \quad
P^{(s)}_m =
\begin{pmatrix}
0 & l^{(s)}_m \\
0 & 0
\end{pmatrix}
,
\ee
where the $l^{(s)}_m$ are the generators of $\sl(N,\mathbb R)$ in the defining representation. Let us stress that the upper-triangular form of the generators with $\so(1,2)$ representations on the diagonal blocks is in agreement with general results on the structure of finite-dimensional indecomposable representations of the Poincar\'e algebra \cite{AIHPA_1984__40_1_35_0}.

\subsubsection{From $\cU(\so(2,2))$ to $\cU(\iso(1,2))$}\label{sec:contraction-ideal_3D}

The previous discussion suffices to recover the algebra $\ihs_3[\l]$ as a quotient of the UEA of $\iso(1,2)$. Our goal in the following sections will be to generalise this construction to arbitrary space-time dimensions and, to this end, it is convenient to track how the results of section~\ref{sec:3D_limit-from-quotient} can be obtained also by  analysing the effects of the limit \eqref{galilean_contraction_3D} on the AdS$_3$ coset construction discussed in section~\ref{sec:rel_3D}.

To begin with, let us consider the explicit rewriting of the $P^{(s)}_m$ and $L^{(s)}_m$ in terms of Poincar\'e generators induced by \eqref{galilean_contraction_3D}. To understand how to handle the limit $\e \to 0$, it is enough to consider the highest-weight components for $s=3$ and $s=4$. Considering that \eqref{galilean_contraction_3D} also implies $P_m \to \e^{-1} P_m$, one obtains
\begin{subequations} \label{examples-L,P}
\begin{alignat}{5}
L_{2}^{(3)} &= \frac{1}{2} \left(L_1 L_1 + \e^{-2} \, P_1 P_1\right) , & \qquad 
P_{2}^{(3)} &= \e \left(\e^{-1} \, L_1 P_1 \right) , \\
L_{3}^{(4)} &= \frac{1}{4} \left(L_1 L_1 L_1 + 3\, \e^{-2} \, L_1 P_1 P_1 \right) , & \qquad
P_{3}^{(4)} &= \frac{\e}{4} \left(3\, \e^{-1}\,L_1 L_1 P_1 + \e^{-3} \, P_1 P_1 P_1 \right) .
\end{alignat}
\end{subequations}
Notice that there are divergent terms in the limit $\e \to 0$. Still, we can obtain sensible expressions by eliminating them using the ideal that we quotiented out in section \ref{sec:rel_3D} before taking the limit. By performing the same rescaling $P_m \to \e^{-1} P_m$ on the elements of the AdS$_3$ ideal, defined in eqs.~\eqref{rel_3D:ideal_1}, \eqref{rel_3D:ideal_2} and \eqref{rel_3D:C2}, one indeed obtains
\begin{subequations} \label{3D-rel:ideal}
\begin{align}
\label{3D-rel:ideal_1} \e^{-2} P_m P_n - L_m L_n &\sim 0\,, \\[5pt]
\label{3D-rel:ideal_2} \e^{-1}\left( L_m P_n - P_m L_n \right) &\sim 0 \,, \\
\label{3D-rel:ideal_3} \frac12\left(L^2 + \e^{-2} P^2\right) - \frac{\l^2-1}{4}\,id &\sim 0 \,,
\end{align}
\end{subequations}
where we also used the $\so(2,2)$ commutation relations \eqref{comm:lorentz} to rewrite these expressions without anticommutators.
One can eliminate, for instance, the divergent term $\e^{-2} P_1 P_1$ in the expression for $L_2^{(3)}$ by using the substitution rule $\e^{-2} P_1 P_1 \sim L_1 L_1$ obtained from \eqref{3D-rel:ideal_1}. The same substitution can be used in the other divergent expressions in eq.~\eqref{examples-L,P} to recover the highest-weight generators introduced in eqs.~\eqref{3D-flat:Ansatz_L} and \eqref{3D-flat:Ansatz_P}. One can then get the other $P^{(s)}_m$ and $L^{(s)}_m$ with the same value of the spin via the adjoint action of the Lorentz subalgebra and similar considerations apply to higher values of $s$.

Using the AdS$_3$ ideal thus allows one to recover our Ansatz \eqref{3D-flat:Ansatz_L} and \eqref{3D-flat:Ansatz_P} even without the need to take any limit. On the other hand, multiplying each relation in \eqref{3D-rel:ideal} by the suitable positive power of $\e$, one can also obtain a finite $\e \to 0$ limit for them:
\begin{subequations}
\begin{align}
\label{3D-flat:ideal_1} P_m P_n &\sim 0  \,,\\
\label{3D-flat:ideal_2} L_m P_n - P_m L_n &\sim 0 \,.
\end{align}
Substituting the trace of eq. \eqref{3D-rel:ideal_1} into eq. \eqref{3D-rel:ideal_3} also gives
\be
\label{3D-flat:ideal_3} L^2 \sim \frac{\l^2-1}{4}\, id \,,
\ee
\end{subequations}
thus recovering all consistency conditions listed in section~\ref{sec:3D_limit-from-quotient}.

\subsection{Another Carrollian contraction} \label{sec:3D-other}

We now show that the algebra obtained by considering the limit $\l \to \infty$ of  $\ihs_3[\l]$ can also be recovered as a subalgebra of a bigger coset algebra constructed from  $\cU(\iso(1,2))$. We also show that $\ihs_3[\infty]$ coincides with the space of Minkowski traceless Killing tensors equipped with the Schouten bracket. The latter construction thus gives a natural interpretation of this algebra as a flat-space contraction of an AdS$_3$ higher-spin algebra or, equivalently, as a Carrollian contraction of a $2D$ conformal higher-spin algebra. This analysis will be extended to generic space-time dimensions in section~\ref{sec:other-carroll}, while it is rather orthogonal to the developments in sections~\ref{sec:carrollian} and \ref{sec:galilean}. It may thus be skipped by readers more interested in higher-spin algebras with the same spectrum as the Eastwood-Vasiliev ones.

\subsubsection{Alternative Ansatz for higher translations}\label{sec:3D-other_ansatz}

Since the $P^{(s)}_m$ form an Abelian ideal, instead of starting from the Ansatz \eqref{3D-flat:Ansatz_L} one may try to recover the algebra \eqref{ihs3} by building the $P^{(s)}_m$ as products of translation generators.
In this way one can easily obtain an Abelian subalgebra transforming correctly under the adjoint action of $\so(1,2)$, but introducing higher-Lorentz generators eventually leads to the algebras introduced in \cite{Ammon:2020fxs}. These have a wider set of generators compared to $\ihs_3[\l]$, but we shall show that they contain a subalgebra corresponding to its $\l \to \infty$ limit.

\paragraph{Higher-spin generators}

We now consider the following Ansatz for the higher-translation sector, that automatically gives the correct adjoint action for the Lorentz subalgebra:
\be \label{3D-other:Ansatz_P}
P_{\pm(s-1)}^{(s)} \equiv \left(P_{\pm 1}\right)^{s-1} \,, \quad P_{m\mp 1}^{(s)} \equiv \frac{\mp 1}{s\pm m-1} \left[ L_{\mp 1} \,,\, P_m^{(s)} \right] .
\ee
The commutators
\be
\left[P_m \,,\, L_n^{(s)} \right] = ((s-1)\,m-n) P_{m+n}^{(s)} \,,
\ee
are then satisfied by defining
\be \label{3D-other:Ansatz_L}
L_{\pm(s-1)}^{(s)} \equiv (s-1)\left(P_{\pm 1} \right)^{s-2}L_{\pm 1} \,,\quad L_{m\mp 1}^{(s)} \equiv \frac{\mp 1}{s\pm m-1} \left[L_{\mp 1}\,,\,L_m^{(s)}\right] .
\ee
For instance, for $s=3$, one obtains
\be
P_{\pm 2}^{(3)} \equiv P_{\pm 1} P_{\pm 1} \,, \qquad
P_{\pm 1}^{(3)} \equiv P_0 P_{\pm 1} \,, \qquad
P_0^{(3)} \equiv P_0 P_0 - \frac{1}{3}\, P^2 \,,
\ee
where $P^2 \equiv \g^{mn} P_m P_n$ is the $\iso(1,2)$ quadratic Casimir operator, and
\be
L_{\pm 2}^{(3)} \equiv 2 P_{\pm 1} L_{\pm 1} \,, \qquad
L_{\pm 1}^{(3)} \equiv P_{\pm 1} L_0 + P_0 L_{\pm 1} \,, \qquad
L_0^{(3)} \equiv 2 P_0 L_0 - \frac{2}{3}\,W \,,
\ee
where $W$ is the other $\iso(1,2)$ quadratic Casimir operator that we introduced in eq.~\eqref{Pauli-Lubanski_3D}. We also impose that $P^2$ is proportional to the identity.

\paragraph{Commutation relations}

With the definition \eqref{3D-other:Ansatz_P} and \eqref{3D-other:Ansatz_L} the $P^{(s)}_m$ clearly form an Abelian ideal since, schematically,
\be
\left[ L^{(s)} \,,\, P^{(t)} \right] \propto \left [P^{s-2} L \,,\, P^{t-1} \right] \propto P^{s+t-3} + \text{traces} \propto P^{(s+t-2)} + P^{(s+t-4)} +\cdots \,.
\ee
The structure of the commutators $[L^{(s)}_m,L^{(t)}_n]$ requires instead a more careful analysis.

We are now going to identify the algebra obtained with the previous construction by computing a set of commutators that suffices to ``bootstrap'' the whole algebra via the Jacobi identities, in analogy with what we did in eqs.~\eqref{comm1}, \eqref{comm2} and \eqref{comm3}. In this case the relevant generators take the form 
\begin{subequations}
\begin{align}
P_{1}^{(4)} & = P_0 P_0 P_1 - \frac{1}{5}\, P^2 P_1 \,,\\
L_{1}^{(4)} & = 2\, P_0 P_1 L_0 + P_0 P_0 L_1 - \frac{1}{5}\, P^2 L_1 - \frac{2}{5}\, W P_1 \,,\\
P_{2}^{(5)} & = P_0 P_0 P_1 P_1 - \frac{1}{7}\,P^2 P_1 P_1 \,.
\end{align}
\end{subequations}
We then obtain the commutators
\begin{subequations}\label{3D-other:commutators}
\begin{align}
\label{3D-other:commutator_1} \left[L_{2}^{(3)}\,,\, P_{-1}^{(3)} \right] &= 6\, P_{1}^{(4)} - \frac{4}{5}\, P^2 P_{1}^{(2)}\,,\\
\label{3D-other:commutator_2} \left[L_{2}^{(3)}\,,\, L_{-1}^{(3)} \right] &= 6\, L_{1}^{(4)} - \frac{4}{5}\, P^2 L_{1}^{(2)} - \frac{8}{5}\, W P_{1}^{(2)}\,,\\
\label{3D-other:commutator_3} \left[L_{-1}^{(3)}\,,\, P_{3}^{(4)} \right] &= -9\, P_{2}^{(5)} + \frac{12}{7}\, P^2 P_{2}^{(3)}\,.
\end{align}
\end{subequations}
From \eqref{3D-other:commutator_2}, we observe that the $L^{(s)}_m$ form a subalgebra if
\be \label{W=0}
W \sim 0 \, .
\ee
Imposing this condition, one can verify that the structure constants in \eqref{3D-other:commutators} coincide with those obtained by performing in $\ihs_3[\l]$ the spin-dependent rescaling
\be
P^{(s)}_m \to \l^{s-2} P^{(s)}_m \,, \quad L^{(s)}_m \to \l^{s-2} L^{(s)}_m \,,
\ee
and sending $\l \to \infty$ \cite{Campoleoni:2014tfa}, provided that one additionally sets 
\be
P^2 \sim \frac14 \,id \, .
\ee
Notice that the value of $P^2$ is actually irrelevant in the limit $\l \to \infty$. This is manifest if one recalls that all terms entering, e.g., the commutators \eqref{3D-other:commutators} contain the same number of $P_m$'s thanks to the Ansatz \eqref{3D-other:Ansatz_P} and \eqref{3D-other:Ansatz_L}. In other terms, for any value of $P^2>0$ we are left with an algebra isomorphic to $\ihs_3[\infty]$.

\paragraph{Impossibility of expressing the result as a quotient of $\cU(\iso(1,2))$}

The algebra spanned by the generators \eqref{3D-other:Ansatz_P} and \eqref{3D-other:Ansatz_L} (that is the algebra of products of generators of $\iso(1,2)$ including at most one $L_m$) closes, but if one tries to recover it as a quotient of $\cU(\iso(1,2))$, one has to face a contradiction. Indeed, the ideal to be factored out should contain the products $\{L_m, L_n\}$ ($[L_m , L_n]$ is already fixed by the commutation relations). However, we cannot identify $\{ L_m, L_n\} \sim 0$ because the entire algebra would be trivial, as can be seen, for instance, from the commutator
\begin{equation}
\left[L_1 L_1, P_0\right] = 2\, L_1 P_1 = L_{+2}^{(3)} \,.
\end{equation}
On the other hand, one can identify $\ihs_3[\infty]$ as a subalgebra of the much bigger algebras $\cU(\iso(1,2))/\langle W \oplus (P^2 - \n\,id) \rangle$. In the notation of \cite{Ammon:2020fxs}, it thus corresponds to the ``right slice'' of the algebras $\ihs(\cM^2 = \n,\cS = 0)$.

\paragraph{Recovering $\ihs(\cM^2 , 0)$  from $\cU(\so(2,2))$}

Let us notice that the generators defined through eqs.~\eqref{3D-other:Ansatz_P} and \eqref{3D-other:Ansatz_L} and their algebra can be recovered from the following spin-dependent contraction of the AdS$_3$ higher-spin generators
\be \label{other-contraction_3D}
P_m^{(s)} \equiv \frac{(2\e)^{s-1}}{2} \left(\cL_m^{(s)} + \bar \cL_{-m}^{(s)} \right) , \quad L_m^{(s)} \equiv (2\e)^{s-2} \left(\cL_m^{(s)} - \bar \cL_{-m}^{(s)} \right) ,
\ee
combined with the rescaling $\l \to \e^{-1} \l$, therefore effectively sending $\l \to \infty$ when $\e \to 0$.

For any finite value of the parameter $\e$, the first highest-weight generators read
\begin{subequations}\label{3D-rel:generators}
\begin{alignat}{5}
L_{2}^{(3)} &= 2\,\e \left(\e^{-1}\,L_1 P_1 \right) , & \qquad
P_{2}^{(3)} &= \e^2 \left(L_1 L_1 + \e^{-2}\,P_1 P_1 \right) , \\
L_{3}^{(4)} &= \e^2 \left(L_1 L_1 L_1 + 3 \,\e^{-2}\,L_1 P_1 P_1 \right) , & \qquad
P_{3}^{(4)} &= \e^3\, \left(3\, \e^{-1}\,L_1 L_1 P_1 + \e^{-3}\,P_1 P_1 P_1 \right) , 
\end{alignat}
\end{subequations}
while from the rescaling of eq.~\eqref{rel_3D:C2} one obtains
\be \label{3D-other:ideal} 
L^2 + \e^{-2} P^2  - \frac{1}{2} \left( \e^{-2} \l^2 - 1 \right) id \sim 0 \,.
\ee
Notice that, thanks to the $s$-dependent rescalings, there are no divergent terms in the expressions \eqref{3D-rel:generators}, even without the need to invoke any additional relations in the UEA as in section~\ref{sec:contraction-ideal_3D}. 
Taking the (smooth) limit $\e \to 0$, one recovers the same expressions for the higher-spin generators as in \eqref{3D-other:Ansatz_P} and \eqref{3D-other:Ansatz_L}, while rescaling the combination \eqref{3D-other:ideal} by $\e^2$ in order to get a smooth limit one also obtains $P^2 \propto id$.

To guarantee that the $L^{(s)}_m$ form a subalgebra, one further has to add the condition $W \sim 0$ that can be imposed even before taking the contraction limit. Since, as recalled in \eqref{W_3D}, $W = \frac{1}{8} \,\ve^{ABCD} \cI_{ABCD}$, this is equivalent to factor out the combination $\cI_{ABCD} \equiv J_{[AB}\,J_{CD]}$. As a result, the algebra $\ihs(\cM^2, 0)$  can be obtained as the In\"on\"u-Wigner contraction of $ \cU(\so(2,2))/\langle \cI_{ABCD} \oplus (C_2-\n\,id) \rangle$. This is nothing but the three-dimensional instance of the family of global algebras for partially-massless fields reviewed in section \ref{sec:PM}. As discussed above, as far as the subalgebra generated by the $P^{(s)}_m$ and $L^{(s)}_m$ is concerned, the value of $\cM^2$ is actually irrelevant after the limit is taken since we can always rescale it. In three dimensions one can also leave $W$ free: being a singlet, it is not necessary to quotient it out as we discussed in section~\ref{sec:rel_3D}. Moreover, in the partially-massless case its square is not constrained by eq.~\eqref{rel_3D:W} because one does not impose $\cI_{AB} \sim 0$. The resulting algebras, in both three-dimensional AdS and Minkowski backgrounds, were considered in \cite{Ammon:2020fxs}.

\subsubsection{Geometric realisation}\label{sec:3D-geometry}

As recalled in section~\ref{sec:isometries}, the global symmetries of Fronsdal's fields are generated by traceless Killing tensors of the background metric. For Killing vectors, one can use the Lie bracket to define a Lie algebra. For Killing tensors, a natural generalisation of this operation is provided by the Schouten bracket  \cite{Schouten:1940, Nijenhuis:1955a, Nijenhuis:1955b, Dubois-Violette:1994tlf}. Given two symmetric contravariant tensors $v$ (of rank $p$) and $w$ (of rank $q$) it yields the following contravariant tensor of rank $p+q-1$:
\be \label{schouten}
[v,w]^{\m_1 \cdots \m_{p+q-1}} \equiv \frac{(p+q-1)!}{p!q!} \left( p\, v^{\a(\m_1 \cdots} \pr_\a w^{\cdots \m_{p+q-1})} - q\, w^{\a(\m_1 \cdots} \pr_\a v^{\cdots \m_{p+q-1})} \right) .
\ee
For $p=1$ or $q=1$ it thus coincides with the Lie bracket. Moreover, the Schouten bracket of two Killing tensors is again a Killing tensor (see Appendix~\ref{sec:properties-Killing}). On the other hand, the bracket of two traceless tensors is, in general, not traceless and this prevents the use of \eqref{schouten} to introduce a Lie bracket on the vector space of traceless Killing tensors.

In Appendix~B of \cite{Campoleoni:2014tfa} an exception to this rule was already noticed: in three space-time dimensions the Schouten bracket of two AdS traceless Killing tensors is not traceless, but it can be decomposed into a sum of traceless Killing tensors.\footnote{When $D > 3$ the right-hand side of \eqref{schouten} can also be decomposed in traceless components, but in general they do not satisfy the Killing equation \eqref{killing} even when both $v$ and $w$ are traceless Killing tensors.} The algebra obtained in this way is the $
\l \to \infty$ limit of $\hs[\lambda] \,\oplus\, \hs[\lambda]$, which is isomorphic to two copies of the algebra of area-preserving diffeomorphisms of the $2D$ hyperboloid. The traceless Killing tensors of Minkowski space can be obtained in a $\L \to 0$ limit from the AdS ones and the bracket \eqref{schouten} is not affected by the curvature of the background. It is therefore not surprising that the algebra $\ihs_3[\infty]$ is isomorphic to the algebra of traceless Killing tensors of three-dimensional Minkowski space equipped with the Schouten bracket.

To clarify this statement, one can start from the standard expression for the Killing vectors of Minkowski space in Cartesian coordinates,
\be
p_a{}^\m = \d_a{}^\m \, , \qquad
m_{ab}{}^\m = 2\, x_{[a} \d_{b]}{}^\m \, ,
\ee
with $a,b \in \{0,1,2\}$, and perform the changes of basis \eqref{dual-j} and \eqref{p->P}. These introduce the six vectors $P_m{}^\m$ and $L_m{}^\m$, with $-1 \leq m \leq 1$, satisfying the Poincar\'e algebra \eqref{poincare} (computed via their Lie brackets). To proceed, we recall that all Killing tensors of constant-curvature spaces can be built as symmetrised products of Killing vectors \cite{Thompson:1986}. Moreover, the Lie derivative of a traceless Killing tensor along a Killing vector is again a traceless Killing tensor. In this particular case, one can further observe that the following symmetrised products are traceless:
\begin{subequations} \label{killing-tens-other_3D}
\begin{align}
P^{(s)}_{\pm(s-1)}{}^{\m_1 \cdots \m_{s-1}} & \equiv \frac{(s-1)!}{(2\sqrt{2})^{s-2}}\,P_{\pm1}{}^{\m_1} \cdots P_{\pm1}{}^{\m_{s-1}} , \\
L^{(s)}_{\pm(s-1)}{}^{\m_1 \cdots \m_{s-1}} & \equiv (s-1) \frac{(s-1)!}{(2\sqrt{2})^{s-2}}\,P_{\pm1}{}^{(\m_1} \cdots P_{\pm1}{}^{\m_{s-2}}\,L_{\pm1}{}^{\m_{s-1})} .
\end{align}
\end{subequations}
For instance, $\h_{\m\n} (P_m)^\m (P_n)^\n = \g_{mn}$, where $\g_{mn}$ is the $\sl(2,\mathbb R)$ Killing metric \eqref{sl2-killing}. For any value of $s$, one can then define other $4s-6$ independent traceless Killing tensors via the recursion relations
\begin{subequations}
\begin{align}
P_{m\pm 1}^{(s)}{}^{\m_1 \cdots \m_{s-1}} & \equiv \frac{\pm 1}{s\mp m-1} \left[ L_{\pm1} \,,\, P_m^{(s)} \right]^{\m_1 \cdots \m_{s-1}} , \\
L_{m\pm 1}^{(s)}{}^{\m_1 \cdots \m_{s-1}} & \equiv \frac{\pm 1}{s\mp m-1} \left[L_{\pm1}\,,\,L_m^{(s)}\right]^{\m_1 \cdots \m_{s-1}} .
\end{align}
\end{subequations}
These tensors span the whole vector space of traceless Killing tensors. Notice also that the symmetrised products in \eqref{killing-tens-other_3D} have the same structure as those defining the highest-weight generators in section~\ref{sec:3D-other_ansatz}.

The Schouten algebra of traceless Killing tensors then closes if one adds to it the inverse Minkowski metric, which is a central element since it commutes with all Killing tensors. For instance,
\begin{subequations} \label{[L,P]-killing}
\begin{align}
[ L^{(3)}_m , P^{(3)}_n ]^{\m\n\r} & = (m-n) \left( 2\,P^{(4)}_{m+n}{}^{\m\n\r} - \frac{2m^2+2n^2-mn-8}{20}\, \h^{(\m\n}\,P_{m+n}{}^{\r)} \right) , \label{[L3,P3]-Killing} \\[5pt]
[ L^{(3)}_m , P^{(4)}_m ]^{\m\n\r\s} & =  (3m-2n)\,P^{(5)}_{m+n}{}^{\m\n\r\s} \nn \\
& + \frac{3}{35}\,(5m^3-n^3-5m^2n+3mn^2-17m+9n)\,\h^{(\m\n}\,P^{(3)}_{m+n}{}^{\r\s)} \, , \label{[L3,P4]-Killing}
\end{align}
\end{subequations}
and the same results hold true if one substitutes everywhere the $P$'s with the corresponding $L$'s. Identifying $\h^{\m\n}$ with the identity, one recovers the same structure constants as in eqs.~\eqref{3D-other:commutators},\footnote{To be precise, the structure constants match if one defines the symmetrised products with $\h^{\m\n}$ using the minimum number of terms needed for the symmetrisation and without any normalisation factor. Our different convention for the symmetrisations is thus at the origin of the apparent mismatch in the relative factors in the comparison between \eqref{3D-other:commutator_1} and \eqref{[L3,P3]-Killing} and between \eqref{3D-other:commutator_3} and \eqref{[L3,P4]-Killing}.} thus proving the isomorphism with $\ihs_3[\infty]$.\\

\section{Flat-space/Carrollian-conformal higher-spin algebras in any dimensions} \label{sec:carrollian}

We now move to the generic case involving $D \geq 4$ space-time dimensions. We build higher-spin extensions of the Poincar\'e algebra as coset algebras, obtained by factoring out a suitable ideal from the UEA of $\iso(1,D-1)$. Reversing the logic we followed in section~\ref{sec:3D}, we first identify this ideal by looking at how the limit of vanishing cosmological constant affects the ideal that one factors out in the AdS$_D$ coset construction. We then check its consistency and track how the resulting algebras can also be recovered as In\"on\"u-Wigner contractions of Eastwood-Vasiliev algebras. We also prove that, under reasonable assumptions, the ideal we obtain in the limit is the only one whose factorisation gives a coset algebra defined on the same vector space as the Eastwood-Vasiliev one. Let us recall once again that, in any $D \geq 4$, the contractions presented below can be interpreted either as flat limits of AdS$_D$ higher-spin algebras or as ultra-relativistic, Carrollian limits of conformal higher-spin algebras in $D-1$ dimensions.

\subsection{Generic bulk dimension $D \geq 4$} \label{sec:carrollian-anyD}

To study the flat-space limit of the AdS coset construction we first have to express the algebra $\so(2,D-1)$ in a basis adapted to the limit. We shall later use the same basis to classify all cosets of the UEA of $\iso(1,D-1)$ that give the same set of generators as in Eastwood-Vasiliev algebras.

\subsubsection{Minkowski/Carrollian-conformal basis for the $\hs_D$ algebra}\label{sec:carrollian-basis_anyD}

We recall that the algebra $\so(2,D-1)$ reads
\be \label{so(2,D-1)}
[J_{AB} \,,\, J_{CD}] = \tilde \h_{AC}\, J_{BD} - \tilde \h_{AD}\, J_{BC} - \tilde \h_{BC}\, J_{AD} + \tilde \h_{BD}\, J_{AC}\,,
\ee
with $A,B \in \{0, \ldots, D\}$ and $\tilde\h = \text{diag}(-,+,\ldots,+,-)$. Within the $J_{AB}$ one can select the generators of transvections and Lorentz transformations in $D$ dimensions:
\be
\cP_a \equiv \e\,J_{a D}\,,\qquad
\cJ_{ab} \equiv J_{ab}\,,
\ee
where $a,b \in \{0, \ldots, D-1\}$. In this basis, the isometry algebra of AdS space reads
\begin{subequations} \label{Poincare}
\begin{align}
[\cJ_{ab} \,,\, \cJ_{cd}] &= \h_{ac}\, \cJ_{bd} - \h_{ad}\, \cJ_{bc} - \h_{bc}\, \cJ_{ad} + \h_{bd} \, \cJ_{ac} \,,\\[5pt]
[\cJ_{ab} \,,\, \cP_c ] &= \h_{ac} \, \cP_b - \h_{bc} \, \cP_a \,,\\[5pt]
[\cP_a \,,\, \cP_b ] &= -\,\e^2\,\cJ_{ab}\,,
\end{align}
\end{subequations}
with $\h = \text{diag}(-,+,\ldots,+)$. For the interpretation of this basis in terms of conformal transformations in $D-1$ dimensions we refer to Appendix~\ref{app:diff}. We assign the same dimensions to the generators $\cP_a$ and $\cJ_{ab}$, so that $\e$ is a dimensionless parameter.
The Poincar\'e algebra $\iso(1,D-1)$ is recovered by sending $\e \to 0$.

\paragraph{Annihilator of the scalar singleton}

In section~\ref{sec:global_anyD} we factored out from the  UEA of $\so(2,D-1)$ an ideal generated by quadratic combinations of the $J_{AB}$, corresponding to the annihilator of the scalar singleton representation. In the basis \eqref{Poincare}, linearly-independent quadratic combinations of the generators can be conveniently classified according to their properties under permutations of their free indices. One has two independent scalars,
\be \label{scalars}
\cP^2 \equiv \cP_a \cP^a \,, \qquad
\cJ^2 \equiv \frac12\, \cJ_{ab} \cJ^{ba} \,,
\ee
one vector,
\be \label{vector}
\cI_a \equiv \{\cP^b, \cJ_{ba} \} \,,
\ee
two traceless symmetric tensors of rank two,
\be \label{symmetric}
\cQ_{ab} \equiv \{ \cP_a , \cP_b \} - \frac{2}{D}\, \h_{ab} \cP^2 \,, \qquad
\cS_{ab} \equiv  \{{\cJ^c}_a, \cJ_{bc}\} - \frac{4}{D}\, \h_{ab} \cJ^2 \,,
\ee
one irreducible and traceless tensor transforming as a hook Young diagram,
\be \label{hook}
\cM_{ab|c} \equiv \{ \cP_{(a} , \cJ_{b)c} \} + \frac{1}{D-1} \left( \h_{ab} \{ \cP^d, \cJ_{cd} \} - \h_{c(a} \{ \cP^d , \cJ_{b)d} \} \right) ,
\ee
one tensor transforming as a traceless two-row rectangular Young diagram,
\be \label{window}
\begin{split}
& \cK_{ab|cd} \equiv \{ \cJ_{a(c} , \cJ_{d)b} \} + \frac{4}{(D-2)(D-1)} \left(\h_{ab} \h_{cd} - \h_{a(c} \h_{d)b} \right) \cJ^2 \\
& \hspace{20pt} - \frac{1}{D-2} \Big( \h_{ab} \{ \cJ_{ce}, \cJ^e{}_d \} +  \h_{cd} \{ \cJ_{ae} , \cJ^e{}_b \} - \h_{c(a} \{ \cJ_{b)e}, \cJ^e{}_{d} \} - \h_{d(a} \{ \cJ_{b)e}, \cJ^e{}_{c} \} \Big) \,, \\
\end{split}
\ee
and two antisymmetric tensors,
\be \label{antisymmetric}
\cI_{abc} \equiv
\{ \cJ_{[ab} , \cP_{c]} \} \, , \qquad 
\cI_{abcd} \equiv \{ \cJ_{[ab} , \cJ_{cd]} \}  \, .
\ee
The tensors in eqs.~\eqref{scalars}--\eqref{window} correspond to the branching in $\so(1,D-1)$ components of the product $J_{A(B} \odot J_{C)D}$, while those in \eqref{antisymmetric} correspond to the branching of $J_{[AB} \odot J_{CD]}$.
Notice that it is not necessary to symmetrise explicitly the indices in $\{\cP_{(a}, \cP_{b)}\}$ and $\{{\cJ^c}_{(a}, \cJ_{b)c}\}$ because the anticommutator automatically projects on the symmetric component. 

The ideal \eqref{rel_anyD:ideal} that we factored out in the AdS$_D$ coset construction is generated by the following combinations (see also \cite{Iazeolla:2008ix}):
\begin{subequations} \label{ideal_D_carrollian}
\begin{align}
\label{ideal_D_carrollian:1} \cJ^2 - \frac{D-1}{2}\, \e^{-2}\,\cP^2 &\sim 0 \,,\\[1pt]
\label{ideal_D_carrollian:2} \e^{-1}\,\{ \cP^b ,\, \cJ_{ba} \} &\sim 0 \,,\\[6pt]
\label{ideal_D_carrollian:3} \cS_{ab} + \e^{-2}\,\cQ_{ab} &\sim 0 \,,\\[6pt]
\label{ideal_D_carrollian:4} \e^{-1}\,\{ \cJ_{[ab} \,, \cP_{c]} \} &\sim 0 \,,\\[6pt]
\label{ideal_D_carrollian:5} \{ \cJ_{[ab} \,,\, \cJ_{cd]} \} &\sim 0 \,,\\
\label{ideal_D_carrollian:6} C_2 \equiv \cJ^2 + \e^{-2}\,\cP^2 &\sim -\frac{(D+1)(D-3)}{4}\, id \,.
\end{align}
\end{subequations}
The first three expressions come from (the branching of) the symmetric traceless product $\cI_{AB}$ defined in eq.~\eqref{rel_anyD:ideal} and the next two come from (the branching of) the completely antisymmetric one, $\cI_{ABCD} = J_{[AB} J_{CD]}$. Finally, $C_2$ is the quadratic Casimir of $\so(2,D-1)$, which is fixed by the factorisation of the previous expressions (see, e.g., eq.~\eqref{I-C2}). Taking linear combinations of the first and the last equations we get
\begin{subequations} \label{central-el_AdS}
\begin{align}
\cJ^2 &\sim \frac{D-1}{D+1}\, C_2 \sim - \frac{(D-1)(D-3)}{4}\,id \,, \\
\e^{-2}\,\cP^2 &\sim \frac{2}{D+1}\, C_2 \sim - \frac{D-3}{2}\,id \,,
\end{align}
\end{subequations}
which means that both $\cP^2$ and $\cJ^2$ and are central elements in the scalar singleton representation.

\paragraph{Higher-spin generators} 

Among the quadratic combinations listed above, all are fixed or factorised except for $\cQ_{ab}$ (or $\cS_{ab}$), $\cM_{ab|c}$ and $\cK_{ab|cd}$, which we identify as the spin-three generators of the Eastwood-Vasiliev algebra.
All these tensors are fully traceless and irreducible, so that they transform as the following $\so(1,D-1)$ irreps:
\be \label{spin-3_diagrams}
\cQ_{ab} \sim \cS_{ab} \simeq\, {\scriptsize \yng(2)} \ , \qquad \cM_{ab|c} \simeq\, {\scriptsize \yng(2,1)} \ , \qquad \cK_{ab|cd} \simeq\, {\scriptsize \yng(2,2)} \ .
\ee
In the rest of this section, we shall use either the name of the generators or their associated Young diagrams to denote them. We shall also denote the latter by a list of the lengths of their rows: for instance, we shall denote the Young diagrams in eq.~\eqref{spin-3_diagrams} as $\{2\}$, $\{2,1\}$ and $\{2,2\}$.
Note that, at this stage, we may choose the spin-three fully-symmetric generator as the traceless part of either $\{\cP_a , \cP_b\}$ or $\{{\cJ^c}_a, \cJ_{bc}\}$ since the two expressions are identified by the relations that define the coset algebra $\mfk{hs}_D$. One has instead to be careful when working in the Poincar\'e UEA, since we shall see that the two expressions cannot be identified anymore. 

The structure of the other generators of the algebra results from the branching rules of two-row Young diagrams of $\so(2,D-1)$ into $\so(1,D-1)$ diagrams: for $s \geq 2$, spin-$s$ generators are associated to all two-row diagrams of $\so(1,D-1)$ with length $s-1$ and depth ranging from 0 to $s-1$,
\be \label{Z^(s,t)}
\cZ^{(s,t)}_{a_1\cdots a_{s-1}|b_1\cdots b_{s-t-1}} \simeq\,
\begin{tabular}{cc} \cline{1-2}
\multicolumn{2}{|c|}{$\quad s-1 \quad$} \\ \cline{1-2}
\multicolumn{1}{|c|}{$\ s-t-1 \ $} & \\ \cline{1-1}
\end{tabular}
\quad \text{with } t \in \{0, \ldots, s-1 \} \,.
\ee
Once again, in analogy with section~\ref{sec:contraction-ideal_3D}, there are multiple expressions for these generators as products of $\cJ$'s and $\cP$'s which are equivalent in AdS (modulo relations of the ideal), but that will become inequivalent in the limit $\e \to 0$.

\subsubsection{Higher-spin algebras from quotients of $\cU(\iso(1,D-1))$} \label{sec:carrollian-anyD-quotient}

We now first consider the $\e \to 0$ limit of the ideal \eqref{ideal_D_carrollian} and check that it defines an ideal in the Poincar\'e UEA. We then show that requiring the same set of generators as in \eqref{Z^(s,t)} also identifies uniquely that ideal.

\paragraph{Flat/Carrollian-conformal ideal from the contraction limit}

By multiplying each expression of the ideal \eqref{ideal_D_carrollian} by the suitable power of $\e$ so as to keep only the leading part, one can take a smooth limit $\e \to 0$ and get
\begin{subequations}
\begin{align}
\{ \cP^b ,\, \cJ_{ba} \} &\sim 0 \,, \label{intermediate_ideal_Carroll:1}\\
\{ \cP_a , \cP_b \} - \frac{2}{D}\, \h_{ab} \cP^2 &\sim 0 \,, \label{intermediate_ideal_Carroll:2} \\
\{ \cJ_{[ab} \,, \cP_{c]} \} &\sim 0 \,, \label{intermediate_ideal_Carroll:3} \\[5pt]
\{ \cJ_{[ab} \,,\, \cJ_{cd]} \} &\sim 0 \,, \label{intermediate_ideal_Carroll:4}
\end{align}
\end{subequations}
together with
\be \label{intermediate_ideal_Carroll:5}
\cP^2 \sim 0 \, , \qquad
\cJ^2 \sim -\frac{(D-1)(D-3)}{4}\, id \,.
\ee
Combining eqs.~\eqref{intermediate_ideal_Carroll:2} and \eqref{intermediate_ideal_Carroll:5} one can eventually recast these expressions as
\begin{subequations} \label{carrollian_ideal}
\begin{align}
\cP_a \cP_b & \sim 0 \,, \label{carrollian_ideal:1} \\[5pt]
\cI_a \equiv \{ \cP^b ,\, \cJ_{ba} \} & \sim 0 \,, \label{carrollian_ideal:2} \\[5pt]
\cI_{abc} \equiv \{ \cJ_{[ab} \,, \cP_{c]} \} & \sim 0 \,,  \label{carrollian_ideal:3} \\[5pt]
\cI_{abcd} \equiv \{ \cJ_{[ab} \,,\, \cJ_{cd]} \} & \sim 0 \,, \label{carrollian_ideal:4} \\
\cJ^2 + \frac{(D-1)(D-3)}{4}\, id & \sim 0 \,. \label{carrollian_ideal:5}
\end{align}
\end{subequations}
We verify in Appendix \ref{app:carrollian-ideal} that these relations span an ideal, that we denote by $\cI_\mfk c$. 

Notice that we recovered the condition $\cP_a \cP_b \sim 0$ that was already manifest in \mbox{$D=3$} and that emerged in the flat limit of the scalar singleton in $D=4$ proposed in \cite{Ponomarev:2021xdq}. On the other hand, we do not have to impose the stronger constraint $\cP_a \sim 0$ that characterises the flat limit proposed in \cite{Flato:1978qz}. Compared to the three-dimensional case, the eigenvalue of $\cJ^2$ is instead fixed. Moreover, both the quadratic Casimir $\cP^2$ and the Pauli-Lubanski tensor \cite{Kuzenko:2020ayk}
\be \label{pauli-lubanski}
\mathbb{W}_{a_1 \cdots a_{D-3}} \equiv \frac{1}{2}\, \e_{a_1 \cdots a_{D-3}bcd} \cJ^{bc} \cP^d
\ee
vanish on account of the relations \eqref{carrollian_ideal}. This implies that all Casimir operators of the Poincar\'e algebra are set to zero in any representation satisfying eqs.~\eqref{carrollian_ideal:1} and \eqref{carrollian_ideal:3}, as one can appreciate by looking at their explicit expressions reported, e.g., in \cite{Poincare_Casimirs}. Eq.~\eqref{carrollian_ideal:1}  also tells us that we are not dealing with any Poincar\'e irreps.\ in Wigner's classification (cf.~\cite{Bekaert:2006py}).

We can now take the quotient of the Poincar\'e UEA by the two-sided ideal $\langle \cI_\mfk c \rangle$ and consider the resulting coset algebra as a flat-space higher-spin algebra in any dimension $D$ or, equivalently, as a Carrollian conformal higher-spin algebra in $D-1$ dimensions:
\be \label{ihs_D}
\ihs_D \equiv \cU(\iso(1,D-1))/\langle \cI_\mfk c \rangle \,.
\ee
The generators of this algebra can still be labelled as the $\cZ^{(s,t)}$ of the AdS$_D$ one and \mbox{spin-$s$} generators are given by products of $s-1$ spin-two generators.
As it is manifest from eq.~\eqref{carrollian_ideal:1}, these products contain at most one translation generator and, more precisely, none if $t$ is even and one if $t$ is odd. Since the $t$-even subalgebra only contains products of $\cJ$'s, it can be viewed as a coset of the UEA of the Lorentz subalgebra. Moreover, the completely antisymmetric projection $\{ \cJ_{[ab} , \cJ_{cd]} \}$ is factorised, so that this subalgebra is isomorphic to one of the higher-spin algebras for partially-massless fields (in $D-1$ dimensions and with de Sitter signature) that we reviewed in section~\ref{sec:PM}. In particular, comparing \eqref{eigenvalue_PM} and \eqref{carrollian_ideal:5}, one can see that it corresponds to the $\m = \frac{1}{2}$ point of the one-parameter family of algebras of \cite{Joung:2015jza}. The $t$-odd part can then be recovered by the adjoint action of $\cP_a$ on the allowed products of $\cJ$'s, with the prescription that $\{\cJ_{[ab}, \cP_{c]}\} \sim 0$ and $\{\cP^b, \cJ_{ba}\} \sim 0$. Computing an additional commutator with $\cP_a$ does not produce new generators nor extra consistency conditions because $\cP_a \cP_b \sim 0$.

Since the generators of $\ihs_D$ are realised as products of Poincar\'e generators, they all transform as Lorentz tensors. For instance, for $s=3$ one has 
\begin{subequations} \label{[J,3]}
\begin{align}
[\cJ_{ab}, \cS_{cd} ] & = \h_{ac} \cS_{bd} + \h_{ad} \cS_{bc} - \h_{bc} \cS_{ad} - \h_{bd} \cS_{ac} \,, \\[5pt]
[\cJ_{ab}, \cM_{cd|e} ] & = 2\,\h_{a(c} \cM_{d)b|e} + \h_{ae} \cM_{cd|b} - 2\,\h_{b(c} \cM_{d)a|e} - \h_{be} \cM_{cd|a} \,, \\[5pt]
[\cJ_{ab} , \cK_{cd|ef}] &= 2 \left(\,\h_{a(c}\,\cK_{d)b|ef} + \h_{a(e}\,\cK_{f)b|cd} - \h_{b(c}\,\cK_{d)a|ef} - \h_{b(e}\,\cK_{f)a|cd} \right) \,,
\end{align}
\end{subequations}
where we used the fact that $\cK_{ab|cd} = \cK_{cd|ab}$ to write the commutators in a compact form. On the other hand, their commutators with translations take a more ``exotic'' form:
\begin{subequations} \label{[P,3]}
\begin{align}
[\cP_a, \cS_{bc} ] & = -2\, \cM_{bc|a} \,, \label{[P,3]:1} \\[5pt]
[\cP_a, \cM_{bc|d} ] & = 0 \,, \label{[P,3]:2} \\[5pt]
[\cP_a , \cK_{bc|de}] &= -\, \h_{ab}\,\cM_{de|c} - \h_{ac}\,\cM_{de|b} - \h_{ad}\,\cM_{bc|e} - \h_{ae}\,\cM_{bc|d} \nn \\
&\quad - \frac{2}{D-2} \left(\h_{d(b} \cM_{c)e|a} + \h_{e(b} \cM_{c)d|a} - \h_{bc} \cM_{de|a} - \h_{de} \cM_{bc|a} \right) . \label{[P,3]:3}
\end{align}
\end{subequations}
This structure generalises to any value of $s$ according to the following schematic rules:
\begin{subequations} \label{[P,Z]}
\begin{align}
\left[\cP, \cZ^{(s,t)} \right] & \propto \cZ^{(s,t-1)} + \h\,\cZ^{(s,t+1)} & \textrm{for}\ t \ \textrm{even} \,, \label{[P,Z]_even} \\
\left[\cP, \cZ^{(s,t)} \right] & = 0 & \textrm{for}\ t \ \textrm{odd} \,, \label{[P,Z]_odd}
\end{align}
\end{subequations}
where $\h\, \cZ^{(s,t+1)}$ denotes a sum of terms involving various permutations of the indices as, e.g., in \eqref{[P,3]:3}. To better appreciate these rules, one can also look at commutators involving spin-four generators:
\begin{subequations}
\begin{align}
\cZ^{(4,0)}_{abc|def} &=  - \cJ_{d(a} \odot \cJ_{b|e|} \odot \cJ_{c)f} + \cdots \simeq\, {\scriptsize \yng(3,3)} \ , \\[3pt]
\cZ^{(4,1)}_{abc|de} &= -  \cJ_{d(a} \odot \cJ_{b|e|} \odot \cP_{c)} + \cdots \simeq\, {\scriptsize \yng(3,2)} \ , \\[3pt]
\cZ^{(4,2)}_{abc|d} &= \cJ^e{}_{(a} \odot \cJ_{b|e|} \odot \cJ_{c)d} + \cdots \simeq\, {\scriptsize \yng(3,1)} \ , \\[3pt]
\cZ^{(4,3)}_{abc} &= \cJ^e{}_{(a} \odot \cJ_{b|e|} \odot \cP_{c)} + \cdots \simeq\, {\scriptsize \yng(3)} \ ,
\end{align}
\end{subequations}
where $\odot$ denotes the symmetrised product and we omitted the terms needed to implement a traceless projection.
Also in this case, the commutators with the Lorentz subalgebra just state that the previous generators transform as tensors, while
\begin{subequations} \label{[P,4]}
\begin{align}
&\left[\cP^{\phantom{(4,0)}}_a\hspace{-13pt} , \cZ^{(4,0)}_{bcd|efg}\right] = 2\left( \cZ^{(4,1)}_{bcd|(ef} \h^{\phantom{(4,0)}}_{g)a}\hspace{-6pt} - \cZ^{(4,1)}_{efg|(bc} \h^{\phantom{(4,0)}}_{d)a}\hspace{-6pt} \right) - \frac{4}{D} \left( \cZ^{(4,1)}_{bcd|a(e} \h^{\phantom{(4,0)}}_{fg)}\hspace{-6pt} - \cZ^{(4,1)}_{efg|a(b} \h^{\phantom{(4,0)}}_{cd)}\hspace{-6pt} \right) \\
& - \frac{2}{D} \left(\cZ^{(4,1)}_{abc|(ef} \h^{\phantom{(4,0)}}_{g)d}\hspace{-7pt} + \cZ^{(4,1)}_{adb|(ef} \h^{\phantom{(4,0)}}_{g)c}\hspace{-7pt} + \cZ^{(4,1)}_{acd|(ef} \h^{\phantom{(4,0)}}_{g)b}\hspace{-7pt} - \cZ^{(4,1)}_{aef|(bc} \h^{\phantom{(4,0)}}_{d)g}\hspace{-7pt} - \cZ^{(4,1)}_{age|(bc} \h^{\phantom{(4,0)}}_{d)f}\hspace{-7pt} - \cZ^{(4,1)}_{afg|(bc} \h^{\phantom{(4,0)}}_{d)e}\hspace{-7pt} \right) , \nn \\[5pt]
&\left[\cP^{\phantom{(4,0)}}_a\hspace{-13pt} , \cZ^{(4,1)}_{bcd|ef}\right] = 0 \,, \\[5pt]
&\left[\cP^{\phantom{(4,0)}}_a\hspace{-13pt} , \cZ^{(4,2)}_{bcd|e}\right] = 2\, \cZ^{(4,1)}_{bcd|ae} + \left( \h^{\phantom{(4,0)}}_{ae}\hspace{-8pt} \cZ^{(4,3)}_{bcd} - \h^{\phantom{(4,0)}}_{a(b}\hspace{-7pt} \cZ^{(4,3)}_{cd)e} \right) - \frac{2}{D} \left(\h^{\phantom{(4,0)}}_{e(b}\hspace{-7pt} \cZ^{(4,3)}_{cd)a} - \h^{\phantom{(4,0)}}_{(bc}\hspace{-7pt} \cZ^{(4,3)}_{d)ae} \right) , \\[5pt]
&\left[\cP^{\phantom{(4,0)}}_a\hspace{-13pt} , \cZ^{(4,3)}_{bcd}\right] = 0 \,,
\end{align}
\end{subequations}
where we used again the property $\cZ^{(4,0)}_{abc|def} = -\cZ^{(4,0)}_{def|abc}$. The vanishing commutators in \eqref{[P,Z]_odd} imply that the curvatures of the algebra \eqref{ihs_D} do not reproduce upon linearisation the linear curvatures introduced in \cite{Vasiliev:1986td} to describe the free dynamics of higher-spin particles. As such, these algebras may require a new paradigm for the formulation of the free dynamics in order to lead to an interacting higher-spin gauge theory in Minkowski space via their gauging. We comment more on this issue in the Conclusions. On the other hand, we obtained a higher-spin algebra containing a Poincar\'e subalgebra and featuring non-Abelian commutators involving higher-spin generators. This is already clear recalling that it contains a subalgebra $\cA_{D-1}[1/2]$ (defined in eq.~\eqref{PM-algebra}) and we discuss the structure of generic commutators in section~\ref{sec:carrollian_inonu-wigner}.

\paragraph{Classification of the ideals compatible with the  Eastwood-Vasiliev spectrum}

In the previous pages we identified an ideal that allows one to obtain a higher-spin extension of the Poincar\'e algebra with the same spectrum as the Eastwood-Vasiliev algebra. We now prove that, under certain assumptions, this algebra is the only coset of the Poincar\'e UEA with the desired set of generators. In particular, we work under the hypothesis that all spin-$s$ generators are built out of products of $s-1$ generators of $\iso(1,D-1)$. This implies, by consistency, that all elements of the ideal to be factored out are homogeneous in the number of generators they contain. We shall thus ignore here the option to add dimension-dependent terms as, e.g., those entering the ideal \eqref{5D-ideal_2} in $D=5$ and we shall discuss the peculiarities of the latter case in section~\ref{sec:carrollian-5D}.

As a first step, we have to identify an ideal such that, after its factorisation, out of the quadratic combinations of Poincar\'e generators listed in eqs.~\eqref{scalars}--\eqref{antisymmetric} only those transforming as the $\so(1,D-1)$ Young diagrams $\{2\}$, $\{2,1\}$ and $\{2,2\}$ in eq.~\eqref{spin-3_diagrams} are left. As discussed in section~\ref{sec:isometries}, this is a necessary condition to interpret the resulting coset algebra as the global symmetries of a set of Fronsdal's fields. Achieving this goal requires to factor out the fully antisymmetric combinations $\cI_{abc} = \{ \cJ_{[ab} , \cP_{c]} \}$ and $\cI_{abcd} = \{ \cJ_{[ab} , \cJ_{cd]} \}$. From the commutation relations presented in eqs.~\eqref{check-iso-ideal_Lorentz} and \eqref{check-iso-ideal_P} it can be seen they form in itself an ideal, consistently with their interpretation as the components of $\cI_{ABCD}$. Therefore, we can consistently factor out these two combinations. 

Among the remaining quadratic combinations, only $\cK_{ab|cd}$ transforms as a $\{2,2\}$ Young diagram, so that we have to keep it. Similarly, only $\cM_{ab|c}$ fits the role of the $\{2,1\}$ generator. The delicate point is that both $\cQ_{ab} = \{ \cP_a , \cP_b \} + \cdots$ and $\cS_{ab} = \{ {\cJ^c}_a , \cJ_{bc} \} + \cdots$ display the correct Lorentz transformations to fill the role of the remaining spin-three generator. 
However, we still have to handle the vector $\cI_a = \{ \cP^b , \cJ_{ba} \}$, that cannot belong to the set of generators of the higher-spin algebra since the vector $\cP_a$ already plays this role. Keeping $\cI_a$ would thus both introduce an unwanted multiplicity and violate our hypothesis on the structure of the generators. Requiring $\cI_a \sim 0$ then implies that  both $\cQ_{ab}$ and $\cP^2$ have to vanish as well when quotienting the ideal since
\be \label{fix-carroll-P2}
0 \sim [\cP_a\,,\, \cI_b ] = - 2 \left( \cP_a \cP_b - \h_{ab} \cP^2 \right) .
\ee
Summarising, factoring out $\cI_a$, $\cI_{abc}$ and $\cI_{abcd}$ (as required to match the Eastwood-Vasiliev spectrum) from the UEA of $\iso(1,D-1)$ implies as well the condition $\cP_a \cP_b \sim 0$. 

What remains to be determined is the fate of $\cJ^2$. As it is manifest in \eqref{check-iso-ideal_J2}, it becomes a central element thanks to the previous conditions. It is thus natural to set it proportional to the identity so as to avoid multiplicities in the spectrum. Its eigenvalue is then fixed by
\be \label{fix-carroll-L2}
0 \sim \cI_{abc} \cJ^{bc} + \frac{2}{3}\, \cJ_{ab} \cI^b + \frac{D-3}{3}\, \cI_a = -\frac43 \left( \cJ^2 + \frac{(D-1)(D-3)}{4}\, id \right) \cP_a  \,.
\ee
In conclusion, if one wants to build a higher-spin extension of the Poincar\'e algebra with the Eastwood-Vasiliev spectrum \eqref{EV-spectrum} as a quotient of its UEA by a two-sided ideal, one can only obtain the coset algebra \eqref{ihs_D}. 

Notice that one can proceed along the same lines to recover the Eastwood-Vasiliev algebra as a coset of the UEA of $\so(2,D-1)$, but eq.~\eqref{fix-carroll-P2} is substituted by
\be
0 \sim \e^{-2}\, [\cP_a\,,\, \cI_b ] = - \left( \cS_{ab} + \e^{-2} \cQ_{ab} \right) - \frac{4}{D}\, \h_{ab} \left(\cJ^2 - \frac{D-1}{2}\,\e^{-2} \cP^2\right) \,,
\ee
and thus implies \eqref{ideal_D_carrollian:1} and \eqref{ideal_D_carrollian:3}. 

\subsubsection{In\"on\"u-Wigner contractions of $\hs_D$}\label{sec:carrollian_inonu-wigner}

Following the approach used in sections~5 and 6 of \cite{Fradkin:1986ka} to classify infinite-dimensional subalgebras of the AdS$_4$ higher-spin algebra, we notice that the parity of $s$ and $t$ is conserved by the Lie bracket in the Anti de Sitter algebra $\hs_D$. More explicitly
\be
\left[\cZ^{(s_1,t_1)}, \cZ^{(s_2,t_2)} \right] \propto \sum_{s_3,t_3} \cZ^{(s_3,t_3)} \,,
\ee
with $s_1 + s_2 - s_3 \text{ mod } 2 = 0$ and $t_1 + t_2 - t_3 \text{ mod } 2 = 0$. The term with highest spin in the decomposition is $s_3 = s_1 + s_2 - 2$, the one with lowest spin is $s_3 = |s_1 - s_2| + 2$ (this is a consequence of the spin addition rules guaranteed by the UEA construction) and we always have $t_1 + t_2 \text{ mod } 2 = t_3 \text{ mod } 2$, since the terms with even $t$ can be written as products of an even number of $\cP$'s, while those with odd $t$ can be written as products of an odd number of $\cP$'s and the number of $\cP$'s is conserved modulo 2 both by the Lie bracket and by the factorisation of the ideal.\footnote{Equivalently, notice that the number of free indices on the left- and right-hand sides of a commutator clearly has to be the same and, since for generic $D$ only the metric tensor can appear besides the generators, one can only use generators with the same number of indices modulo a multiple of two.} We can use these observations to identify different subalgebras that we can use to define Inönü-Wigner contractions:
\begin{itemize}
\item $s \text{ mod } 2 = 0$, which corresponds to the subalgebra of even spin;
\item $s + t \text{ mod } 2 = 0$, which corresponds to the generalisation to any dimensions of (the bosonic part of) the $h_2$ subalgebra of \cite{Fradkin:1986ka};
\item $s \text{ mod } 2 = 0$ and $t \text{ mod } 2 = 0$, which corresponds to the generalisation to any dimensions of (the bosonic part of) the $f_{22}$ subalgebra of \cite{Fradkin:1986ka};
\item $t \text{ mod } 2 = 0$, which corresponds to the generalisation to any dimensions of (the bosonic part of) the $k$ subalgebra of \cite{Fradkin:1986ka}.
\end{itemize}
For each of the previous subalgebras one can define an Inönü-Wigner contraction by rescaling by $\e^{-1}$ the generators not comprised into them and taking the limit $\e \to 0$. Still, not all of them are of direct relevance for our purposes, since the corresponding contracted algebras may not be realised as quotients of the Poincar\'e UEA.
The first item in the list leads to the truncation to minimal higher-spin algebras and can also be used to define a contraction which ``Abelianises'' all generators of odd spins, while leaving the $\so(2,D-1)$ subalgebra untouched. The last item leads instead to an Inönü-Wigner contraction of the Eastwood-Vasiliev algebra $\hs_D$ that reproduces our $\ihs_D$ algebra \eqref{ihs_D}. The third item can be used to define a minimal truncation of $\ihs_D$ to its even-spin subalgebra, while it does not seem possible to realise the second item as a coset of $\cU(\iso(1,D-1))$.

From this point of view, it is clear that the structure constants of $\ihs_D$ are exactly the same as those of $\hs_D$, except in the commutators of two generators with $t$ odd which vanish.
Notice also that in $D = 3$ most of the higher-spin generators are absent because of the low dimensionality, except for $\cZ^{(s,s-2)}$ and $\cZ^{(s,s-1)}$ which are rescaled, respectively, as $\e^{- (s \text{ mod } 2)}$ and $\e^{- (s -1 \text{ mod } 2)}$ in the limit that leads to $\ihs_D$. One can then dualise $\cZ^{(s,s-2)}_{a_1\cdots a_{s-1}|b}$ into the completely symmetric tensor $\cZ^{(s,s-2)}_{a_1\cdots a_{s-1}}$ using the Levi-Civita tensor $\ve_{abc}$. In order to make contact with the generators $L^{(s)}$ and $P^{(s)}$ defined in section \ref{sec:3D-flat}, the last step is to swap the interpretation of $\cZ^{(s,s-2)}$ and $\cZ^{(s,s-1)}$ as higher-spin rotations and translations for every odd value of the spin. Then, the $L^{(s)}_m$ and $P^{(s)}_m$ are recovered as the list of independent components of the previous $\so(1,2)$ tensors in a $\sl(2,\mathbb R)$-adapted basis.

\subsection{The particular case $D=5$} \label{sec:carrollian-5D}

As reviewed in section~\ref{sec:5D_rel}, when $D=5$ there exists a one-parameter family of higher-spin algebras $\hs_5[\l]$ with the same spectrum of generators as in the Eastwood-Vasiliev algebra, sitting at $\l = 0$. The algebras with $\l \neq 0$ are obtained by factorising an ideal in $\cU(\so(2,4))$ that is the same as in eq.~\eqref{ideal_D_carrollian} aside from the substitution of $\cI_{abc}$ and $\cI_{abcd}$ with 
\be \label{ideal_5D_carrollian}
\cI^\l{}_{\!\!\!abc} \equiv \{ \cJ_{[ab} , \cP_{c]} \} - i\,\frac{2\l}{3}\,\ve_{abcde} \cJ^{de} \,, \qquad
\cI^\l{}_{\!\!\!abcd} \equiv \{ \cJ_{[ab} , \cJ_{cd]} \} -  i\,\frac{2\l}{3}\,\ve_{abcde} \cP^e \, .
\ee
This implies that the quadratic Casimir of $\so(2,4)$ is not fixed anymore (see eq.~\eqref{C2_5D}) and
\be
\cJ^2 \sim 2(\l^2-1)\, id \,, \qquad
\cP^2 \sim (\l^2-1)\, id \,,
\ee
which, for $\l = 0$ and $D = 5$, consistently display the same eigenvalues as in eq.~\eqref{central-el_AdS}.

\subsubsection{Coset construction}

The flat/ultra-relativistic limit of the previous ideal for $\l = 0$ corresponds to the singleton case studied before and leads to the higher-spin algebra $\ihs_5$ defined in \eqref{ihs_D}. For any finite $\l$, keeping the leading order in $\e$ in the limit of all expressions of the ideal would lead to the too restrictive condition $\cP_a \sim 0$. However, it is still possible to deform the construction of section~\ref{sec:carrollian-anyD} by rescaling
$\l \to \e \, \l$, which leads in the limit $\e \to 0$ to the ideal
\begin{subequations} \label{carrollian_ideal_mod}
\begin{align}
\cP_a \cP_b & \sim 0 \,, \label{carrollian_ideal_mod:1} \\[5pt]
\cI_a \equiv \{ \cP^b ,\, \cJ_{ba} \} & \sim 0 \,, \label{carrollian_ideal_mod:2} \\[5pt]
\cI_{abc} \equiv \{ \cJ_{[ab} \,, \cP_{c]} \} & \sim 0 \,,  \label{carrollian_ideal_mod:3} \\[5pt]
\cI^\l{}_{\!\!\!abcd} \equiv \{ \cJ_{[ab} \,,\, \cJ_{cd]} \} - i\,\frac{2\l}{3}\, \ve_{abcde} \cP^e & \sim 0 \,, \label{carrollian_ideal_mod:4} \\
\cJ^2 + 2\, id & \sim 0 \,. \label{carrollian_ideal_mod:5}
\end{align}
\end{subequations}
Notice that the actual value of $\l$ in \eqref{carrollian_ideal_mod} has no significance because it can be reabsorbed by a rescaling of $\cP_a$. We thus have two non-isomorphic algebras, one corresponding to the limit of the scalar singleton and the other to the limit of $\hs_5[\l \to 0]$ in which $\e^{-1} \l$ is kept fixed. The presence of the term $\ve_{abcde} \cP^e$ in the ideal induces different commutation relations (see Appendix~\ref{sec:carrollian-commutators}) and it does not seem possible to absorb these differences with a change of basis. We denote the algebra obtained by quotienting out from the Poincar\'e UEA the ideal \eqref{carrollian_ideal_mod} rather than \eqref{carrollian_ideal} as $\widetilde{\ihs}_5$.

\subsubsection{In\"on\"u-Wigner contractions of $\hs_5[\l]$}\label{sec:carrollian_contractions_5D}

The In\"on\"u-Wigner contraction of the algebra with $\l = 0$ follows the procedure detailed in the case of general dimension. In Appendix \ref{sec:carrollian-commutators} we also argue that the limit $\l \to 0$ with $\e^{-1} \l$ fixed converges and leads to the algebra $\widetilde{\ihs}_5$.

The one-parameter family of AdS$_5$ higher-spin algebras admits however also  finite-dimensional truncations, while this option is absent in the flat/Carrollian limit we discussed above. Still, it is possible to classify all possible In\"on\"u-Wigner contractions of the AdS$_5$ truncations, involving only fields up to a given spin. For instance, if we fix the value $\l = 2$, we obtain a finite-dimensional algebra featuring only generators of spin 2 and 3. Looking at the commutation relations, there are several possible contractions including a Poincar\'e subalgebra. Among them we focus on
\be \label{exotic-contraction_5D}
\cP_a \to \e^{-1} \cP_a \,,\quad \cK_{ab|cd} \to \e^{-1} \cK_{ab|cd} \,,\quad \cM_{ab|c} \to \e^{-2} \cM_{ab|c} \,,\quad \cQ_{ab} \to \e^{-3} \cQ_{ab} \,,
\ee
leading, in the limit $\e \to 0$, to an algebra in which the commutators with $\cJ_{ab}$ remain the same as in \eqref{[J,3]}, while
\begin{subequations} \label{good-[P,Z]}
\begin{align}
[\cP_a , \cK_{bc|de}] &= -\, \h_{ab}\,\cM_{de|c} - \h_{ac}\,\cM_{de|b} - \h_{ad}\,\cM_{bc|e} - \h_{ae}\,\cM_{bc|d} \nn \\
&\quad\, - \frac{2}{3} \left(\h_{d(b} \cM_{c)e|a} + \h_{e(b} \cM_{c)d|a} - \h_{bc} \cM_{de|a} - \h_{de} \cM_{bc|a} \right) , \\[5pt]
[\cP_a, \cM_{bc|d} ] &= -\frac{4}{3} \left( \h_{ad}\,\cS_{bc} - \h_{a(b}\,\cS_{c)d} \right) - \frac{1}{3} \left(\h_{bc}\,\cS_{da} - \h_{d(b}\,\cS_{c)a} \right) , \\[5pt]
[\cP_a, \cS_{bc} ] &= 0 \,.
\end{align}
\end{subequations}
With the contraction \eqref{exotic-contraction_5D} one thus keeps untouched the commutators leading to the linearised curvatures of \cite{Vasiliev:1986td} (although, as discussed in \cite{Boulanger:2013zza}, this example is not suited to describe the unitary propagation of higher-spin particles due to the finite-dimensional truncation). In an arbitrary space-time dimensions rescalings of the type \eqref{exotic-contraction_5D} would lead to a Poisson algebra like those discussed in \cite{Ponomarev:2017nrr}, while in this case one keeps some additional terms in the higher-spin commutators:
\begin{subequations}
\begin{align}
[\cK_{ab|cd} \,, \cK_{mn|pq}] &\propto \Pi_1 \left[\, 4\, \h_{am}\,{\ve_{bdnq}}^i\,\cM_{i[c|p]} + \h_{am}\,\h_{cp}\,({\ve_{bdn}}^{ij}\,\cM_{qi|j} - {\ve_{nqb}}^{ij}\,\cM_{di|j}) \right] , \\[5pt]
[\cK_{ab|cd} \,, \cM_{mn|p} ] &\propto \Pi_2 \left[\,\h_{am}\,{\ve_{bdnp}}^i\,\cS_{ic} \,\right] , 
\end{align}
\end{subequations}
where it is understood that the projectors $\Pi_k$ implement the symmetries of the left-hand side (for instance, in the first equation both the indices $a, b, c, d$ and $m, n, p, q$ must be Young projected to transform as a $\{2,2\}$ diagram and a traceless projection must be implemented, see also Appendix~\ref{sec:carrollian-commutators}). The remaining commutators are
\be
[\cK_{ab|cd} \,, \cS_{mn} ] = 0 \,,\quad
[\cM_{ab|c} \,, \cM_{mn|p}] = 0 \,,\quad
[\cM_{ab|c} \,, \cS_{mn} ] = 0 \,,\quad
[\cS_{ab} \,, \cS_{mn}] = 0 \,.
\ee

Contrary to $\ihs_5$ and $\widetilde{\ihs}_5$, the algebra obtained in this limit does not imply any higher-spin gravitational back-reaction (in other words, commutators of spin-three generators do not contain any spin-two contribution), which means that the spin-three field can be consistently decoupled. In algebraic terms, factorisation by the ideals $\{\cS\} \subset \{\cS, \cM\} \subset \{\cS, \cM, \cK\}$ are allowed. Nevertheless, it provides an example of a higher-spin algebra including a Poincar\'e subalgebra, respecting the commutators \eqref{good-[P,Z]} and containing higher-spin self-interactions, although it does not seem possible to realise it as a quotient of the Poincar\'e UEA by a two-sided ideal.\\

\section{Galilean-conformal higher-spin algebras in any dimensions} \label{sec:galilean}

We now introduce, in any space-time dimension $D -1 \geq 3$, a Lie algebra with the same spectrum as the Eastwood-Vasiliev one and containing a Galilean conformal subalgebra. In analogy with the previous section, we first obtain this higher-spin algebra as a quotient of the UEA of $\gca_{D-1}$ and then we recover it as a suitable In\"on\"u-Wigner contraction of the conformal higher-spin algebra in $D-1$ dimensions.

\subsection{Generic bulk dimension $D \geq 4$} \label{sec:galilean-anyD}

We first have to write the conformal algebra \mbox{$\so(2,D-1)$} in a basis adapted to its Galilean contraction, in order to define a limit of the ideal that one factors out in \mbox{$\cU(\so(2,D-1))$}. We eventually define a higher-spin Galilean conformal algebra by factoring out the resulting ideal from $\cU(\gca_{D-1})$.

\subsubsection{Galilean-conformal basis for the $\hs_D$ algebra}

We start once again from the presentation of $\so(2,D-1)$ in terms of the generators $J_{AB}$ as in \eqref{so(2,D-1)} and we proceed by implementing the branching rules $\so(2,D-1) \to \so(1,D-2) \to \so(D-2)$  to single out first the Lorentz tensors in $D-1$ dimensions that generate the conformal algebra and then, e.g., the generators of time translations and boosts. In addition, we regroup the generators into different tensors according to their transformations under the adjoint action of the $\so(D-2)$ subalgebra of spatial rotations. In the basis adapted to the Galilean contraction of the conformal algebra a $\sl(2,\mathbb R)$ subalgebra also naturally emerges \cite{Bagchi:2009my}. Besides the $J_{ij}$ components, with $i,j \in \{1, \ldots, D-2\}$, one can indeed group the other generators as
\begin{subequations} \label{gca_generators}
\begin{alignat}{6}
\Lb_{-1} & = J_{0,D} + J_{0,D-1} \,,\qquad & \Lb_0 & = J_{D-1,D} \,,\qquad  & \Lb_1 & = J_{0,D} - J_{0,D-1} \,, \\[5pt]
T_{i,-1} & = \e^{-1}(J_{i,D-1} + J_{i,D}) \,,\qquad  & T_{i,0} & = \e^{-1} J_{i0} \,,\qquad  & T_{i,1} & = \e^{-1}(J_{i,D-1} - J_{i,D}) \,.
\end{alignat}
\end{subequations}
The $J_{ij}$ generate spatial rotations; the $T_{i,-1}$ generate spatial translations, the $T_{i,0}$ generate spatial boosts and the $T_{i,1}$ generate spatial conformal transformations (spatial accelerations); $\Lb_{-1}$ generates time translations, $\Lb_0$ generates dilations and $\Lb_1$ generates the zeroth component of conformal transformations (time acceleration). See Appendix~\ref{app:diff} for more details. Both $\Lb_m$ on the one hand and $T_{i,m}$ for any $i$ on the other hand transform as a  $\sl(2,\mathbb R)$ vector and the $\so(2,D-1)$ commutation relations read in this basis\footnote{When $D=3$, the adjoint representation of $\so(D-2)$ is trivial and the $J_{ij}$ are therefore not present. One can however introduce the generators $L_m \equiv \Lb_m$, $P_m \equiv T_{1,m}$ and recover the algebra \eqref{comm:lorentz}, whose contraction limit can thus be neatly interpreted as a two-dimensional Galilean conformal algebra.}
\begin{subequations} \label{gca_algebra}
\begin{alignat}{5}
[J_{ij},J_{kl}] & = \d_{ik} J_{jl} - \d_{jk} J_{il} - \d_{il} J_{jk} + \d_{jl} J_{ik} \,, \qquad & 
[\Lb_m,\Lb_n] & = (m-n)\Lb_{m+n} \,, \\[5pt]
[J_{ij},T_{k,m}] &= \d_{ik} T_{j,m} - \d_{jk} T_{i,m} \,, \qquad & [\Lb_m,T_{i,n}] & = (m-n) T_{i,m+n} \,, \\[4pt]
[T_{i,m}, T_{j,n}] & = \e^2 \Big( \d_{ij}(m-n) \Lb_{m+n} - \g_{mn} J_{ij} \Big) \,, \qquad & 
[J_{ij},\Lb_m] & = 0 \,, 
\end{alignat}
\end{subequations}
where $\g_{mn}$ denotes the $\sl(2,\mathbb R)$ Killing metric \eqref{sl2-killing}, while $\d_{ij}$ is the Kronecker symbol in $D-2$ dimensions.
From the viewpoint of the isometries of AdS$_D$, the $\sl(2,\mathbb R)$ subalgebra has an interesting interpretation as the isometries of an AdS$_2$ factor that one can identify in the AdS$_D$ metric \cite{Bagchi:2009my}.
The non-relativistic contraction of the conformal algebra is obtained by sending $\e \to 0$, thus making the $T_{i,m}$ the generators of an Abelian ideal. We denote the resulting algebra as $\gca_{D-1}$.

Both the elements of the ideal to be factored out from $\cU(\gca_{D-1})$ and the higher-spin generators are built as symmetric products of the ``spin-two'' generators in \eqref{gca_generators} and we group them in tensors transforming irreducibly under both $\so(D-2)$ and $\sl(2,\mathbb R)$. We label $\so(D-2)$ irreps by their associated Young diagram $Y = \{\l_1,\l_2,\ldots\}$ and $\sl(2,\mathbb R)$ irreps by their spin $\mathbf \qb$ (corresponding to an irrep.\ of dimension $2\mathbf \qb+1$). For instance, spin-two generators are labelled by
\be
J_{ij} \simeq \left(\,{\scriptsize \yng(1,1)}\,,\mathbf{0}\right) , \quad 
\Lb_m \simeq \left(\,\bullet\,,\mathbf{1}\right) , \quad 
T_{i,m} \simeq \left(\,{\scriptsize \yng(1)}\,,\mathbf{1}\right) .
\ee
The index structure of any combination can be easily reconstructed from these labels. For instance:
\be
\left( \{\l,\m\}, \mathbf \qb \right) \to \cM_{i_1 \cdots i_\l | j_1 \cdots j_\m,\, m} \,,
\ee
where $|m| \leq \mathbf \qb$, a vertical bar separates two sets of symmetrised indices and
\be
\cM_{(i_1 \cdots i_\l | j_1)j_2 \cdots j_\m,\, m} = 0 \, ,
\qquad 
\d^{kl} \cM_{kli_1\cdots i_{\l-2} | j_1 \cdots j_\m,\, m} = 0 \, .
\ee

\paragraph{Annihilator of the scalar singleton}

As in section~\ref{sec:carrollian-anyD}, we begin by listing all independent quadratic symmetric products of spin-two generators.
Since several combinations in this list transform in the same way under the adjoint action of both \mbox{$\so(D-2)$} and $\sl(2,\mathbb R)$, we also introduce an extra label that we call \emph{translation number} and that counts how many generators $T_{i,m}$ enter the combination: we thus denote each symmetrised product by $\galanyD{\!Y\!}{\qb}{s}{t}$,
where $s-1 \geq 1$ is the total number of $\gca_{D-1}$ generators in the product (so that it correspond to the spin label of section~\ref{sec:carrollian}), $0 \leq t \leq s-1$ is the translation number, $Y$ is a Young diagram of $\so(D-2)$ and $\mathbf{\qb}$ is the $\sl(2,\mathbb R)$ spin. With this notation, the products of two $J$'s are 
\begin{subequations}
\begin{alignat}{2}
\galanyD{\yng(2,2)}{0}{3}{0} & = J_{i(j} J_{k)l} - \text{tr.} \,, & \qquad
\galanyD{\yng(2)}{0}{3}{0} & = {J^k}_{(i} J_{j)k} - \text{tr.} \,, \\
\galanyD{\bullet}{0}{3}{0} &= J^2 \equiv \frac12 J_{ij} J^{ji} \,, & \qquad
\galanyD{\yng(1,1,1,1)}{0}{3}{0} &= J_{[ij} J_{kl]} \,, \label{J^2_galilei}
\end{alignat}
\end{subequations}
where ``$-$ tr.'' denotes the terms needed to implement a traceless projection (cf.~\eqref{symmetric} and \eqref{window}). The product of one $J$ and one $\Lb$ gives
\be
\galanyD{\yng(1,1)}{1}{3}{0} = J_{ij} \Lb_m \,,
\ee
while there are two independent products of two $\Lb$'s:
\begin{subequations}
\begin{alignat}{2}
\galanyD{\bullet}{2}{3}{0} &= \left\{\Lb_m, \Lb_n\right\} -  \frac{2}{3}\,\g_{mn} \Lb^2 \,, & \qquad
\galanyD{\bullet}{0}{3}{0} &= \Lb^2 \equiv \g^{mn} \Lb_m \Lb_n \,. \label{L^2_galilei}
\end{alignat}
\end{subequations}
The products of $J$'s and $T$'s give the following independent structures,
\begin{subequations}
\begin{alignat}{3}
\galanyD{\yng(2,1)}{1}{3}{1} &= \left\{J_{i(j}, T_{k),m} \right\} - \text{tr.} \,, & \qquad
\galanyD{\yng(1)}{1}{3}{1} &= \left\{ {J_i}^j , T_{j,m} \right\} ,\\
\galanyD{\yng(1,1,1)}{1}{3}{1} &= \left\{ J_{[ij} , T_{k],m} \right\} , & &
\end{alignat}
\end{subequations}
the products of $\Lb$'s and $T$'s give 
\begin{subequations}
\begin{alignat}{2}
\galanyD{\yng(1)}{2}{3}{1} &= \left\{\Lb_m, T_{i,n}\right\} + \left\{\Lb_n, T_{i,m}\right\} - \text{tr.} \,, & \qquad
\galanyD{\yng(1)}{0}{3}{1} &= \g^{mn} \left\{ \Lb_m , T_{i,n} \right \} , \label{T^2_galilei} \\[5pt]
\galanyD{\yng(1)}{1}{3}{1} &= \g^{kn} (m-n) \left\{ \Lb_k , T_{i,m+n} \right \} , & &
\end{alignat}
\end{subequations}
and the products of two $T$'s give
\begin{subequations}
\begin{alignat}{2}
\galanyD{\yng(2)}{2}{3}{2} &= \left\{ T_{(i,|m|} , T_{j),n} \right\} - \text{tr.} \,, & \qquad
\galanyD{\yng(2)}{0}{3}{2} &= \g^{mn} \left\{ T_{i,m} , T_{j,n} \right\} - \text{tr.} \,,\\[5pt]
\galanyD{\bullet}{2}{3}{2} &= \d^{ij} \left\{ T_{i,m} , T_{j,n} \right\} - \text{tr.} \,, & \qquad
\galanyD{\bullet}{0}{3}{2} &= T^2 \equiv \d^{ij} \g^{mn} T_{i,m} T_{j,n} \,,\\
\galanyD{\yng(1,1)}{1}{3}{2} &= \left\{ T_{[i,|m|} , T_{j],n} \right\} . & &
\end{alignat}
\end{subequations}
Notice that there are still multiple generators with the same quantum numbers (e.g.\ $\galanyD{\bullet}{0}{3}{0}$ in \eqref{J^2_galilei} and \eqref{L^2_galilei}), together with multiple generators that differ only by their translation number (e.g.\ $\galanyD{\yng(2)}{0}{3}{0}$ and $\galanyD{\yng(2)}{0}{3}{2}$). On the other hand, the relativistic ideal identifies linear combinations of the generators with the same $\sl(2,\mathbb R)$ and $\so(D-2)$ quantum numbers to zero in a series of relations. Eventually, we shall see that this allows one to identify uniquely higher-spin generators using the labels we introduced.

The relativistic ideal, corresponding to the annihilator of the scalar singleton, in the Galilean-conformal basis reads
\begin{subequations} \label{AdS-Galilei-ideal_1}
\begin{align}
\g^{mn} \e^{-2} \left\{T_{i,m}, T_{j,n}\right\} + 2\, J_{k(i} {J_{j)}}^k - \frac{2}{D-2}\, \d_{ij} \left(\e^{-2} \, T^2 + 2\, J^2\right) &\sim 0 \,, \label{AdS-Galilei-ideal:1} \\
\d^{ij} \e^{-2} \left\{T_{i,m}, T_{j,n}\right\} - \left\{ \Lb_m, \Lb_n \right\} - \frac23\, \g_{mn} \left(\e^{-2} \, T^2 - \Lb^2\right) &\sim 0 \,, \label{AdS-Galilei-ideal:2} \\[2pt]
6\, J^2 - 2(D-2)\, \Lb^2 - \e^{-2} \, (D-5)\, T^2 &\sim 0 \,, \label{AdS-Galilei-ideal:scalar} \\[6pt]
\e^{-1} \left\{{J_i}^j, T_{j,m} \right\} - \e^{-1} \, \g^{kn} (m-n) \left\{ \Lb_k, T_{i,m+n} \right\} &\sim 0 \,,
\end{align}
\end{subequations}
together with
\begin{subequations} \label{AdS-Galilei-ideal_2}
\begin{align}
\e^{-1} \left\{ J_{[ij}, T_{k],m} \right\} &\sim 0 \,,\\[5pt]
\e^{-1} \, \g^{mn} \left\{ \Lb_m, T_{i,n} \right\} &\sim 0 \,,\\[5pt]
2 \, \e^{-2} \left\{ T_{[i,|m|}, T_{j],n} \right\} - (m-n) \left\{ J_{ij}, \Lb_{m+n} \right\} &\sim 0 \,,\\[5pt]
J_{[ij} J_{kl]} &\sim 0 \,,
\end{align}
\end{subequations}
and 
\be \label{AdS-Galilei-ideal_3}
C_2 \equiv J^2 + \Lb^2 + \e^{-2} \, T^2 \sim - \frac{(D+1)(D-3)}{4}\, id \,.
\ee
The relations \eqref{AdS-Galilei-ideal_1} correspond to the $\frac12 D(D+3)$ independent components of the symmetric and traceless condition $\cI_{AB} \sim 0$ imposed in section~\ref{sec:global_anyD}. The relations \eqref{AdS-Galilei-ideal_2} correspond instead to the  $\frac{1}{6!}(D-2)(D-1)D(D+1)$ independent components of the condition \mbox{$\cI_{ABCD} \sim 0$}. Finally, \eqref{AdS-Galilei-ideal_3} is the counterpart of \eqref{rel_anyD:Casimir} and expresses the quadratic Casimir of $\so(2,D-1)$, whose value is fixed by the factorisation of the previous combinations, in terms of the scalars $J^2$, $\Lb^2$ and $T^2$ defined in eqs.~\eqref{J^2_galilei}, \eqref{L^2_galilei} and \eqref{T^2_galilei}. 

\paragraph{Higher-spin generators}

As in the bases considered in the previous sections, the relations \eqref{AdS-Galilei-ideal_1}--\eqref{AdS-Galilei-ideal_3} reduce the number of possible independent spin-three generators to the $\frac{1}{12}(D-2)(D+1)(D+2)(D+3)$ independent components of the traceless projection of $\{ J_{A(B} , J_{C)D}\}$. The spectrum of generators up to $s=3$ in the Galilean basis is given in Table~\ref{table1}, where the spin-one generator corresponds to the identity.
\begin{table}[h]
\centering
\begin{tabular}{ccccc}
\multicolumn{3}{c}{spin $s$} \\ 
\multicolumn{1}{|c|}{1} & \multicolumn{1}{c|}{2} & \multicolumn{1}{c|}{3} & & \multirow{4}{*}{\rotatebox{90}{translation $t$}} \\ \cline{1-4}
\multicolumn{1}{|c|}{$\left(\bullet,\mathbf 0\right)$} & \multicolumn{1}{c|}{$\left(\tiny\yng(1,1),\mathbf 0\right), \left(\bullet,\mathbf 1\right)$} & \multicolumn{1}{c|}{$\left(\tiny \yng(2,2),\mathbf 0\right), \left(\tiny \yng(2),\mathbf 0\right), \left(\tiny \yng(1,1),\mathbf 1\right), \left(\bullet,\mathbf 2\right), \left(\bullet,\mathbf 0\right)$} & 0 & \\ \cline{1-4}
\multicolumn{1}{|c|}{-} & \multicolumn{1}{c|}{$\left(\tiny\yng(1),\mathbf 1\right)$} & \multicolumn{1}{c|}{$\left(\tiny\yng(2,1),\mathbf 1\right), \left(\tiny\yng(1),\mathbf 2\right), \left(\tiny\yng(1),\mathbf 1\right)$} & 1 & \\ \cline{1-4}
\multicolumn{1}{|c|}{-} & \multicolumn{1}{c|}{-} & \multicolumn{1}{c|}{$\left(\tiny\yng(2),\mathbf 2\right)$} & 2 & \\ \cline{1-4}
\end{tabular}
\caption{Generators of $\hs_D$ with $s \leq 3$ in the Galilean basis.} \label{table1}
\end{table}\\
Note that some of the Young diagrams in this list, as for instance $\tiny \yng(2,2)$, might vanish identically for $D-2 < 4$. Moreover, the quantum numbers $\galanyD{\!Y\!}{\qb}{s}{t}$ suffice to uniquely identify a given generator of the algebra $\hs_D$. We shall see shortly that this property extends to arbitrary values of $s$.

Similarly, taking all possible products on the left- and right-hand side of the relations included in the ideal with elements of $\cU(\so(2,D-1))$ allows one to reduce the number of independent components of the $n^\text{th}$ power of spin-two generators to that of a $\so(2,D-1)$ rectangular Young diagram with two rows of $n$ boxes.
To proceed, we also adopt the prescription that whenever a higher-spin generator admits multiple expressions in \mbox{$\cU(\so(2,D-1))$} that can be identified in the relativistic ideal \eqref{AdS-Galilei-ideal_1}--\eqref{AdS-Galilei-ideal_3}, we take as representative the expression with the lowest $t$ number. The reason for this choice will become clear in the discussion about the Galilean In\"on\"u-Wigner contraction: whenever a combination admits several rewritings, that with the lowest $t$ survives the contraction whereas those with higher $t$ are identified to zero in the Galilean ideal. Whenever several rewritings with the same value of $t$ are available, it does not matter which expression we choose and we select the one with the most number of $J$'s (e.g.\ for $\galanyD{\bullet}{0}{3}{0}$ we use $J^2$ instead of $\Lb^2$).
Following this prescription, one can identify, e.g., the set of spin-four generators in the Galilean basis presented in Table~\ref{table2}.
\begin{table}[h]
\centering
\begin{tabular}{ccc}
\multicolumn{1}{c}{spin $s$} \\ 
\multicolumn{1}{|c|}{4} & & \\ \cline{1-2}
\multicolumn{1}{|c|}{$\left(\tiny\yng(3,3),\mathbf 0\right),\left(\tiny\yng(3,1),\mathbf 0\right),\left(\tiny\yng(2,2),\mathbf 1\right),\left(\tiny\yng(2),\mathbf 1\right),\left(\tiny\yng(1,1),\mathbf 2\right),\left(\bullet,\mathbf 3\right),\left(\tiny\yng(1,1),\mathbf 0\right), \left(\bullet,\mathbf 1\right)$} & 0 & \multirow{4}{*}{\rotatebox{90}{translation $t$}} \\ \cline{1-2}
\multicolumn{1}{|c|}{$\left(\tiny\yng(3,2),\mathbf 1\right),\left(\tiny\yng(3),\mathbf 1\right),\left(\tiny\yng(2,1),\mathbf 2\right),\left(\tiny\yng(1),\mathbf 3\right),\left(\tiny\yng(2,1),\mathbf 1\right), \left(\tiny\yng(1),\mathbf 2\right), \left(\tiny\yng(1),\mathbf 1\right)$} & 1 & \\ \cline{1-2}
\multicolumn{1}{|c|}{$\left(\tiny\yng(3,1),\mathbf 2\right),\left(\tiny\yng(2),\mathbf 3\right),\left(\tiny\yng(2),\mathbf 2\right)$} & 2 & \\ \cline{1-2}
\multicolumn{1}{|c|}{$\left(\tiny\yng(3),\mathbf 3\right)$} & 3 & \\ \cline{1-2}
\end{tabular}
\caption{Generators of $\hs_D$ with $s = 4$ in the Galilean basis.}
\label{table2}
\end{table}\\

For a generic $s \geq 2$, the set of generators fulfilling the previous prescription contains a single copy of all  $\galanyD{Y}{\qb}{s}{t}$ with translation number $0 \leq t \leq s-1$ and Young diagram $Y = \{\l,\m\}$ such that $t$, $\l$, $\m$ and $\mathbf \qb$ verify the following conditions:
\begin{itemize}
\item $t \leq \l \leq s-1$ and $t \leq \mathbf \qb \leq s-1$,
\item $2\,t \leq \l + \mathbf \qb \leq s+t-1$,
\item $0 \leq \m \leq \l-t$ and $\l + \m \equiv t$ mod 2,
\item in addition if $t=0$ then $\l + \mathbf \qb \equiv s-1$ mod 2.
\end{itemize}
This statement can be proved in three steps. First, we prove that the number of independent components of all admissible symbols $\galanyD{Y}{\qb}{s}{t}$ matches that of a two-row Young diagram $\so(2,D-1)$ with shape $\{s-1,s-1\}$. Then we show how to associate to each symbol a certain product of spin-two generators and, finally, we argue that this decomposition is the one which minimises the overall $t$ number.

$i)$ For a given spin $s$,
\be
\begin{split}
&\sum_{t=1}^{s-1} \sum_{u=2t}^{t+s-1} \underset{\l+\mathbf \qb = u}{\sum_{\l,\mathbf \qb=t}^{s-1}} \sum_{v=0}^{\left\lfloor \frac{\l-t}{2} \right\rfloor} \text{dim}_{\so(D-2)}[\{\l,\l-t-2v\}] (2\,\mathbf \qb+1) \\
&\quad + \sum_{u=0}^{\left\lfloor \frac{s-1}{2}\right\rfloor} \underset{\l+\mathbf \qb=s-1-2u}{\sum_{\l,\mathbf \qb=0}^{s-1}} \sum_{v=0}^{\left\lfloor \frac{\l}{2} \right\rfloor} \text{dim}_{\so(D-2)}[\{\l,\l-2v\}] (2\,\mathbf \qb+1) \\
&= \text{dim}_{\so(2,D-1)}[\{s-1,s-1\}] \,.
\end{split}
\ee

$ii)$ For given $s$, $t$, and admissible quantum numbers $Y=\{\l,\m\}$ and $\mathbf \qb$ we can write
\be \label{gal_D_expression}
\galanyD{Y}{\qb}{s}{t} \simeq \mathbb Y_Y \underbrace{(J \ldots J)}_{s-1-\mathbf \qb} \underbrace{(\Lb \ldots \Lb)}_{\mathbf \qb-t} \underbrace{(T \ldots T)}_{t} \,,
\ee
where all $\so(D-2)$ indices sitting on the $T$'s and $J$'s are Young projected so as to match the shape of $Y$ and a traceless projection is also understood. Similarly, all $\sl(2,\mathbb R)$ indices on $T$ and $\Lb$ have to be symmetrised and a traceless projection has to be imposed also on them (to fit the spin $\mathbf \qb$ representation).

$iii)$ The expressions schematically introduced in \eqref{gal_D_expression} are manifestly those with the lowest $t$. For instance, for $t=s-1$ we have only one generator $\galanyD{\{s-1\}}{s-1}{s}{s-1}$ and the only way to get it while writing a symmetrised product of only $s-1$ generators is to take a product of $T$'s only. Then for $t=s-2$ we have three generators $\galanyD{\{s-1,1\}}{s-2}{s}{s-2}$, $\galanyD{\{s-2\}}{s-1}{s}{s-2}$ and $\galanyD{\{s-2\}}{s-2}{s}{s-2}$ and the only way to generate them is to take the product of $(s-2)$ $T$'s and a single $J$ or $\Lb$.

\subsubsection{Higher-spin algebras from quotients of $\cU(\gca_{D-1})$} \label{sec:galilean-anyD-quotient}

Again in analogy with section~\ref{sec:carrollian-anyD}, we now first consider the $\e \to 0$ limit of the ideal \eqref{AdS-Galilei-ideal_1}--\eqref{AdS-Galilei-ideal_3} and then we check that it defines an ideal in the UEA of $\gca_{D-1}$. Given the more involved structure of the set of higher-spin generators in the Galilean basis, in this case we refrain from giving a detailed analysis of the (im)possibility to obtain other non-isomorphic algebras with the same spectrum, but we prove at least that the factorisation of the Galilean ideal fixes the central elements $T^2$ and $J^2 - \Lb^2$.

\paragraph{Galilean limit of the annihilator of the scalar singleton}

We begin by combining eqs.~\eqref{AdS-Galilei-ideal:scalar} and \eqref{AdS-Galilei-ideal_3} to get
\be \label{gal_D_singleton_central}
J^2 - \Lb^2 \sim - \frac{(D-3)(D-5)}{4}\, id \,,
\ee
so as to deal with expressions with the same $t$ numbers. In general, in the Galilean contraction $\e \to 0$ keeps only the leading part in $T$ (i.e.\ the part of highest $t$) of each expression in the ideal. After multiplying each expression by the suitable power of $\e$ to cancel divergences, we have the smooth limit
\begin{subequations} \label{gal_D_singleton_ideal}
\begin{align}
\label{gal_D_singleton_ideal:1} \g^{mn} \left\{T_{i,m}, T_{j,n}\right\} &\sim 0 \,,\\[3pt]
\label{gal_D_singleton_ideal:2} \d^{ij} \left\{T_{i,m}, T_{j,n}\right\}  &\sim 0 \,,\\[3pt]
\label{gal_D_singleton_ideal:4} \cI_{i,m} \equiv \left\{{J_i}^j, T_{j,m} \right\} - \g^{kn} (m-n) \left\{ \Lb_k, T_{i,m+n} \right\} &\sim 0 \,,\\[3pt]
\label{gal_D_singleton_ideal:5} \cI_{ijk,m} \equiv \left\{ J_{[ij}, T_{k],m} \right\} &\sim 0 \,,\\[3pt]
\label{gal_D_singleton_ideal:6} \cI_i \equiv \g^{mn} \left\{ \Lb_m, T_{i,n} \right\} &\sim 0 \,,\\[3pt]
\label{gal_D_singleton_ideal:7} \cI_{ij,mn} \equiv \left\{ T_{[i,|m|}, T_{j],n} \right\} &\sim 0 \,,\\[3pt]
\label{gal_D_singleton_ideal:8} \cI_{ijkl} \equiv J_{[ij} J_{kl]} &\sim 0 \,,
\end{align}
\end{subequations}
together with the condition \eqref{gal_D_singleton_central} that remains untouched in the limit. Eqs.~\eqref{gal_D_singleton_ideal:1} and \eqref{gal_D_singleton_ideal:2} are obtained by combining the limit of eqs.~\eqref{AdS-Galilei-ideal:1} and \eqref{AdS-Galilei-ideal:2} with the condition $T^2 \sim 0$ that one recovers in the limit from \eqref{AdS-Galilei-ideal_3}. We verify explicitly in Appendix \ref{sec:galilean-ideal} that these relations span an ideal, that we denote by $\cI_\mfk g$. 

We can now take the quotient of $\cU(\gca_{D-1})$ by the two-sided ideal $\langle \cI_\mfk g \rangle$ and consider the resulting coset algebra as a  higher-spin Galilean conformal algebra in $D-1$ dimensions:
\be
\gchs_{D-1} \equiv \cU(\gca_{D-1})/\langle\cI_\mfk g\rangle \,.
\ee
Once again, the advantage of using the lowest $t$ representatives for the higher-spin generators is that those are untouched in the limit, whereas their higher $t$ counterparts are identified to zero in the quotient of the ideal.\footnote{In two dimensions, this amounts to choose the representative with the smallest number of $P_m$'s in the basis \eqref{comm:lorentz}. In this way, the generators of $\hs[\l] \oplus \hs[\l]$ are presented directly as in eqs.~\eqref{3D-flat:Ansatz_L} and \eqref{3D-flat:Ansatz_P} and they admit a smooth $\e \to 0$ limit without the need to first resort to the ideal as we did in section~\ref{sec:contraction-ideal_3D}.} Therefore, $\gchs_{D-1}$ clearly has the same spectrum as $\hs_D$ and we can keep the same expressions for the generators in the limit.

One can also check that the eigenvalue of the central combination $J^2-\Lb^2$ is fixed by the quotienting of the other elements of the ideal \eqref{gal_D_singleton_ideal}. Indeed, consistently with eq.~\eqref{gal_D_singleton_central}, one has
\be \label{fix-central-Galilei}
\begin{split}
0 & \sim \cI_{ijk,m} J^{ij} - (D-5)\, \cI_{k,m} - 2\, J_k{}^j \, \cI_{j,m} - \g^{np} (m-n) \{\Lb_p, \cI_{k,m+n}\} + \{\Lb_m, \cI_k\} \\
&= -4\left(J^2 - \Lb^2 + \frac{(D-3)(D-5)}{4} \,id\right) T_{k,m} \, .
\end{split}
\ee
%

\subsubsection{In\"on\"u-Wigner contractions of $\hs_D$}\label{sec:inonu-wigner_galilean}

We now argue that the previously defined $\gchs_{D-1}$ algebras can be obtained as an In\"on\"u-Wigner contraction of the Eastwood-Vasiliev algebra $\hs_D$. To begin with, we observe the following useful inequality on the translation numbers of $\hs_D$ generators:
\be
\underset{Z \in [X,Y]}{\text{max}} t[Z] \leq t[X] + t[Y] \,,
\ee
where the left- and right-hand sides are not equal in general because the commutator of two $T$'s is proportional to $\Lb$ and $J$ (see eq.~\eqref{gca_algebra}) and $t[J] = t[\Lb] = 0 < 2\,t[T]$. The subadditivity of the translation number $t$ under the Lie bracket is clearly satisfied in the $\so(2,D-1)$ subalgebra and naturally extends to $\cU(\so(2,D-1))$ thanks to the Leibniz rule.  
Upon factorisation of the relativistic ideal this might not be true anymore, because different expressions in $\cU(\so(2,D-1))$ with different $t$ numbers are related to each other by expressions in the ideal. On the other hand, this property is conserved if one adopts our prescription of using the expression with the minimal $t$ for the representatives. The generic form of a commutator is then
\be \label{comm_Galilei}
\left[\galanyD{Y_1}{\qb_1}{s_1}{t_1}\!, \galanyD{Y_2}{\qb_2}{s_2}{t_2}\right] = \underset{s_3 \leq s_1+s_2-2}{\underset{t_3 \leq t_1+t_2}{\sum_{Y_3, \mathbf \qb_3}}} \! f\!\left[ \begin{array}{cc|c} (s_1, t_1) & (s_2, t_2) & (s_3, t_3) \\ (Y_1, \mathbf \qb_1) & (Y_2, \mathbf \qb_2) & (Y_3, \mathbf \qb_3) \end{array} \right] \galanyD{Y_3}{\qb_3}{s_3}{t_3} \,,
\ee
where the bound on $s_3$ comes from the Leibniz rule and the bound on $t_3$ is the direct manifestation of the subadditivity of $t$ under the Lie bracket. We can thus safely rescale the generators in the ``natural'' way induced by the previous coset construction, i.e.\ according to the number of $T$'s contained in their representative (which is equal to $t$): since $T_{i,m} \to \e^{-1}\, T_{i,m}$ in the Galilean limit, we thus rescale a generator $\galanyD{Y}{\qb}{s}{t}$ as
\be
\galanyD{Y}{\qb}{s}{t} \to \e^{-t} \galanyD{Y}{\qb}{s}{t} \,.
\ee
The Galilean contraction is then defined by sending $\e \to 0$. The limit is well-defined since one can cancel all negative powers of $\e$ appearing in the commutators \eqref{comm_Galilei} multiplying them by $\e^{t_1+t_2}$ on both sides. After the limit, the commutators retain the same structure as in the relativistic case, except that the limit select the generator(s) on the right-hand side with the maximal $t_3 = t_1 + t_2$. If this value of $t_3$ is not attained then the commutator vanishes, as in the case of $[T_{i,m}, T_{j,n}]$. In particular, for the maximal values of $t$ on the left-hand side (i.e. $t_1 = s_1 - 1$ and $t_2 = s_2 - 1$), the commutators always vanish because $t_3 = s_1 + s_2 - 2$ is not an admissible translation number for any $s_3 \leq s_1 + s_2 - 2$. Still, several higher-spin generators remain non-Abelian in the limit.

\subsection{The particular case $D = 5$} \label{sec:galilean-5D}

We now focus on the special case $D=5$, with its one-parameter family of relativistic conformal higher-spin algebras $\hs_5[\l]$ (see section~\ref{sec:5D_rel}). To study the Galilean limit in this case it is convenient to introduce yet another basis, in which we dualise the $J_{ij}$ so as to make manifest that $\so(2,4)$ contains two $\sl(2)$ subalgebras.

\subsubsection{Galilean-conformal basis for the $\hs_5[\l]$ family}

\paragraph{The spin-two sector}

When $D=5$, we can introduce the generators
\be \label{Lm}
L_m = \{J_{31}+i J_{12}\,,\, i J_{23}\,,\, J_{31}
-i J_{12} \} \quad \text{for } m \in \{-1, 0, 1\} \,,
\ee
together with the $\Lb_m$ defined in \eqref{gca_generators}. Notice that we used the imaginary unit in the previous combinations so as to move from the $\so(3)$ subalgebra spanned by the $J_{ij}$ to a $\sl(2,\mathbb R)$ subalgebra. This will allow us to present the relativistic conformal algebra in a more symmetric way that simplifies the ensuing analysis. Similarly, the Galilean translations can be collected in the tensor $T_{m,n}$ with $m,n \in \{-1,0,1\}$ defined as
\begin{subequations}
\begin{align} \label{Tmn}
T_{m,-1} &= (-iP_2 + P_3, P_1, -i P_2 - P_3) \,, \\
T_{m,0} &= (-iB_2 + B_3, B_1, -i B_2 - B_3) \,, \\
T_{m,+1} &= (-iK_2 + K_3, K_1, -i K_2 - K_3) \,,
\end{align}
\end{subequations}
where we resorted to the labelling of the generators introduced in Appendix~\ref{app:diff} to identify the various components. In this basis the commutation relations of $\so(2,4)$ (or, more precisely, of the real form of $D_3$ obtained via the changes of basis \eqref{Lm} and \eqref{Tmn}) read
\begin{subequations} \label{galilean_basis_5D}
\begin{alignat}{5}
[L_m, L_n] &= (m-n)\,L_{m+n} \,, & \qquad
[\Lb_m, \Lb_n] &= (m-n)\,\Lb_{m+n} \,, \label{gal_5D_rel:1} \\[5pt]
[L_m, T_{n,k}] &= (m-n)\,T_{m+n,k} \,, & \qquad
[\Lb_m, T_{k,n}] &= (m-n)\,T_{k,m+n} \,, \label{gal_5D_rel:2} \\[5pt]
[T_{m,k}, T_{n,l}] &= (m-n)\,\g_{kl}L_{m+n} + (k-l)\,\g_{mn}\Lb_{k+l} \,, \hspace{-12pt} &  [\Lb_m, L_m] & = 0 \, , \label{gal_5D_rel:3}
\end{alignat}
\end{subequations}
where $\g_{mn}$ is the $\sl(2,\mathbb R)$ Killing metric \eqref{sl2-killing}.

\paragraph{Annihilator of the scalar singleton}

In the basis \eqref{galilean_basis_5D}, the portion \eqref{AdS-Galilei-ideal_1} of the annihilator of the scalar singleton reads
\begin{subequations} \label{rel_5D_ideal_A}
\begin{align}
\label{rel_5D_ideal:1} \g^{kl} \left\{T_{m,k}, T_{n,l}\right\} - \left\{ L_m, L_n \right\} - \frac23\, \g_{mn} \left(T^2 - L^2\right) &\sim 0 \,,\\
\label{rel_5D_ideal:2} \g^{mn} \left\{T_{m,k}, T_{n,l}\right\} - \left\{ \Lb_k, \Lb_l \right\} - \frac23\, \g_{kl} \left(T^2 - \Lb^2\right) &\sim 0 \,,\\[2pt]
\label{rel_5D_ideal:3} L^2 - \Lb^2 &\sim 0 \,,\\[5pt]
\label{rel_5D_ideal:4} \g^{kl} (m-k) \left\{L_l, T_{m+k,n} \right\} - \g^{kl} (n-k) \left\{ \Lb_l, T_{m,n+k} \right\} &\sim 0 \,,
\end{align}
\end{subequations}
while the portion \eqref{AdS-Galilei-ideal_2} reads\footnote{We further select here the real form of $D_3$ such that the coefficient in front of $\l$ be real. Starting from the conventions of section~\ref{sec:galilean-anyD}, this can be achieved by sending expressions involving $s-1$ generators to $i^s$ times themselves.}
\begin{subequations} \label{rel_5D_ideal_B}
\begin{align}
\label{rel_5D_ideal:5} \g^{mn} \left\{ L_m, T_{n,k} \right\} - 2 \l\, \Lb_k &\sim 0 \,,\\[5pt]
\label{rel_5D_ideal:6} \g^{mn} \left\{ \Lb_m, T_{k,n} \right\} - 2 \l\, L_k &\sim 0 \,,\\[5pt]
\label{rel_5D_ideal:7} \left\{ T_{m,k}, T_{n,l} \right\} - \left\{ T_{n,k}, T_{m,l} \right\} + (m-n)(k-l) \left\{ L_{m+n}, \Lb_{k+l} \right\} & \nn \\
- 2 \l\, (m-n)(k-l)\,T_{m+n,k+l} &\sim 0 \,,
\end{align}
\end{subequations}
where $L^2 = \g^{mn} \, L_m \, L_n$, $\Lb^2 = \g^{mn} \, \Lb_m \, \Lb_m$ and $T^2 = \g^{mn} \, \g^{kl} \, T_{m,k} \, T_{n,l}$. Moreover, as discussed in section~\ref{sec:5D_rel}, the quadratic Casimir is not fixed anymore but is rather a function of $\l$:
\be
\label{rel_5D_ideal:8} C_2 \equiv L^2 + \Lb^2 + T^2 \sim 3(\l^2-1)\, id \,.
\ee
The ideal thus displays the same structure as in the generic $\l = 0$ case in eqs.~\eqref{rel_5D_ideal_A}, while eqs.~\eqref{rel_5D_ideal_B} now mix terms involving a different number of generators. The latter can be introduced only when $D=5$ because they crucially involve a dualisation (see section~\ref{sec:5D_rel}). 

\paragraph{Spin-three generators}

Even when $D=5$, we could use the method discussed in section~\ref{sec:galilean-anyD} to identify higher-spin generators. In the following, we resort however to an alternative labelling inspired by the analysis of the three-dimensional case.
We begin by introducing the spin-three generators $W_m$ and $\Wb_m$ that correspond, respectively, to $\galanyD{\tiny\yng(2)}{0}{3}{0}$ and $\galanyD{\bullet}{2}{3}{0}$ in Table~\ref{table1} and that are defined as
\begin{alignat}{4}
W_{2} &\equiv \left(L_1\right)^2 \,, & \quad W_{m-1} &\equiv -\frac{1}{m+2} \left[L_{-1} , W_m\right] \,, & \quad m = -2,\ldots,2 \,, \label{W} \\
\Wb_{2} &\equiv \left(\Lb_1\right)^2 \,, & \quad \Wb_{m-1} &\equiv -\frac{1}{m+2} \left[\Lb_{-1} ,\Wb_m\right] \,, & \quad m = -2,\ldots,2 \,. \label{Wb}
\end{alignat}
The mixed products $L_m \Lb_n$ with $m,n \in \{-1, 0, 1\}$ in this case are not bound to vanish as in three dimensions. We shall discuss them after having introduced other generators that will prove useful to clarify their role. The traces of the products $L_m L_n$ and $\Lb_m \Lb_n$ are also not anymore proportional to the identity. They can be combined to give an additional $U(1)$ factor with $t=0$ besides the identity, corresponding to $\galanyD{\bullet}{0}{3}{0}$ in Table~\ref{table1}:
\be \label{A}
A \equiv L^2 + \Lb^2 - \frac15\, C_2 \,,
\ee
where we subtracted a multiple of $C_2$ so as to avoid that certain commutators of spin-two generators with spin-three ones contain a contribution proportional to the identity (see Appendix~\ref{sec:galilean-5D-commutators}). 

The products of an $L$ or $\Lb$ with a $T$ gives rise to the $t=1$ quadratic combinations.
There are three such products. Schematically, $\t \equiv \left\{ L_{(\cdot} \,,\, T_{\cdot),\cdot} \right\} - \text{tr.}$, $\tb \equiv \left\{ \Lb_{(\cdot} \,,\, T_{|\cdot|,\cdot)} \right\} - \text{tr.}$ and $\Tt \equiv \left\{ L_{[\cdot} \,,\, T_{\cdot],\cdot} \right\} \sim \left\{ \Lb_{[\cdot} \,,\, T_{|\cdot|,\cdot]} \right\}$ 
where the last identification comes from eq.~\eqref{rel_5D_ideal:4} of the ideal, and we also introduced a traceless projection in the first two definitions. All in all, we can introduce recursively the generators $\t_{m,k}$ and $\tb_{k,m}$ with $m \in \{-2,\ldots,2\}$ and $k \in \{-1,0,1\}$,
\begin{alignat}{6}
\t_{2,1} &\equiv \frac{\left\{T_{1,1}\,,\, L_1 \right\}}{2} \,, \quad & \ \t_{m-1,k} &\equiv -\frac{\left[L_{-1} , \t_{m,k}\right]}{m+2} \,, \quad & \ \t_{m,k-1} &\equiv -\frac{\left[\Lb_{-1} , \t_{m,k}\right]}{k+1}  \,, \\
\tb_{1,2} &\equiv \frac{\left\{T_{1,1}\,,\, \Lb_1 \right\}}{2} \,, \quad & \ \tb_{k,m-1} &\equiv -\frac{\left[\Lb_{-1} , \tb_{k,m}\right] }{m+2} \,, \quad & \ \tb_{k-1,m} &\equiv -\frac{\left[L_{-1} , \tb_{k,m}\right]}{k+1} \,,
\end{alignat}
and the generator $\Tt_{i,j}$ with $i,j \in \{-1,0,1\}$:
\be
\Tt_{i,j} \equiv \g^{kn}(i-k)\left\{L_n , T_{i+k,j} \right\} \sim \g^{kn} (j-k)\left \{\Lb_n , T_{i,j+k}\right\} .
\ee
For the product of two $T$'s giving the generator $\galanyD{\tiny \yng(2)}{\mathbf 2}{3}{2}$ in Table~\ref{table1}, we introduce the tensor $U_{m,n}$ with $m, n \in \{-2,\ldots,2\}$:
\be
U_{2,2} \equiv \frac{\left(T_{1,1}\right)^2}{2} \,, \quad U_{m-1,n} \equiv -\frac{\left[L_{-1} , U_{m,n}\right]}{m+2} \,, \quad U_{m,n-1} \equiv -\frac{\left[\Lb_{-1} , U_{m,n}\right]}{n+2} \,.
\ee

We can now go back to the mixed products $L_m \Lb_n$. This expression has the same $\sl(2)$ quantum numbers as $T_{m,n}$ but, in the notation of section~\ref{sec:galilean-anyD}, it does not have neither the same translation number $t$, nor the same spin $s$. However, due to the $\l$-dependent terms in the relations \eqref{rel_5D_ideal_B}, one has
\be \label{lambda-commutator}
\begin{split}
\left[T_{m,k} , L_n \Lb_l \right] &\sim (k-l)\, \t_{m+n,k+l} + (m-n) \, \tb_{m+n,k+l} + (m-n)\,(k-l)\, \Tt_{m+n,k+l} \\
& \quad + \frac{\l}{3}\,(m-n)\, \g_{kl}\, L_{m+n} + \frac{\l}{3}\,(k-l)\,\g_{mn}\, \Lb_{k+l} \, .
\end{split}
\ee
The presence of spin-two contributions on the right-hand side implies that it is not appropriate to identify the combination $L_m \Lb_n$ with a spin-three generator. For instance, in a matrix representation implementing the truncation in which all higher-spin generators become trivial, one necessarily has $L_m \Lb_n \sim \k\, T_{m,n}$ rather than $L_m \Lb_n \sim 0$, since otherwise eq.~\eqref{lambda-commutator} would imply that the whole algebra be trivial. Requiring that the commutator above takes the same form as in the $\l = 0$ case, leads to define the leftover $\galanyD{\tiny \yng(1)}{1}{3}{0}$ generator in Table~\ref{table1} as 
\be \label{T-hat}
\Th_{m,n} \equiv L_m \, \Lb_n - \frac{\l}{3}\, T_{m,n} \,.
\ee
This expression is not anymore homogeneous both in the $t$ number and in the number of spin-two generators it contains. However, we shall treat terms with a lower number of generators as ``corrections'' around $\l = 0$. There is also a way to return to a homogeneous expression by making use eq.~\eqref{rel_5D_ideal:7}, but we shall not resort to it since it does not respect our convention selecting the minimal number of $T$'s in the expressions for the generators.

\paragraph{Higher-spin generators} \label{gal_5D_field_content}

The general procedure to identify the generators of $\hs_D$ in the Galilean basis proposed in section~\ref{sec:galilean-anyD} can be used in this context to derive the $\sl(2,\mathbb R)$ quantum numbers. Any non-vanishing two-row Young diagram $\{k,l\}$ of $\so(3)$ can indeed be dualised to a one-row diagram $\{k\}$ using the Levi-Civita tensor, and its length serves as the definition for the $\sl(2,\mathbb R)$ charge $\mathbf q$ which is the analogue of the previously defined $\mathbf \qb$. For instance, the spectrum of generators up to $s=4$ is presented in Table~\ref{table3}.
\begin{table}[h]
\centering
\begin{tabular}{ccccc}
\multicolumn{3}{c}{spin $s$} \\ 
\multicolumn{1}{|c|}{2} & \multicolumn{1}{c|}{3} & \multicolumn{1}{c|}{4} & & \\ \cline{1-4}
\multicolumn{1}{|c|}{$L^{(\mathbf 1,\mathbf 0)}, \Lb^{(\mathbf 0,\mathbf 1)}$} & \multicolumn{1}{c|}{$W^{(\mathbf 2,\mathbf 0)},\Th^{(\mathbf 1,\mathbf 1)}, \Wb^{(\mathbf 0,\mathbf 2)}, A^{(\mathbf 0,\mathbf 0)}$} & \multicolumn{1}{c|}{$X^{(\mathbf 3,\mathbf 0)},Y^{(\mathbf 2,\mathbf 1)},\Yb^{(\mathbf 1,\mathbf 2)},\Xb^{(\mathbf 0,\mathbf 3)},B^{(\mathbf 1,\mathbf 0)},\Bb^{(\mathbf 0,\mathbf 1)}$} & 0 & \multirow{4}{*}{\rotatebox{90}{translation $t$}} \\ \cline{1-4}
\multicolumn{1}{|c|}{$T^{(\mathbf 1,\mathbf 1)}$} & \multicolumn{1}{c|}{$\t^{(\mathbf 2,\mathbf 1)}, \tb^{(\mathbf 1,\mathbf 2)},\Tt^{(\mathbf 1,\mathbf 1)}$} & \multicolumn{1}{c|}{$\s^{(\mathbf 3,\mathbf 1)},\sh^{(\mathbf 2,\mathbf 2)},\sb^{(\mathbf 1,\mathbf 3)},\pi^{(\mathbf 2,\mathbf 1)},\pib^{(\mathbf 1,\mathbf 2)},\pit^{(\mathbf 1,\mathbf 1)}$} & 1 & \\ \cline{1-4}
\multicolumn{1}{|c|}{-} & \multicolumn{1}{c|}{$U^{(\mathbf 2,\mathbf 2)}$} & \multicolumn{1}{c|}{$\r^{(\mathbf 3,\mathbf 2)},\rb^{(\mathbf 2,\mathbf 3)},\rt^{(\mathbf 2,\mathbf 2)}$} & 2 & \\ \cline{1-4}
\multicolumn{1}{|c|}{-} & \multicolumn{1}{c|}{-} & \multicolumn{1}{c|}{$V^{(\mathbf 3,\mathbf 3)}$} & 3 & \\ \cline{1-4}
\end{tabular}
\caption{Generators of $\hs_5[\l]$ with $s \leq 4$ in the Galilean basis, with their $\mathbf{q}$ and $\mathbf{\qb}$ charges.} \label{table3}
\end{table}
Notice that the actual expression of a given field may be in general more complicated than in the $\l = 0$ case because of the same considerations that led us to introduce a $\l\,T$ term in the expression for the generator $\Th_{m,n}$ in eq.~\eqref{T-hat}. Upon introducing the change of basis leading to commutators respecting the schematic rule on the spin label $[2,s] \propto s$, the translation number $t$ may no longer be univocally defined but by convention we retain the original $t$, thus ignoring low-spin corrections.

For a given spin $s \geq 2$ and translation number $0 \leq t \leq s-1$, the generators are thus labelled by the $\galfiveD{q}{\qb}{s}{t}$ that verify the following conditions:
\begin{itemize}
\item both charges individually verify $t \leq \mathbf q \leq s-1$ and $t \leq \mathbf \qb \leq s-1$,
\item the sum of charges verifies $2\,t \leq \mathbf q + \mathbf \qb \leq s+t-1$,
\item in addition for $t=0$ the sum of charges must also verify $\mathbf q + \mathbf \qb \equiv s-1$ mod 2,
\end{itemize}
with all admissible $\galfiveD{q}{\qb}{s}{t}$ appearing only once. The proof is completely analogous to that leading to the classification of the generators in the Galilean basis in any $D$.

\subsubsection{Coset construction}\label{sec:5D-mod_coset}

The limit $\e \to 0$ of the relativistic ideal \eqref{rel_5D_ideal:1}--\eqref{rel_5D_ideal:8} induced by the rescaling $T_{m,n} \to \e^{-1} T_{m,n}$ gives rise to the same Galilean ideal $\cI_\mfk g$ as in \eqref{gal_D_singleton_ideal} for any finite value of $\l$. In the current basis it reads
\begin{subequations} \label{gal_5D_ideal}
\begin{align}
\label{gal_5D_ideal:1} \g^{kl} \left\{T_{m,k}, T_{n,l}\right\} &\sim 0 \,,\\
\label{gal_5D_ideal:2} \g^{mn} \left\{T_{m,k}, T_{n,l}\right\} &\sim 0 \,,\\
\label{gal_5D_ideal:3} L^2 - \Lb^2 &\sim 0 \,,\\
\label{gal_5D_ideal:4} \g^{kl} (m-k) \left\{L_l, T_{m+k,n} \right\} - \g^{kl} (n-k) \left\{ \Lb_l, T_{m,n+k} \right\} &\sim 0 \,,\\
\label{gal_5D_ideal:5} \g^{mn} \left\{ L_m, T_{n,k} \right\} &\sim 0 \,,\\
\label{gal_5D_ideal:6} \g^{mn} \left\{ \Lb_m, T_{k,n} \right\} &\sim 0 \,,\\
\label{gal_5D_ideal:7} \left\{ T_{m,k}, T_{n,l} \right\} - \left\{ T_{n,k}, T_{m,l} \right\} &\sim 0 \,.
\end{align}
\end{subequations}
On the other hand, if we also send $\l \to \e^{-1} \l$, the limit $\e \to 0$ yields a different ideal, in analogy with our findings in the Carrollian case (although here $\l \to \infty$ in the limit, while there $\l \to 0$). We denote it by $\tilde \cI_{\mfk g}$:
\begin{subequations} \label{gal_5D_ideal_modified}
\begin{align}
\g^{kl} \left\{T_{m,k}, T_{n,l}\right\} - \frac23\, \g_{mn} T^2 &\sim 0 \,, \label{gal_5D_ideal_modified:1}\\
\g^{mn} \left\{T_{m,k}, T_{n,l}\right\} - \frac23\, \g_{kl} T^2 &\sim 0 \,, \label{gal_5D_ideal_modified:2}\\
L^2 - \Lb^2 &\sim 0 \,, \\
\g^{kl} (m-k) \left\{L_l, T_{m+k,n} \right\} - \g^{kl} (n-k) \left\{ \Lb_l, T_{m,n+k} \right\} &\sim 0 \,,\\
\g^{mn} \left\{ L_m, T_{n,k} \right\} - 2 \l\, \Lb_k &\sim 0 \,, \label{gal_5D_ideal_modified:3} \\
\g^{mn} \left\{ \Lb_m, T_{k,n} \right\} - 2 \l\, L_k &\sim 0 \,, \label{gal_5D_ideal_modified:4} \\
\left\{ T_{m,k}, T_{n,l} \right\} - \left\{ T_{n,k}, T_{m,l} \right\} - 2 \l (m-n)(k-l)\, T_{m+n,k+l} &\sim 0 \,,\\
T^2 &\sim 3\l^2\,id \,. \label{gal_5D_ideal_modified:5}
\end{align}
\end{subequations}
As in the modified five-dimensional Carrollian ideal \eqref{carrollian_ideal_mod}, the asymptotic value of $\l$ is not relevant since it can always be reabsorbed by a rescaling of $T_{m,n}$ and one can check that these relations span an ideal. We denote the Galilean conformal higher-spin algebras obtained by factoring out $\langle \cI_{\mfk g} \rangle$ and $\langle \tilde{\cI}_{\mfk g} \rangle$ from $\cU(\gca_4)$ by $\gchs_4$ and $\widetilde{\gchs}_4$.

\subsubsection{In\"on\"u-Wigner contractions reproducing the coset algebras} \label{sec:galilean-5D_IW}

Similarly to the case of arbitrary dimension, the previous higher-spin algebras can be obtained as In\"on\"u-Wigner contractions of $\hs_5[\l]$ by introducing the ``natural'' rescaling
\be
\galfiveD{q}{\qb}{s}{t} \to \e^{-t} \galfiveD{q}{\qb}{s}{t} .
\ee
We already discussed the case $\l = 0$ in section~\ref{sec:inonu-wigner_galilean}. Rescaling $\l \to \e \,\l$ gives the same ideal as in \eqref{gal_5D_ideal} in the limit $\e \to 0$, while keeping the same expression \eqref{T-hat} for $\Th_{m,n}$. Consistently with this observation, in the examples collected in Appendix~\ref{sec:galilean-5D-commutators} we checked that it is enough to have $\l \to \e \,\l$ in order to guarantee convergence of the limit in the commutators for any $\l \neq 0$, provided that one selects the basis in which the rule $[2,s] \propto s$ is respected. A $\l$ dependence naively seems to survive in the limit, but all algebras obtained in this way are isomorphic as it is clear from the fact that they can all be obtained as the quotient of the UEA of $\gca_4$ by the same two-sided ideal \eqref{gal_5D_ideal}. The $\l$ dependence can indeed be absorbed by undoing the change of basis \eqref{T-hat} in order to return to the original $L\,\Lb$ expression after the limit, and similarly for its higher-spin cousins. Notice that in the Carrollian case one can also similarly define a smooth $\e \to 0$ limit of the commutators by introducing the rescaling $\l \to \e \,\l$ but, due to the simpler spectrum, it does not seem possible to go back to the $\ihs_5$ algebra obtained from the contraction of the $\l = 0$ member of the family with a change of basis. This is in agreement with the observation that in the Carrollian limit the rescaling $\l \to \e \,\l$ leads to a different ideal with respect to the strict $\l = 0$ case.

The limit $\l \to \infty$ behaves instead rather differently. Looking at commutators involving spin two and three generators (see, e.g., \eqref{lambda-commutator}), one can see that if one defines $\Th_{m,n} = L_m\,\Lb_n$ then the limit $\l \to \e^{-1} \l$ is smooth, although spin-three fields would not couple minimally to gravity, in the sense that $[2, 3] \propto 3 + 2$.

\subsubsection{Alternative contractions and finite-dimensional algebras} \label{sec:galilean-5D_finite}

We now show that in addition to the contractions with $\l = 0$ (or $\l \to 0$) on the one hand, and $\l \to \infty$ on the other hand, there are other possible In\"on\"u-Wigner contractions which leave $\l$ untouched and therefore admit finite-dimensional truncations in the same way as the original algebra does. These require however a scaling of the higher-spin generators that does not seem compatible with their realisation as products of $\gca_4$ generators. In other words, while one can obtain in this way a one-parameter family of Lie algebras that may be interpreted as four-dimensional Galilean-conformal higher-spin algebras, these cannot be realised as a quotient of the UEA of $\gca_4$.

\paragraph{In\"on\"u-Wigner contraction}

All $\hs_5[\l]$ algebras contain a subalgebra spanned by the generators with at least one zero charge $\mathbf q$ or $\mathbf \qb$. The reason is that they are built using products of $L$'s only or $\Lb$'s only (plus, possibly, the identity) and that $[L_m,\Lb_n] = 0$. This subalgebra contains a $U(1) \oplus \hs[2\l-1] \oplus \hs[2\l-1]$ subalgebra, together with additional generators so as to get all $\galfiveD{q}{0}{s}{0}$ and $\galfiveD{0}{\qb}{s}{0}$ with $1 \leq s \leq 2\l-1$ when $\l \in \frac{\mathbb{N}}{2}$ with the admissible quantum numbers (see, e.g., Table~\ref{table3}).

The rest of the generators, i.e.\ those that have non-zero charges in both sectors, can then be rescaled by $\e^{-1}$ and turned into a set of Abelian generators in the limit $\e \to 0$. The adjoint action of the previous subalgebra on its Abelian complement is preserved, so that a $\l$ dependence is kept. For instance: 
\begin{subequations}
\begin{align}
[W_m,\Th_{n,k}] &= \frac{\l}{6}\left((m-2\,n)\, \t_{m+n,k} + \frac13 \left(6\,n^2- 3\,n\,m + m^2 - 4\right) \, \Tt_{m+n,k}\right) + \cdots \,, \\
[W_m,\t_{n,k}] &= - \frac{(m-n)}{12} \left(2\,m^2 - m\,n + 2\,n^2 - 8\right)\left(\frac{\l}{2}\,\Th_{m+n,k} + \frac{2(4\l^2-9)}{21}\,T_{m+n,k}\right) + \cdots \,, \\
[W_m,\Tt_{n,k}] &= \frac{1}{12} \left(6\,n^2- 3\,n\,m + m^2 - 4\right) \left(\l\, \Th_{m+n,k} - \frac{2(4\l^2-9)}{21}\, T_{m+n,k} \right) + \cdots \, , \\
[W_m,\tb_{n,k}] &= \frac{\l}{2}\,(m-2\,n)\, U_{m+n,k} + \cdots \,,
\end{align}
\end{subequations}
where we omitted terms with $s > 3$. This means that the contracted algebra still admits finite-dimensional truncations and one can study the same limit directly for the finite-dimensional algebras occurring for \mbox{(half-)integer} $\l$.

\paragraph{Relation with cosets of $\cU(\gca_4)$} \label{gal_5D_limit_finite}

We now focus on the simplest example in which we keep only spin-two and three generators, corresponding to the $\l = 2$ truncation of the previous contracted family of algebras, and we show that it can only be realised as a subalgebra of the UEA of $\gca_4$ evaluated on a specific finite-dimensional representation. To begin with, we observe that a class of matrix representations of the $\gca_4$ algebra is given by the block upper-triangular matrices 
\be \label{matrix_5D-mod}
L^{\mfk g}_m \equiv \begin{pmatrix} L_m & 0 \\ 0 & L_m \end{pmatrix} , \quad \Lb^{\mfk g}_n \equiv \begin{pmatrix} \Lb_n & 0 \\ 0 & \Lb_n \end{pmatrix} , \quad T^{\mfk g}_{m,n} \equiv \begin{pmatrix} 0 & T_{m,n} \\ 0 & 0 \end{pmatrix} ,
\ee
where $L_m$, $\Lb_m$ and $T_{m,n}$ denote here the representatives of the corresponding generators of the \emph{relativistic} conformal algebra $\so(2,4)$ in a given finite-dimensional truncation (see Appendix~\ref{app:finite-dim_irreps_5D}). In particular, for concreteness we select here the matrix representation of $\so(2,4)$ giving the higher-spin algebra $\hs_5[2] \simeq \sl(10,\mathbb R)$ when computing all possible matrix products of the generators \cite{Manvelyan:2013oua}.

The generators $W^{\mfk g}$, $\Wb^{\mfk g}$ and $A^{\mfk g}$ can then be recovered using the same formulae \eqref{W}--\eqref{A} as in the relativistic case, but with $L$ and $\Lb$ replaced with $L^{\mfk g}$ and $\Lb^{\mfk g}$. The same procedure applies to the set of Abelian generators spanned by $\{\t^{\mfk g}, \tb^{\mfk g}, \Tt^{\mfk g}\}$. The remaining generators ($\Th^{\mfk g}$ and $U^{\mfk g}$) require instead more work. Since we build the matrices \eqref{matrix_5D-mod} using a matrix representation of the relativistic conformal algebra $\so(2,4)$ within each block, eqs.~\eqref{rel_5D_ideal:5} and \eqref{rel_5D_ideal:6} imply
\be
\g^{mn}\left\{L^{\mfk g}_m,T^{\mfk g}_{n,k}\right\} = \begin{pmatrix} 0 & 2\l\,\Lb_k \\ 0 & 0 \end{pmatrix} \quad \text{and} \quad \g^{mn}\left\{\Lb^{\mfk g}_m,T^{\mfk g}_{k,n}\right\} = \begin{pmatrix} 0 & 2\l\,L_k \\ 0 & 0 \end{pmatrix} .
\ee
The right-hand sides are independent from $L^{\mfk g}_m$, $\Lb^{\mfk g}_n$ and $T^{\mfk g}_{m,n}$. They do not fit the properties of any of the generators we obtained in the contraction, but we can use them to define the generator $\Th^{\mfk g}$ as
\be
\Th^{\mfk g}_{m,n} \equiv \frac{1}{2\l}\, L^{\mfk g}_m \g^{pq}\left\{L^{\mfk g}_p,T^{\mfk g}_{q,n}\right\} - \frac{\l}{3}\, T^{\mfk g}_{m,n} \,,
\ee
This unusual cubic formula gives us the expression of the relativistic generator $\Th_{m,n} = L_m \Lb_n - \frac{\l}{3}\, T_{m,n}$ placed in the upper-right block, and thus provides us with a generator with the desired properties, at least for $\l = 2$ (and in fact for any half-integer value of $\l$ inducing a finite-dimensional truncation). For $U^{\mfk g}$, recalling that in $\hs_5[\l]$ one has
\be
[W_m,\tb_{n,k}] = \frac{\l}{2}\,(m-2\,n)\,U_{m+n,k} + \ldots \,,
\ee
where we omitted generators with $s > 3$, one can define
\be
U^{\mfk g}_{2,2} \equiv \frac{1}{\l}\, [W^{\mfk g}_{2},\tb^{\mfk g}_{0,2}] \,.
\ee
In this case the definition is valid only for $\l = 2$ (for different values of $\l$ there are more terms on the right-hand side) and we can recover the other components of $U^{\mfk g}_{m,n}$ by the adjoint action of $L^{\mfk g}$ and $\Lb^{\mfk g}$.

In this way, we managed to reproduce the algebra obtained with the In\"on\"u-Wigner contraction introduced above, but there are other non-vanishing matrix products that we ignored (e.g.\ $\g^{mn}\{L^{\mfk g}_m,T^{\mfk g}_{n,k}\}$ or $L^{\mfk g}_m \Lb^{\mfk g}_n$). Therefore, we realised that algebra as a subalgebra of the full UEA of $\gca_4$ evaluated on the module \eqref{matrix_5D-mod}. We expect a similar construction to hold for higher values of $\l$, but we may need to introduce some additional generators by hand besides those in \eqref{matrix_5D-mod}, thus making this construction more and more artificial.\\

\section{Flat-space/Carrollian-conformal algebras from Schouten's bracket} \label{sec:other-carroll}

In the previous sections we focussed on higher-spin algebras defined as quotients of the UEA of either $\iso(1,D-1)$ or $\gca_{D-1}$ and having the same set of generators as Eastwood-Vasiliev algebras. On the other hand, we already encountered examples of higher-spin algebras containing either a Poincar\'e or a Galilean conformal subalgebra that do not satisfy these requirements, like those obtained directly via an In\"on\"u-Wigner contraction in sections~\ref{sec:carrollian_inonu-wigner}, \ref{sec:carrollian_contractions_5D} and \ref{sec:galilean-5D_finite}. Here we show that the Schouten bracket \cite{Schouten:1940, Nijenhuis:1955a, Nijenhuis:1955b, Dubois-Violette:1994tlf} allows one to define higher-spin algebras including a Poincar\'e subalgebra on the vector space of Minkowski Killing tensors, thus extending to any value of $D$ the considerations of section~\ref{sec:3D-other}. While this option has long been noticed in the literature (see, e.g., \cite{Berends:1984rq} and the recent overview \cite{Bekaert:2021sfc}), the resulting algebras did not find yet useful applications because they have a wider set of generators compared to those discussed above. In the following, we propose two interpretations for the Schouten bracket algebra of Minkowski Killing tensors, as the global symmetries of certain higher-derivative gauge theories in Minkowski space and as flat/Carrollian contractions of the global symmetries of partially-massless fields reviewed in section~\ref{sec:PM}.

\subsection{The Schouten bracket algebra of Killing tensors}\label{sec:schouten-algebra}

We already defined the Schouten bracket in eq.~\eqref{schouten}. We recall here its definition introducing a notation in which repeated indices denote a symmetrisation (again with the convention that dividing by the number of terms used in the symmetrisation is understood), while, e.g., $\m(n)$ is a shorthand for a set of $n$ symmetrised indices. For instance, if $v^{\m\n}$ is a symmetric tensor then $\pr^\m v^{\m(2)} \equiv \frac{1}{3} \left( \pr^{\m_1} v^{\m_2\m_3} + \pr^{\m_2} v^{\m_3\m_1} + \pr^{\m_3} v^{\m_1\m_2} \right)$. Given two symmetric contravariant tensors $v$ (of rank $p$) and $w$ (of rank $q$), their Schouten bracket is the contravariant tensor of rank $p+q-1$ defined as follows:
\be \label{schouten_2}
[v,w]^{\m(p+q-1)} \equiv \frac{(p+q-1)!}{p!q!} \left( p\, v^{\a\m(p-1)} \pr_\a w^{\m(q)} - q\, w^{\a\m(q-1)} \pr_\a v^{\m(p)} \right) .
\ee
For $p = 1$ or $q = 1$ the Schouten bracket reduces to the Lie bracket and the bracket of two Killing tensors is again a Killing tensor (see eq.~\eqref{killing-to-killing}). Therefore, one can use it to define a Lie algebra on the infinite-dimensional vector space of Killing tensors of a constant curvature spacetime \cite{Thompson:1986}, containing the isometry subalgebra generated by Killing vectors. Starting from the Killing tensors of Minkowski space, one readily obtains a ``higher-spin'' algebra including a Poincar\'e subalgebra, that we denote by $\mfk{Sch}_D$.\footnote{Starting from the non-relativistic scaling limit of Anti de Sitter space discussed in \cite{Gomis:2005pg, Bagchi:2009my} and from the associated contraction of its isometries, it should be possible to use the same strategy to build other Galilean-conformal higher-spin algebras in any dimensions. We however defer this analysis to future work.} On the other hand, as already mentioned in section~\ref{sec:3D-other}, the Schouten bracket does not map neither traceless tensors into traceless tensors nor divergenceless tensors into divergenceless tensors. As a result, it cannot be used to define non-Abelian global symmetries for Fronsdal or Maxwell-like fields, that involve either traceless or divergenceless Killing tensors (see section~\ref{sec:isometries}). 

To bypass the problem, we recall that free, higher-spin, higher-derivative gauge theories in Minkowski space involving fully unconstrained gauge parameters were proposed in \cite{Joung:2012qy, Francia:2012rg}. It is thus natural to interpret the Schouten bracket algebra of Killing tensors as a candidate global symmetry for possible non-linear completions thereof. We recall the relevant models in section~\ref{sec:higher-derivative}. Before that, we show that expressing the Killing tensors of Minkowski space in a specific basis also suggests to interpret the very same algebra, after the identifications of certain generators, as a flat/Carrollian contraction of the algebra $\cA_D[\m]$ describing the global symmetries of a certain set of partially-massless fields (see section~\ref{sec:PM}). Even if partially-massless fields can only be defined on curved backgrounds, in section~\ref{sec:PM-like} we present free field equations on $D$-dimensional Minkowski space that display the same global symmetries. We instead defer to future work the interpretation of these algebras as global symmetries of specific Carrollian-conformal field theories in $D-1$ dimensions.  

\subsubsection{Killing tensors of Minkowski space}

We begin our analysis of the Schouten algebra $\mfk{Sch}_D$ and of its relation with $\cA_D[\m]$ by exhibiting a basis of Killing tensors of rank $s-1$ in a Minkowski space of dimension $D$. The number of basis elements is the same as the number of independent components of a rectangular $\mfk{gl}(D+1)$ Young diagram with two rows of length $s-1$ and they can be constructed as symmetrised products of Killing vectors \cite{Thompson:1986}. The latter can be presented as
\be \label{killing-vectors}
\cJ_{ab}{}^\m = \d^\m{}_a x_b - \d^\m{}_b x_a \, , \qquad \cP_a{}^\m = \d^\m{}_a \,,
\ee
and their Lie brackets give the Poincar\'e algebra, that is \eqref{Poincare} with $\e = 0$. Notice that Lorentz transformations and translations have homogeneity (i.e.\ scaling dimension in $x$) 1 and 0 respectively. We can group the independent symmetric Killing tensors of rank $s-1$ into objects labelled by indices $a, b \in \{0,\ldots,D-1\}$, that correspond to the branching of a two-row rectangular $\mfk{gl}(D+1)$ Young diagram into its $\so(D)$ components. For each Young diagram obtained from the branching we introduce the tensor $(\cM^{(\l)}_{a_1 \cdots a_m | b_1 \cdots b_n})^{\m_1 \cdots \m_{s-1}}$ with $m \geq n$ satisfying 
\be
\begin{split}
\cM^{(\l)}_{a(m) | ab(n-1)}{}^{\m(s-1)} & = 0 \, , \qquad
\h^{cd}\,\cM^{(\l)}_{cd a(m-2) | b(n)}{}^{\m(s-1)} = 0 \, , \\[5pt]
\pr^{\m}\,\cM^{(\l)}_{a(m) | b(n)}{}^{\m(s-1)} & = 0 \, , 
\end{split}
\ee
where the indices $a,b$ within each group are symmetrised, while $\l$ is an additional label that accounts for possible multiplicities.

Explicitly, for $s=3$ we have, to begin with,
\begin{subequations} \label{massless-3_sch}
\begin{align}
\cK_{ab|cd}{}^{\m\n} & \equiv \cJ_{ac}{}^{(\m} \cJ_{db}{}^{\n)} + \cJ_{ad}{}^{(\m} \cJ_{cb}{}^{\n)} + \cdots \,, \label{JJ_sch} \\[5pt]
\cM_{ab|c}{}^{\m\n} & \equiv \cP_a{}^{(\m} \cJ_{bc}{}^{\n)} + \cP_b{}^{(\m} \cJ_{ac}{}^{\n)} + \cdots \,, \label{PJ_sch} \\
\cQ_{ab}{}^{\m\n} & \equiv 2 \left( \cP_a{}^{(\m} \cP_b{}^{\n)} - \frac{1}{D}\, \h_{ab} \h^{cd} \cP_c{}^{(\m} \cP_d{}^{\n)} \right) , \label{PP_sch}
\end{align}
\end{subequations}
where in the first two definitions we omitted the terms that are necessary to implement a traceless projection in the Latin indices (as in eq.~\eqref{PP_sch}) and that can be read off from eqs.~\eqref{hook} and \eqref{window}. These rank-two tensors are also traceless in the Greek indices (\mbox{$\h_{\m\n}(\cQ_{ab})^{\m\n} = 0$} etc.) and they actually form a basis for the subspace of \emph{traceless} Killing tensors of rank two. They can thus be interpreted as the generators of the global symmetries of a massless spin-three field and they correspond to the tensors introduced in $D=3$ in eq.~\eqref{killing-tens-other_3D} (in that case $\cK_{ab|cd}$ vanishes identically due to the low dimension). They have homogeneity 2, 1 and 0 respectively. To complete the basis, we also have to consider the traces (in the Latin indices) of all admissible symmetrised products of Killing vectors, i.e.\
\begin{subequations} \label{PM-3_sch}
\begin{align}
\cS_{ab}{}^{\m\n} & \equiv 2 \left(\h^{cd} \cJ_{ac}{}^{(\m} \cJ_{db}{}^{\n)} - \frac{1}{D}\, \h_{ab} \h^{cd}\h^{ef} \cJ_{ce}{}^{(\m} \cJ_{fd}{}^{\n)} \right)  \,,\\
\cI_{a}{}^{\m\n}& \equiv 2\,\h^{bc} \cP_b{}^{(\m} \cJ_{ca}{}^{\n)} \,, \\[4pt]
(\cJ^2)^{\m\n} & \equiv \frac{1}{2} \, \h^{ab} \h^{cd} \cJ_{ac}^{(\m} \cJ_{db}{}^{\n)} \,,
\end{align}
\end{subequations}
with homogeneity 2, 1 and 2, together with
\be \label{P2_sch}
(\cP^2)^{\m\n} \equiv \h^{ab} \cP_a{}^{(\m} \cP_b{}^{\n)} = \h^{\m\n} \,,
\ee
with homogeneity 0. Notice that we listed almost all combinations that appear at the beginning of section~\ref{sec:carrollian-basis_anyD}, with the exception of those in eq.~\eqref{antisymmetric}. The reason is that, due to the specific realisation of the Poincar\'e algebra introduced in \eqref{killing-vectors}, one has
\be \label{ideal_schouten}
\cI_{abc}{}^{\m\n} \equiv 2\, \cJ_{[ab}{}^{(\m}\,\cP_{c]}{}^{\n)} = 0 \, , \quad
\cI_{abcd}{}^{\m\n} \equiv 2\,\cJ_{[ab}{}^{(\m}\,\cJ_{cd]}{}^{\n)} = 0 \, .
\ee

The key observation to proceed is that $\h^{\m\n}$ behaves like the identity: its Schouten bracket with any Killing tensor vanishes, and this is actually an alternative way to characterise Killing tensors \cite{Thompson:1986}. Moreover ---~given the constraints \eqref{P2_sch} and \eqref{ideal_schouten}~--- if one interprets a symmetrised product with $\h^{\m\n}$ (as, e.g., $\h^{(\m\n} \cP_a{}^{\r)}$) as a product with the identity (as we already did in eq.~\eqref{[L,P]-killing}) then the spectrum of independent generators is manifestly the same as that of the algebras $\cA_{D}[\m]$ defined in eq.~\eqref{PM-algebra}. More precisely, with the identification $\h^{\m\n} \simeq id$ and using the associative product induced by the symmetrised product of tensors, the Schouten algebra corresponds to the coset algebra obtained by evaluating the Poincar\'e UEA on a representation satisfying $\cP^2 \sim \n\, id$, $\cI_{abc} \sim 0$, $\cI_{abcd} \sim 0$ and with all higher-order Casimir operators vanishing (because they are all proportional to the Pauli-Lubanski tensor \eqref{pauli-lubanski} that is built out of $\cI_{abc}$). We denote this algebra by $\widetilde{\mfk{Sch}}_D$ and in section~\ref{sec:contraction-PM} we shall see explicitly that these conditions also emerge in an In\"on\"u-Wigner contraction of $\cA_{D}[\m]$. This is the same setup that we already encountered in $D=3$ in section~\ref{sec:3D-other} and, like in that case, we also remark that one can always rescale at will $\cP_a$, because the number of translations on the left- and right-hand sides of any non-Abelian commutator is always the same (in $\iso(1,D-1)$ the adjoint action of $\cP_a$ either gives another translation or zero, differently from what happens in $\so(2,D-1)$). Therefore $\n$ is not a true parameter in the algebra $\widetilde{\mfk{Sch}}_D$ and it can be fixed to an arbitrary value.
The previous construction also agrees with the definition in \cite{Bekaert:2008sa} of a higher-spin algebra in Minkowski space, named ``off-shell'', as the centraliser of $\cP^2$ in the Poincar\'e UEA, realised through Weyl-ordered polynomials of the operators $\hat{x}^\m$ and $\hat{\cP}^\m$.

To have a better grasp on the structure of the algebras $\mfk{Sch}_D$ and $\widetilde{\mfk{Sch}}_D$, one can compute the Schouten bracket of the tensors \eqref{massless-3_sch} and \eqref{PM-3_sch} among themselves or with the Killing vectors \eqref{killing-vectors}. For the massless sector one obtains
\begin{subequations} \label{comm_sch_1}
\begin{align}
\label{comm_sch_1:1} [\cP_a, \cK_{bc|de}]^{\m\n} &= - \left( \h_{ab}\,\cM_{de|c}{}^{\m\n} + \h_{ac}\,\cM_{de|b}{}^{\m\n} + \h_{ad}\,\cM_{bc|e}{}^{\m\n} + \h_{ae}\,\cM_{bc|d}{}^{\m\n} \right) \nn \\
&\hspace{-20pt} - \frac{2}{D-2} \left(\h_{d(b}\,\cM_{c)e|a}{}^{\m\n} + \h_{e(b}\,\cM_{c)d|a}{}^{\m\n} - \h_{bc}\,\cM_{de|a}{}^{\m\n} - \h_{de}\,\cM_{bc|a}{}^{\m\n} \right) \,, \\
\label{comm_sch_1:2} [\cP_a, \cM_{bc|d}]^{\m\n} &= \left(\h_{ad}\,\cQ_{bc}{}^{\m\n} - \h_{a(b}\,\cQ_{c)d}{}^{\m\n}\right) + \frac{1}{D-1} \left(\h_{bc}\,\cQ_{ad}{}^{\m\n} - \h_{d(b}\,\cQ_{c)a}{}^{\m\n} \right) \,, \\
\label{comm_sch_1:3} [\cP_a, \cQ_{bc}]^{\m\n} &= 0 \,,
\end{align}
\end{subequations}
while the additional generators satisfy, e.g., 
\be
\label{comm_sch_1:4} [\cP_a, \cI_b]^{\m\n} = -\cQ_{ab}{}^{\m\n} + \frac{2(D-1)}{D}\, \h_{ab}\,(\cP^2)^{\m\n} \,.
\ee
Brackets with $\cJ_{ab}{}^\m$ take instead the same form as in \eqref{[J,3]} and \eqref{check-iso-ideal_Lorentz} since all rank-two tensors are given by products of Killing vectors. Notice that, compared to the commutators \eqref{[P,3]} of the algebra $\ihs_D$, one obtains a non-vanishing contribution on the right-hand side of \eqref{comm_sch_1:2}, while now the right-hand side of \eqref{comm_sch_1:3} vanishes. Gauging either $\mfk{Sch}_D$ or $\widetilde{\mfk{Sch}}_D$ thus reproduces the linearised curvatures describing the propagation of a massless field in Minkowski space \cite{Vasiliev:1986td} plus some extra contributions coming, e.g., from the commutator \eqref{comm_sch_1:4}. Notice also the presence of $\cP^2$ on the right-hand side of \eqref{comm_sch_1:4}: in the algebra $\mfk{Sch}_D$ it has to be considered as an additional singlet besides the identity, while in $\widetilde{\mfk{Sch}}_D$ it coincides with the latter.

Commutators of spin-three fields are, in general, non-Abelian. For instance, while $[\cQ_{ab},\cQ_{cd}]^{\m\n\r} = 0$, one has
\begin{align}
&[\cM_{ab|c},\cQ_{de}]^{\m\n\r} = \h_{a(d}\,\cR_{e)bc}{}^{\m\n\r} + \h_{b(d}\,\cR_{e)ac}{}^{\m\n\r} - 2\,\h_{c(d}\,\cR_{e)ab}{}^{\m\n\r} \label{[M,Q]_Sch} \\
&\quad - \frac{2}{D-1} \left(\h_{ab}\,\cR_{cde}{}^{\m\n\r} - \h_{c(a}\,\cR_{b)de}{}^{\m\n\r} \right) + \frac{4}{D} \left(\h_{d(a}\,\h_{b)e}  - \frac{1}{D-1}\,\h_{ab}\,\h_{de} \right) \h^{(\m\n}\,\cP_c{}^{\r)} \nn \\
&\quad + \frac{4}{D} \left(\h_{c(d}\,\h_{e)(a} - \frac{1}{D-1}\,\h_{de}\,\h_{c(a} \right) \h^{(\m\n}\,\cP_{b)}{}^{\r)} + \frac{4}{D(D-1)} \left(\h_{c(a}\,\h_{b)(d} - \h_{ab}\,\h_{c(d} \right) \h^{(\m\n}\,\cP_{e)}{}^{\r)} \,, \nn
\end{align}
and
\be
\begin{split}
&[\cS_{ab}, \cQ_{cd}]^{\m\n\r} = \cB_{ab(c|d)}{}^{\m\n\r} + 2\,\left(\h_{c(a}\,\cB'_{b)d}{}^{\m\n\r} + \h_{d(a}\,\cB'_{b)c}{}^{\m\n\r}\right) \\
&\quad - \frac{4}{D} \left(\h_{ab}\,\cB'_{cd}{}^{\m\n\r} + \h_{cd}\,\cB'_{ab}{}^{\m\n\r}\right) - \frac{2(D-2)}{D} \h^{(\m\n} \left(\h_{c(a}\,\cJ_{b)d}{}^{\r)} + \h_{d(a}\,\cJ_{b)c}{}^{\r)} \right) ,
\end{split}
\ee
where we introduced the spin-four generators
\begin{subequations}
\begin{align}
\cB_{abc|d}{}^{\m\n\r} &= 3 \left(\cP_a{}^{(\m}\,\cP_b{}^\n\,\cJ_{cd}{}^{\r)} + \cP_c{}^{(\m}\,\cP_a\,{}^\n\,\cJ_{bd}{}^{\r)} + \cP_b{}^{(\m}\,\cP_c{}^\n\,\cJ_{ad}{}^{\r)}\right) + \cdots \,,\\
\cR_{abc}{}^{\m\n\r} &= 6\,\cP_a{}^{(\m}\,\cP_b{}^\n\,\cP_c{}^{\r)} + \cdots \,, \label{Rabc}\\
\cB'_{ab}{}^{\m\n\r} &= 3 \h^{cd} \left(\cP_c{}^{(\m}\,\cJ_{da}{}^\n\,\cP_b{}^{\r)} + \cP_c{}^{(\m}\,\cJ_{db}{}^\n\,\cP_a{}^{\r)} \right) \,,
\end{align}
\end{subequations}
omitting again in the definitions the terms that are needed to implement a traceless projection in the Latin indices.
Let us stress that the Killing tensor $\cR_{abc}{}^{\m\n\r}$ defined in \eqref{Rabc} is traceless in its Greek indices and it is the only rank-three tensor entering the decomposition of the bracket \eqref{[M,Q]_Sch} besides $\h^{(\m\n}\,\cP_a{}^{\r)}$. In the algebra $\widetilde{\mfk{Sch}}_D$, in analogy with eq.~\eqref{[L,P]-killing}, the latter is not considered as an independent tensor with respect to $\cP_a{}^{\m}$, so that \eqref{[M,Q]_Sch} only contains contributions from the independent traceless Killing tensors. This observation can be generalised: Killing tensors containing at most one Lorentz generator span a subalgebra of $\widetilde{\mfk{Sch}}_D$, generalising the $\ihs_3[\infty]$ subalgebra that we identified in section~\ref{sec:3D-geometry}. We prove this fact in Appendix~\ref{sec:properties-Killing}.

Let us now go back to the interpretation of $\widetilde{\mfk{Sch}}_D$ as a limit of the algebra $\cA_D[\m]$ of global symmetries of a set of massless and partially-massless fields that we mentioned above. At first sight this proposal may look confusing because we only considered ``ordinary'' Killing tensors, while the global symmetries of partially-massless fields involve higher-order Killing tensors. Still, one can use the Schouten bracket only as a tool to extract consistent  structure constants for generators that can be grouped in the tensors $\cM^{(\l)}_{a_1 \cdots b_m | b_1 \cdots b_n}$ that are appropriate to describe the global symmetries of partially-massless fields. Moreover, the analogy can be made even stronger by noticing that the traces (in the Greek indices) of the tensors in \eqref{PM-3_sch} actually satisfy a third order Killing equation:
\be
\pr_\m \pr_\n \pr_\r (\cS_{ab})_\l{}^\l = 0 \, , \quad
\pr_\m \pr_\n \pr_\r (\cI_{a})_\l{}^\l = 0 \, , \quad
\pr_\m \pr_\n \pr_\r (\cJ^2)_\l{}^\l = 0 \, .
\ee
This is manifest because none of these expressions have homogeneity greater than 2. As a result, while the traceless tensors \eqref{massless-3_sch} can be interpreted as the set of independent gauge parameters of a massless spin-three field preserving the vacuum, the traces of the tensors \eqref{PM-3_sch} are naturally interpreted as the set of independent gauge parameters of a ``partially-massless-like'' spin-three field of depth two preserving the vacuum. Free gauge theories in Minkowski space displaying gauge transformations of this type will be presented in section~\ref{sec:PM-like}, thus allowing one to interpret the tensors \eqref{massless-3_sch} as the components of the Young diagram ${\tiny \yng(2,2)}$ in \eqref{spectrum-PM} and the tensors \eqref{PM-3_sch} as the components of the Young diagram ${\tiny \yng(2)}$ in the same equation. This analogy can be extended also to higher values of $s$, because the $n$-th trace in the Greek indices of a Killing tensor verifies a order $(2n+1)$ Killing equation, while the Killing tensors corresponding to the $\mfk{o}(D+1)$ Young diagram $\{s-1,s-1\}$ are traceless in the Greek indices (see Appendix~\ref{sec:properties-Killing}).

Notice also that the list of independent Killing tensors of rank two in eqs.~\eqref{massless-3_sch}, \eqref{PM-3_sch} and \eqref{P2_sch} also contains $\h^{\m\n}$ that, in $\widetilde{\mfk{Sch}}_D$, we interpreted as the identity. Similarly, among the independent (as elements in a vector space) Killing tensors of rank three there appear also $\h^{(\m\n} \cP_a{}^{\r)}$ and $\h^{(\m\n} \cJ_{ab}{}^{\r)}$ that we have to identify with $\cP_a$ and $\cJ_{ab}$. This mechanism eventually reduces the set of truly independent generators so as to match the last group of terms in the decomposition \eqref{spectrum-PM}. More generally, the set of tensors of rank $s-1$ that are solutions to the Killing equation should include as a subset all tensors of rank $s-3$, $s-5$ etc. that are solutions to the Killing equation, multiplied by the suitable power of $\h^{\m\n}$, to be identified with generators already included into the spectrum.

\subsubsection{Carrollian In\"on\"u-Wigner contraction of $\cA_D[\m]$}\label{sec:contraction-PM}

The algebra $\cA_D[\m]$ was defined in section~\ref{sec:PM} by factoring out the ideal
\be \label{ideal_PM}
\cI_{abc} \equiv
\{ \cJ_{[ab} , \cP_{c]} \} \sim 0 \, , \quad 
\cI_{abcd} \equiv \{ \cJ_{[ab} , \cJ_{cd]} \} \sim 0 \, , \quad
C_2 \equiv \cJ^2 + \cP^2 \sim \n[\m] \, id
\ee
from the enveloping algebra of $\so(2,D-1)$. $\cI_{abc}$ and $\cI_{abcd}$ correspond indeed to the rewriting in the basis \eqref{Poincare} of the tensor $\cI_{ABCD}$ that we factored out in eq.~\eqref{PM-algebra}.

Starting from these conditions, we can define a Lie algebra including a Poincar\'e subalgebra using the same strategy as in sections~\ref{sec:carrollian} and \ref{sec:galilean}: rescaling the $\so(2,D-1)$ generators as $\cP_a \to \e^{-1}\,\cP_a$ while sending $\n \to \e^{-2} \, \n$ (or equivalently $\m \to \e^{-1}\,\m$) and $\e \to 0$ indeed both induces the contraction $\so(2,D-1) \to \iso(1,D-1)$ and gives the following smooth limit for the relations \eqref{ideal_PM}:
\be \label{ideal_PM_flat}
\cI_{abc} \equiv
\{ \cJ_{[ab} , \cP_{c]} \} \sim 0 \, , \quad 
\cI_{abcd} \equiv \{ \cJ_{[ab} , \cJ_{cd]} \} \sim 0 \, , \quad
C_2 \equiv \cP^2 \sim \n \, id 
\ee
The analysis in Appendix~\ref{app:carrollian-ideal} confirms that the relations obtained in the limit still span an ideal. We can thus factor it out from the UEA of $\iso(1,D-1)$ to define abstractly the algebra $\widetilde{\mfk{Sch}}_D$ we introduced above using the Schouten bracket of Minkowski Killing tensors. In this construction it is manifest that one can rescale at will $\cP_a$, since this operation does not modify neither the commutation relations of the Poincar\'e algebra nor the combinations entering the ideal \eqref{ideal_PM_flat}. This confirms once again that $\n$ is not a true parameter distinguishing non-isomorphic algebras. We also recall that the relation $\cI_{abc} \sim 0$ implies that all higher-order Casimir operators vanish. When $D=3$ the previous construction therefore reproduces the algebra $\ihs(\cM^2>0,\cS=0)$ discussed in section~\ref{sec:3D-other}. 

The algebra $\widetilde{\mfk{Sch}}_D$ can also be recovered as an In\"on\"u-Wigner contraction of $\cA_D[\m]$ by shifting $\n \to \e^{-2} \n$ in the structure constants and by rescaling its generators as
\be \label{scaling-PM}
\cM^{(\l)}_{a_1 \cdots b_m | b_1 \cdots b_n} \to \e^{-t} \cM^{(\l)}_{a_1 \cdots b_m | b_1 \cdots b_n} \, ,
\ee
where $t$ denotes the number of $\cP_a$ generators contained in their realisation as elements of $\cU(\so(2,D-1))$. For instance, referring to generators that can be realised as quadratic products, one has to introduce the rescalings
\be \label{scaling-PM-spin3}
\cM_{ab|c} \to \e^{-1} \cM_{ab|c} \, , \quad
\cQ_{ab} \to \e^{-2} \cQ_{ab} \, , \quad
\cI_a \to \e^{-1}\,\cI_a \, .
\ee
In analogy with what we have seen in three dimensions in \eqref{3D-rel:generators}, in this case the identification of the right scaling is even simpler with respect to section~\ref{sec:carrollian_inonu-wigner} because the ideal \eqref{ideal_PM_flat} does not contain any relation mixing terms with a different number of transvections as in \eqref{ideal_D_carrollian:3}. The convergence of the contraction limit is guaranteed by the observation that the number of transvections $t$ is conserved modulo two and cannot increase, both when computing commutators and factorising relations of the ideal. Therefore, the commutator of generators containing $t_1$ and $t_2$ transvections gives rise on the right-hand side to generators containing $(t_1+t_2)$, $(t_1+t_2-2)$ etc.\ transvections. The In\"on\"u-Wigner contraction $\e \to 0$ thus keeps only the terms with $(t_1+t_2)$. Once the contraction has been performed, one can always redefine all generators by multiplying them by $\k^t$ with $\k > 0$, which will not affect the structure constants provided that $\n \to \k^2 \n$. Notice that, losing in the limit the free parameter that distinguishes between non-isomorphic AdS algebras, we lose at the same time the option to have finite-dimensional truncations. However, we checked explicitly that the same scaling as in \eqref{scaling-PM-spin3} together with $\cJ^2 \to \e^{-2} \cJ^2$ gives a smooth contraction of the simplest non-trivial finite-dimensional truncation $\cA_D[\frac{D+1}{2}]$. Consequently, we expect that it is possible to define In\"on\"u-Wigner contractions for all truncated AdS algebras, but without the option of realising the limit as a quotient of the Poincar\'e UEA.

To conclude, let us stress again that the algebra $\widetilde{\mfk{Sch}}_D$ obtained in the limit contains a subalgebra spanned by two classes of generators of the massless type (i.e.\ corresponding to traceless projections of symmetrised products of Poincar\'e generators): those constructed from the product of translations only and those constructed from the product of translations and a single Lorentz generator. The proof of this statement is given in Appendix~\ref{sec:properties-Killing} via the previous differential realisation of the Schouten algebra. When $D=3$ this subalgebra corresponds to the ``right-slice'' of $\ihs(\cM^2,\cS)$ and then to $\ihs_3[\infty]$. When $D \geq 4$ a similar interpretation is missing because these generators cannot be associated to the global symmetries of any massless (or partially-massless) field, but they may admit an interesting interpretation in terms of Carrollian conformal fields in $D-1$ dimensions.

\subsection{Selected non-unitary higher-spin gauge theories in flat space}\label{sec:non-unitary}

In this subsection we present free gauge theories in Minkowski space whose  vacuum-preserving gauge transformations are in one-to-one correspondence with the generators of the algebras discussed in the previous subsection. We focus on the structure of the gauge transformations, without analysing in detail the spectrum of propagating degrees of freedom. This step was however already performed in the literature for some of the examples we are going to present and the available results indicate that these models are not unitary.

\subsubsection{Partially-massless-like gauge theories}\label{sec:PM-like}

In eq.~\eqref{Fierz-PM} we recalled the Fierz system that describes the free propagation of a partially-massless particle on a constant-curvature background. In the limit of vanishing cosmological constant ($L \to \infty$) these field equations reduce to those describing the free propagation of a massless particle on a Minkowski background, so that the higher-order gauge symmetry \eqref{gauge-PM} is enhanced to that of a massless particle. Alternatively, one can start from the Stueckelberg formulation of partially-massless fields and obtain, in the limit, a collection of massless particles \cite{Zinoviev:2001dt}. In this approach the number of propagating degrees of freedom is preserved in the limit, but in both cases the global symmetries of the resulting gauge theories in Minkowski space do not have the same structure as the global symmetries of the original partially-massless gauge theory in (A)dS.\footnote{A similar phenomenon appears in the description of massless mixed-symmetry fields in (A)dS: in the limit $L \to \infty$ the Fierz system displays an enhanced gauge symmetry and the number of propagating d.o.f. is preserved in the limit only if one introduces Stueckelberg fields in (A)dS before sending $L \to \infty$ (see, e.g., \cite{Brink:2000ag, Campoleoni:2012th}). Aside from the special case of massless symmetric Fronsdal's fields, implementing the limit of vanishing cosmological constant on the global symmetries of a given gauge theory may thus be subtle.}

On the other hand, in eq.~\eqref{sol-killing-PM} we showed that the flat limit of the higher-order Killing equation that selects the global symmetries of partially-massless fields admits at least the same number of independent solutions as its (A)dS counterpart. In view of the results discussed in section~\ref{sec:schouten-algebra}, it is thus natural to look for gauge theories in flat space whose vacuum-preserving gauge transformations satisfy the higher-order Killing equation \eqref{killing-PM}, together with the constraints \eqref{constraints-PM}. Free field theories on a constant-curvature background that satisfy this property in the $L \to \infty$ limit were actually already introduced in \cite{Campoleoni:2020ejn}. There it was observed that the equations of motion
\be \label{AdS-PM}
\begin{split}
& \left[\, \Box - \frac{(D+s-2) - (s-t-1)(D+s-t-4)}{L^2} \,\right] \vf_{\m(s)} \\
& - \frac{s(D+2s-4)}{(t+1)(D+2s-t-4)} \left( \bar{\nabla}_{\!\m\phantom{(}\!\!} \bar{\nabla} \cdot \vf_{\m(s-1)} - \frac{s-1}{D+2(s-2)}\, g_{\m\m\phantom{(}\!\!} \bar{\nabla} \cdot \bar {\nabla} \cdot \vf_{\m(s-2)} \right) = 0
\end{split}
\ee
involving the traceless field $\vf_{\m(s)}$
are left invariant by the gauge transformations
\be
\d \vf_{\m(s)} = \binom{s}{t+1}\, \bar \nabla_{\!\m\phantom{(}\!\!} \cdots \bar \nabla_{\!\m\phantom{(}\!\!} \e_{\m(s-t-1)} \, , 
\quad \textrm{with} \quad
\e_{\m(s-t-3)\l}{}^\l = \bar{\nabla} \cdot \e_{\m(s-t-2)} = 0 \, ,
\ee
where we used the same conventions as in eq.~\eqref{schouten_2}.
These equations of motion follow from the quadratic action
\be
S = \int\! \sqrt{-g}\, d^Dx\, \vf^{\m(s)} \left( \left[ \Box - m^2(D,s,t) \right] \vf_{\m(s)} - \frac{s(D+2s-4)}{(t+1)(D+2s-t-4)} \bar{\nabla}_{\!\m\phantom{(}\!\!} \bar{\nabla} \cdot \vf_{\m(s-1)} \right)
\ee
and the mass-like term they contain coincides with that appearing in the Fierz system \eqref{Fierz-PM}. For $t = 0$ one recovers the equations of motion of a massless field in their Maxwell-like form \cite{Campoleoni:2012th}, while for the other values of $t$ one gets more complicated spectra. In particular, for $s = 2$ and $t = 1$ one obtains the equations of motion introduced in \cite{Drew:1980yk} (and discussed in \cite{Deser:1983mm}), describing a partially-massless spin-two field coupled, in a non-unitary way, to an additional spin-one field. In the limit $L \to \infty$ one gets the equations of motion
\be \label{flat-PM}
\Box\,\vf_{\m(s)} - \frac{s(D+2s-4)}{(t+1)(D+2s-t-4)} \left( \pr_{\m\phantom{(}\!\!} \pr\cdot \vf_{\m(s-1)} - \frac{s-1}{D+2(s-2)}\, g_{\m\m\phantom{(}\!\!} \pr \cdot \pr \cdot \vf_{\m(s-2)} \right) = 0
\ee
that still admit gauge symmetries of the form
\be
\d \vf_{\m(s)} = \frac{s!}{(s-t-1)!}\, \pr_{\m\phantom{(}\!\!} \cdots \pr_{\m\phantom{(}\!\!} \e_{\m(s-t-1)} \, , 
\quad \textrm{with} \quad
\e_{\m(s-t-3)\l}{}^\l = \pr \cdot \e_{\m(s-t-2)} = 0 \, .
\ee

The vacuum-preserving gauge symmetries of eqs.~\eqref{flat-PM} thus satisfy the higher-order Killing equation \eqref{killing-PM} with the additional constraints \eqref{constraints-PM}. The gauge theories \eqref{flat-PM} are therefore natural candidates to substitute the partially-massless field theories with the corresponding gauge transformations in the interpretation of the role of the generators of the algebra $\widetilde{\mfk{Sch}}_D$ obtained as a contraction of the algebra $\cA_D[\m]$ in section~\ref{sec:schouten-algebra}. The price to pay is that the spectra of propagating d.o.f.\ of the two models do not coincide. It will be interesting to check more in detail if the algebras $\cA_D[\m]$ may be interpreted also as the global symmetries of a suitable combination of field equations of the form \eqref{AdS-PM} on an Anti de Sitter background and if a more natural interpretation of the algebra $\widetilde{\mfk{Sch}}_D$ is available in the context of field theories with Carrollian-conformal symmetry.

\subsubsection{Higher-derivative theories}\label{sec:higher-derivative}

A series of higher-derivative equations of motion involving a traceful tensor $\vf_{\m(s)}$ that are invariant under gauge transformations
\be \label{gauge_HS-HD}
\d \vf_{\m(s)} = \pr_{\m\phantom{(}\!\!} \e_{\m(s-1)} \,,
\ee
generated by parameters satisfying different trace constraints were introduced in \cite{Francia:2012rg} (see also \cite{Joung:2012qy}). In this context, we are interested in the last equations in that hierarchy: they can be defined starting from the de Wit and Freedmann higher-spin curvatures \cite{deWit:1979sib}
\be \label{HS-curvatures}
\cR_{\r(s)\,,\,\m(s)} \equiv \sum_{k=0}^s (-1)^{k} \binom{s}{k} \pr_\r^{s-k} \pr_\m^{k}\, \vf_{\r(k)\m(s-k)} \,,
\ee
and imposing the conditions
\begin{subequations} \label{high-derivative}
\begin{alignat}{5}
\cR^{[\frac{s}{2}]}{}_{\m(s)} & = 0 \qquad & \textrm{for} & \ s \ \textrm{even} \, , \label{high-derivative_even}\\[5pt]
\pr\cdot \cR^{[\frac{s-1}{2}]}{}_{\m(s)} & = 0 \qquad & \textrm{for} & \ s \ \textrm{odd} \, , \label{high-derivative_odd}
\end{alignat}
\end{subequations}
where the exponent $[n]$ indicates that we computed $n$ traces in the first set of indices. Similarly, the divergence in \eqref{high-derivative_odd} is computed in the first set of indices. The curvatures \eqref{HS-curvatures} are left invariant by gauge transformations of the form \eqref{gauge_HS-HD} with an \emph{unconstrained} gauge parameter and so are the equations of motion \eqref{high-derivative}. 

These equations of motion are the local counterpart of the non-local field equations $\Box^{1-\frac{s}{2}} \cR^{[\frac{s}{2}]} = 0$ for $s$ even and $\Box^{-\frac{s-1}{2}} \pr\cdot \cR^{[\frac{s-1}{2}]} = 0$ for $s$ odd introduced in \cite{Francia:2002aa}. While the latter have been shown to be equivalent to Fronsdal's equations, eqs.~\eqref{high-derivative} always contain a contribution $\Box^{\left\lfloor\frac{s+1}{2}\right\rfloor} \vf_{\m(s)}$ and, consequently, they describe the propagation of a non-unitary spectrum. A detailed analysis of the spectrum of propagated d.o.f.\ was performed in Appendix~B of \cite{Joung:2012qy} in the example of eq.~\eqref{high-derivative_even} with $s=4$ (eqs.~\eqref{high-derivative_even} and \eqref{high-derivative_odd} correspond, respectively, to the equations $\cG_{s} = 0$ and $\cM_{s+1} = 0$ of \cite{Joung:2012qy}), showing that this four-derivative equation propagates spin 4, 3, 2 massless regular modes and spin 4, 2 massless ghosts. With a similar analysis one can check that the $s=3$ instance of eq.~\eqref{high-derivative_odd} is a four-derivative equation propagating spin 3, 2, 1 massless regular modes and spin 3, 1 massless ghosts. In both cases the d.o.f.\ correspond to those of massless fields of spin $s$ and $s-2$ together with a partially-massless field of spin $s$ and depth $2$. The corresponding global symmetries match precisely the decomposition of the $\mfk{gl}(D+1)$ Young diagrams ${\tiny \yng(2,2)}$ and ${\tiny \yng(3,3)}$ in $\mfk{so}(D)$ components (ignoring the identifications among components that we used in section~\ref{sec:schouten-algebra}).

In conclusion, eqs.~\eqref{high-derivative} propagate a wider, but non-unitary spectrum compared to their non-local counterparts. We therefore conjecture that, in the absence of the equivalence to Fronsdal's equations that is induced by the non-local $\Box^{-n}$ overall factors of \cite{Francia:2002aa}, their global symmetries are in one-to-one correspondence with traceful Killing tensors of Minkowski space. Besides the observation that any Killing tensor does not induce any variation of the field via \eqref{gauge_HS-HD}, this claim is also supported by the analysis of the spectrum of propagating degrees of freedom in the first non-trivial cases, that is amenable to be rearranged in objects that display global symmetries in one-to-one correspondence with the independent solutions of the Killing equation.\\

\section{Conclusions and outlook}

In this paper we presented, for any value of $D \geq 3$, a higher-spin algebra containing a Poincar\'e subalgebra $\iso(1,D-1)$ and having the same set of generators as the Lie algebra $\mfk{hs}_D$ entering both Vasiliev's equations in AdS$_D$ \cite{Eastwood:2002su, Vasiliev:2003ev} and conformal higher-spin gravity in $D-1$ dimensions \cite{Segal:2002gd}. We also presented, again for any $D \geq 3$, a higher-spin algebra with the same set of generators and containing a Galilean conformal subalgebra $\gca_{D-1}$ \cite{Bagchi:2009my}. We realised both of them as cosets of the universal enveloping algebra of, respectively, $\iso(1,D-1)$ and $\gca_{D-1}$ and we argued in sections \ref{sec:carrollian-anyD-quotient} and \ref{sec:galilean-anyD-quotient} around eqs.~\eqref{fix-carroll-L2} and \eqref{fix-central-Galilei} that these are the unique Lie algebras with that set of generators which can be realised in this way.
When $D=3$, both algebras correspond to a member of the known one-parameter family of Minkowski higher-spin algebras \cite{Ammon:2017vwt}. In this specific case, we also explained how the previous construction can be extended to encompass a one-parameter family of quotients of $\cU(\iso(1,2))$.

We propose to interpret our Lie algebras as candidate Minkowski or Carrollian conformal higher-spin algebras in the $\iso(1,D-1)$ case and as Galilean conformal higher-spin algebras in the $\gca_{D-1}$ case. This is motivated by the observation that their generators are in one-to-one correspondence with the vacuum-preserving gauge symmetries of Fronsdal's theory. Moreover, we showed that they can be recovered as In\"on\"u-Wigner contractions of the AdS or, equivalently, conformal higher-spin algebras $\mfk{hs}_D$. Let us stress, however, that the curvatures associated to the algebras presented in section~\ref{sec:carrollian} do not reproduce upon linearisation the linearised curvatures that are used to describe the free dynamics of higher-spin particles in Minkowski space \cite{Vasiliev:1986td}. Actually, for $D = 4$ our higher-spin extension of the Poincar\'e algebra was already obtained with an In\"on\"u-Wigner contraction of the AdS$_4$ higher-spin algebra in \cite{Fradkin:1986ka} and the mismatch with the linearised curvatures of \cite{Vasiliev:1986td} was pointed out. In practice, our curvatures miss some of the terms that allow one to eliminate algebraically auxiliary fields. While this feature is clearly disturbing, we remark that similar problems already emerged in the gauging of both ultra- and non-relativistic symmetry algebras and have often been solved by gauging suitable central extensions of the algebras thereof (see, e.g., \cite{Andringa:2010it, Hartong:2015xda, Bergshoeff:2017btm}). We also point out that, compared to the linearised curvatures in Minkowski space of \cite{Vasiliev:1986td}, additional algebraic terms are present. This may suggest the option to eliminate a different set of auxiliary fields so as to describe the same free dynamics via a dualised Fronsdal field \cite{Boulanger:2003vs}. 

The question of how to gauge our algebras in order to obtain sensible interacting physical systems should also be approached from the vantage viewpoint of conformal higher-spin theories \cite{Fradkin:1985am} and their Carrollian or Galilean limits.\footnote{See \cite{Henneaux:2021yzg} for a recent analysis of Carrollian limits of Fronsdal's free theory, while some examples of field theories displaying  Galilean- or Carrollian-conformal symmetries were discussed in \cite{Bagchi:2017yvj, Bagchi:2019xfx, Ciambelli:2018wre}.} We thus defer a detailed investigation of the previous proposals to future work. On the other hand, we stress that the higher-spin extensions of the Poincar\'e algebra presented in section~\ref{sec:other-carroll} reproduce the linearised curvatures of \cite{Vasiliev:1986td} plus some additional contributions. The price to pay in this case is thus the presence of both additional generators, for which we propose an interpretation in terms of non-unitary field theories in section~\ref{sec:non-unitary}, and additional terms in the linearised curvatures involving the corresponding fields. A classification of higher-spin algebras including a Poincar\'e subalgebra but with a wider set of generators compared to those relevant to Vasiliev's equations goes beyond the scope of this work and alternatives reproducing the linearised curvatures of \cite{Vasiliev:1986td} while preserving unitarity may be available. A natural direction to explore may be the addition of mixed symmetry fields along the lines of \cite{Boulanger:2011se}. 

Aside from consolidating the interpretation of the Lie algebras discussed in this paper, our work may be extended to identify other contractions of the known  higher-spin algebras in AdS. A posteriori, the procedure we followed to build higher-spin algebras looks rather general: we took the limit of the ideal that one factors out from the universal enveloping algebra of $\mfk{so}(2,D-1)$ to get $\mfk{hs}_D$ and we factored it out from the universal enveloping algebra of the corresponding contraction of $\mfk{so}(2,D-1)$. We expect that the same procedure produce infinite-dimensional counterparts of all kinematical algebras classified in \cite{Bacry:1968zf, Bacry:1986pm, Figueroa-OFarrill:2018ilb} and it will be interesting to check this explicitly. Furthermore, the ideal to be factored out from $\cU(\mfk{so}(2,D-1))$ to get $\mfk{hs}_D$ corresponds to the annihilator of the scalar singleton module: our procedure thus defines implicitly both an ultra- and a non-relativistic limit of the scalar singleton. In particular, let us remind that the contraction to Poincar\'e leads to an ideal including the condition $\cP_a \cP_b \sim 0$, which also independently emerged in the study of the flat limit of the scalar singleton in \cite{Ponomarev:2021xdq}, but which is weaker than the trivial action of translations that characterise the flat limit proposed in \cite{Flato:1978qz}.

Alternative contractions of higher-spin algebras in three-dimensions were already considered in \cite{Bergshoeff:2016soe} and our techniques should allow one to identify those that admit a realisations as quotients of the universal enveloping algebra of the isometries of a given spatially isotropic spacetime \cite{Figueroa-OFarrill:2018ilb}. Some of these three-dimensional algebras have been already exploited to set up non-AdS higher-spin holographic correspondences \cite{Gary:2012ms, Afshar:2012nk, Gary:2014mca, Prohazka:2017lqb, Chernyavsky:2019hyp} and our results may pave the way to non-AdS holographic scenarios involving higher-spin fields even in four or more space-time dimensions (see, e.g., \cite{Campoleoni:2017mbt, Campoleoni:2020ejn} for an analysis of the symmetries that may be relevant to flat-space higher-spin holography in any dimensions). 

\acknowledgments

We are grateful to I.~Basile, X.~Bekaert, N.~Boulanger, D.~de Filippi, D.~Francia, S.~Fredenhagen, C.~Heissenberg, M.~Henneaux, Y.~Herfray, B.~Oblak, D.~Ponomarev, J.~Rosseel, P.~Salgado-Rebolledo and E.~Skvortsov for useful discussions. This work was supported by the Fonds de la Recherche Scientifique - FNRS under Grants No.\ FC.36447, F.4503.20 (HighSpinSymm) and T.0022.19 (Fundamental issues in extended gravitational theories). We thank the Erwin Scr\"odinger International Institute for Mathematics and Physics (ESI) in Vienna for hospitality during the completion of this work.

\appendix

\section{Properties of $\hs[\l]$ and $\ihs_3[\l]$} \label{app:hs[lambda]}

\paragraph{The $\hs[\l]$ algebra}

In eq.~\eqref{hs[lambda]} we presented the algebra $\hs[\l]$ as
\be
\left[ \cL_m^{(s)} \,,\, \cL_n^{(t)} \right] = \underset{s+t+u \text{ even}}{\sum_{u=|s-t|+2}^{s+t-2}} g^{st}_{s+t-u}(m,n;\l) \cL_{m+n}^{(u)} \,.
\ee
Its structure constants can be decomposed as
\be
\label{3D:structure_1} g^{st}_u(m,n;\l) = \frac{q^{u-2}}{2(u-1)!}\, \f^{st}_{u}(\l) N^{st}_{u}(m,n) \,,
\ee
with
\begin{align}
\label{3D:structure_2} N^{st}_u(m,n) &= \sum_{k=0}^{u-1} (-1)^k \binom{u - 1}{k} [ s-1+m ]_{u-k-1} \, [ s-1-m ]_k \, [ t-1+n ]_k \, [t-1-n]_{u-k-1} \,,\\
\label{3D:structure_3} \f^{st}_u(\l) &= {}_4F_3\left[ \begin{array}{c|} \frac12 + \l \,,\, \frac12 - \l \,,\, \frac{2-u}{2} \,,\, \frac{1-u}{2} \\ \frac32-s \,,\, \frac32-t \,,\, \frac12 + s+t-u \end{array} \ 1 \right] ,
\end{align}
where we fix by convention $q = \frac{1}{4}$, while $[a]_n = a(a-1)\ldots(a-n+1)$ is the descending Pochhammer symbol. These expressions, first conjectured in \cite{Pope:1989sr}, were proved in \cite{Korybut:2014jza,Basile:2016goq}.

Replacing the normalisation factor $q$ by $1/\l$ in the previous definition and taking the limit $\l \to \infty$ one obtains the finite result
\be
\tilde g^{st}_u(n,m) = \frac{N^{st}_u(n,m)}{2(u-1)!}  \frac{\left(\frac{1-u}{2}\right)_r}{\left(\frac32-s\right)_r \left(\frac32-t\right)_r \left(\frac12+s+t-u\right)_r} \,,
\ee
where $(a)_n = a(a+1) \ldots (a+n-1)$ is the raising Pochhammer symbol. We obtained in this way the structure constants of the algebra of area-preserving diffeomorphisms of the two-dimensional hyperboloid recalled in eq.~\eqref{hs[lambda]_structure-consts}, which also appear in the algebra $\ihs[\infty]$ discussed in section~\ref{sec:3D-other}.

\paragraph{The $\ihs_3[\l]$ algebra}

In section~\ref{sec:3D-flat} we often used the fact that the expressions for the higher-spin generators $P^{(s)}_m$ and $L^{(s)}_m$ differ only by the replacement of the single $P_m$ contained in the first ones by a $L_m$. Here we prove this in general starting from the Ansatz \eqref{3D-flat:Ansatz_L} and \eqref{3D-flat:Ansatz_P}.
We begin by recalling that any higher-spin generator in the $L$ sector can be decomposed as follows \cite{Pope:1989sr}:
\be \label{3D-flat:decomposition_L}
L_m^{(s)} = \sum_{\sum_i m_i = m} a_{\{m_i\}}^{(s)} \, L_{m_1} \cdots L_{m_{s-1}} \,,
\ee
with certain coefficients $a^{(s)}_{\{m_i\}}$ and the convention that the identity $id$ corresponds to $L^{(1)}_0$. This decomposition can be proved recursively on $m$: $(i)$ the $m = s-1$ case is a direct consequence of the Ansatz \eqref{3D-flat:Ansatz_L} and $(ii)$ the $m \to m-1$ case is obtained by acting on both sides with $L_{-1}$ and using the Leibniz rule.

A similar decomposition can be performed for the $P$ sector with a similar proof:
\be \label{3D-flat:decomposition_P}
P_m^{(s)} = \sum_{\sum_i m_i = m} \tilde a_{\{m_i\}}^{(s)} \, L_{m_1} \cdots L_{m_{s-2}} P_{m_{s-2}} \,,
\ee
with possibly different coefficients $\tilde a^{(s)}$ and where the unique $P$ generator in the decomposition can always be pushed to the right by using the consistency relation \eqref{3D-flat:consistency_1} without modifying the order of the $m_i$ or the number of terms (this will be our convention). We now prove that $a = \tilde a$. By acting on both sides of eq.~\eqref{3D-flat:decomposition_L} with $P_0$ and using the Leibniz rule, we obtain for $m \neq 0$
\be
\begin{split}
P_m^{(s)} &= - \frac{1}{m} \left[ P_0 \,,\, L_m^{(s)} \right] = - \frac{1}{m} \sum_{j=1}^{s-1} \sum_{\sum_i m_i = m} a_{\{m_i\}}^{(s)} \, L_{m_1} \cdots \left[P_0 \,,\, L_{m_j} \right] \cdots L_{m_{s-1}} \\
&= - \frac{1}{m} \sum_{j=1}^{s-1} \sum_{\sum_i m_i = m} (-m_j) \, a_{\{m_i\}}^{(s)} \, L_{m_1} \cdots P_{m_j} \cdots L_{m_{s-1}} \\
&\sim - \frac{1}{m} \sum_{\sum_i m_i = m} \sum_{j=1}^{s-1} (-m_j) \, a_{\{m_i\}}^{(s)} \, L_{m_1} \cdots L_{m_{s-2}} P_{m_{s-1}} \\
&= \sum_{\sum_i m_i = m} a_{\{m_i\}}^{(s)} \, L_{m_1}  \cdots L_{m_{s-2}} P_{m_{s-1}} \,,
\end{split}
\ee
where we used eq. \eqref{3D-flat:consistency_1} again to write the $P$ generator at the end of the expression while keeping the indices fixed and used the fact that $\sum_i m_i = m$. For $m=0$ instead
\begin{align}
L_0^{(s)} & = \frac{1}{s}\, [ L_1^{(s)} \,,\, L_- ] = \frac{1}{s} \sum_{j=1}^{s-1} \sum_{\sum_i m_i = 1} \!a_{\{m_i\}}^{(s)} (m_j+1)\, L_{m_1} \cdots L_{m_j-1} \cdots L_{m_{s-1}} \,, \\
P_0^{(s)} & = \frac{1}{s}\, [ L_1^{(s)} \,,\, P_- ] = \frac{1}{s} \sum_{j=1}^{s-1} \sum_{\sum_i m_i = 1} \!a_{\{m_i\}}^{(s)} (m_j+1)\, L_{m_1} \cdots L_{m_j-1} \cdots L_{m_{s-2}} P_{m_{s-1}} \,,
\end{align}
therefore we can conclude that the coefficients $\tilde a_{\{m_i\}}^{(s)}$ and $a_{\{m_i\}}^{(s)}$ in the decompositions \eqref{3D-flat:decomposition_L} and \eqref{3D-flat:decomposition_P} are the same.

\section{Differential realisation of the conformal algebra} \label{app:diff}

In this appendix we define the Carrollian ($c \to 0$) and Galilean ($c \to \infty$) limits of the conformal algebra starting from a differential realisation of its generators that allows one to single out their dependence on the time coordinate. We also use it to show that the Carrollian contraction is isomorphic to the Poincar\'e algebra.

\paragraph{Conformal algebra}

The differential realisation of the generators of conformal group in $D-1$ dimensions is
\begin{subequations} \label{conformal-generators}
\begin{alignat}{5}
J_{\m\n} &= -\left(x_\m\pr_\n - x_\n\pr_\m \right) \,, & \qquad
P_\m &= \pr_\m \,,\\[5pt]
K_\m &= 2\,x_\m x^\n \pr_\n -  x^\n x_\n \pr_\m \,, & \qquad
D &= x^\m \pr_\m \,,
\end{alignat}
\end{subequations}
where $\m, \n \in \{0,\ldots,D-2\}$.
These generators obey the following commutation rules
\begin{subequations}
\begin{align}
[J_{\m\n}, J_{\r\s}] &= \h_{\m\r}\,J_{\n\s} - \h_{\n\r}\,J_{\m\s} - \h_{\m\s}\,J_{\n\r} + \h_{\n\s}\,J_{\m\r} \,,\\
[J_{\m\n}, P_\r] &= \h_{\m\r}\,P_\n - \h_{\n\r}\,P_\m \,,\\
[J_{\m\n}, K_\r] &= \h_{\m\r}\,K_\n - \h_{\n\r}\,K_\m \,,\\
[D, P_\m] &= - P_\m \,,\\
[D, K_\m] &= K_\m \,,\\
[K_\m, P_\n] &= 2\,J_{\m\n} - 2\,\h_{\m\n}\,D \,,
\end{align}
\end{subequations}
which define the $\so(2,D-1)$ algebra in the conformal basis. 
One can split the time component from the $D-2$ spatial components to identify the generators
\begin{subequations} \label{conformal_generators_open}
\begin{alignat}{5}
J_{ij} &= -(x_i \pr_j - x_j \pr_i) \,, & \qquad 
B_i & = J_{0i} = t\,\pr_i + x_i\,\pr_t \,,\\[5pt]
P_i &= \pr_i \,, & \qquad
H &= \pr_t \,,\\[5pt]
K_i &= 2\,x_i \left(t\,\pr_t + x^j \pr_j \right) - \left(x^2 - t^2\right)\pr_i \,, & \qquad
K &= - 2\,t \left(t\,\pr_t + x^i \pr_i \right) - \left(x^2 - t^2\right)\pr_t \,, \\[5pt]
D &= t\,\pr_t + x^i\,\pr_i \,, & &
\end{alignat}
\end{subequations}
where $i,j \in \{1, \ldots, D-2\}$.

\paragraph{Carrollian contraction}

In the limit $\e \to 0$, the Carrollian rescaling $t \to \e \, t$, $x_i \to x_i$  sends the speed of light $c \simeq \frac{\text d \mathbf x}{\text d t} \to 0$ and is therefore interpreted as an ultra-relativistic limit. The generators \eqref{conformal-generators} can then be divided into two groups: $\{J_{ij}, P_i, K_i, D\}$ are not rescaled (although $K_i$ changes expression in the limit $\e \to 0$), while $\{B_i, H, K\}$ are rescaled by $\e^{-1}$ (and $K$, $B_i$ change expression as well). The generators in the first group can be regrouped into the antisymmetric tensor $\cJ_{ab}$ with $a,b \in \{ 0,\ldots,D-1 \}$ as follows:
\be \label{J-diff}
\cJ_{ij} \equiv J_{ij} \,,\quad \cJ_{i0} \equiv \frac12 \left(P_i - K_i\right) ,\quad \cJ_{i,D-1} \equiv \frac12 \left(P_i + K_i\right) ,\quad \cJ_{0,D-1} \equiv D \,.
\ee
Those in the second can be regrouped into the vector $\cP_a$ as follows:
\be \label{P-diff}
\cP_i \equiv B_i \,,\quad \cP_0 \equiv \frac12 \left(H - K\right) ,\quad \cP_{D-1} \equiv \frac12 \left(H + K\right) .
\ee
In the limit $\e \to 0$ the generators \eqref{conformal_generators_open} span, by definition, the Carrollian conformal algebra in \mbox{$D-1$} dimensions (see also \cite{Bagchi:2019xfx}). If one looks at the resulting commutators written in terms of the quantities $\cJ_{ab}$ and $\cP_a$ defined in \eqref{J-diff} and \eqref{P-diff}, one realises that this algebra is isomorphic to the Poincar\'e algebra in $D$ dimensions (i.e.\ to the $\e \to 0$ limit of \eqref{Poincare}). This is analogous to the interpretation of $\mfk{so}(2,D-1)$ either as the conformal algebra in $D-1$ dimensions or as the algebra of isometries of AdS$_D$.

\paragraph{Galilean contraction}

The Galilean rescaling of space-time coordinates $t \to t$, $x_i \to \e \, x_i$, together with the limit $\e \to 0$, sends the speed of light to infinity and is therefore interpreted as a non-relativistic limit. The generators \eqref{conformal-generators} can again be divided into two groups: $\{J_{ij}, H, D, K\}$ are not rescaled (although $K$ changes expression in the limit $\e \to 0$), while $\{P_i, B_i, K_i\}$ are rescaled by $\e^{-1}$ (and $K_i$, $B_i$ change expression as well). Defining
\begin{subequations}
\begin{alignat}{8}
\Lb_- & = -H \,, & \quad \Lb_0 & = - D \,, & \quad \Lb_+ & = K \,, \\[5pt]
T_{i,-} & = P_i \,, & \quad T_{i,0} & = B_i \,, & \quad T_{i,+} & = K_i \,,
\end{alignat}
\end{subequations}
the generators satisfy the algebra \eqref{gca_algebra} that in the limit $\e \to 0$ defines the Galilean conformal algebra $\gca_{D-1}$ (see also \cite{Bagchi:2009my}).

\section{More about Minkowski/Carrollian-conformal higher-spin algebras}

\subsection{Carrollian-conformal ideal in any dimension} \label{app:carrollian-ideal}

The ideal \eqref{carrollian_ideal} that we factored out from $\cU(\iso(1,D-1))$ to define Minkowski/Carrollian-conformal higher-spin algebras in section~\ref{sec:carrollian} contains the quadratic combinations $\cP_a \cP_b$, $\cI_a = \{ \cP^b , \cJ_{ba} \}$, $\cI_{abc} = \{ \cJ_{[ab}\, , \cP_{c]} \}$ and $\cI_{abcd} = \{ \cJ_{[ab}\, , \cJ_{cd]} \}$. We also imposed that $\cJ^2$ be proportional to the identity.
We obtained eqs.~\eqref{carrollian_ideal} via a limit of the $\so(2,D-1)$ ideal that one factorises in the coset construction of higher-spin algebras in AdS. Still, we are not aware of any general result guaranteeing that our limiting procedure preserves the property of being an ideal. For this reason, we now check explicitly that this is indeed the case.

To begin with, $\cJ^2$ becomes a central element if one factors out the ideal \eqref{carrollian_ideal}:
\be \label{check-iso-ideal_J2}
[ \cJ_{ab}\,, \cJ^2 ] = 0 \,, 
\qquad
[\cP_a\,, \cJ^2 ] = \cI_a \,, 
\ee
so that it is consistent to impose that it is proportional to the identity.
For the other elements of the ideal, all commutators with Lorentz generators clearly belong to the ideal itself since both $\cP_a$ and $\cJ_{ab}$ transform as Lorentz tensors:
\begin{subequations} \label{check-iso-ideal_Lorentz}
\begin{align}
[ \cJ_{ab}\,, \cP_c \cP_d ] &= 2\h_{a(c} \cP_{d)} \cP_b - 2\h_{b(c} \cP_{d)} \cP_a  \,, \label{check-iso-ideal:3} \\[5pt]
[ \cJ_{ab}\,,\, \cI_c ] &= \h_{ac} \cI_b - \h_{bc} \cI_a \,, \label{check-iso-ideal:4} \\[5pt]
[ \cJ_{ab}\,,\, \cI_{cde} ] &= 3\,\h_{a[c} \cI_{de]b} - 3\,\h_{b[c} \cI_{de]a} \,, \label{check-iso-ideal:5} \\[5pt]
[\cJ_{ab}\,,\, \cI_{cdef} ] &= 4\, \h_{b[c} \cI_{def]a} - 4\, \h_{a[c} \cI_{def]b} \,. \label{check-iso-ideal:6} 
\end{align}
\end{subequations}
The commutators with $\cP_a$, a priori, are not guaranteed to close on the ideal, but we can check they either give zero or an element of the ideal:
\begin{subequations} \label{check-iso-ideal_P}
\begin{align}
[\cP_a\,, \cP_b \cP_c ] &= 0 \,, \label{check-iso-ideal:7} \\[5pt]
[\cP_a\,,\, \cI_b ] &= - 2 \left( \cP_a \cP_b - \h_{ab} \cP^2 \right) , \label{check-iso-ideal:8} \\[5pt]
[\cP_a\,,\, \cI_{bcd} ] &= 0 \,, \label{check-iso-ideal:9} \\[5pt]
[\cP_a\,,\, \cI_{bcde} ] &= -\, 4\, \h_{a[b} \cI_{cde]} \,. \label{check-iso-ideal:10}
\end{align}
\end{subequations}
Notice that the elements of the ideal split in two disconnected groups as in AdS and that in the limit it is not possible to obtain $\cJ^2+\frac{(D-1)(D-3)}{4}\,id$ and $\cI_{abcd}$ by the adjoint action of translations on another element of the ideal. On the contrary, in $\so(2,D-1)$ one can recover any element from any other element in the same group by acting with $[\cP,\cdot\,]$ multiple times.

For what concerns the ideal \eqref{ideal_5D_carrollian} in $D=5$, the proof is essentially the same, with the only difference that now $\cI_{abcd}$ is substituted by
\be
\cI^\l{}_{\!\!\!abcd} \equiv \{ \cJ_{[ab} , \cJ_{cd]} \} - i\,\frac{2\l}{3}\,\ve_{abcde} \cP^e \, .
\ee
The commutator with $\cJ_{ab}$ just manifests that this quantity transforms as Lorentz tensor, while the commutator with $\cP_a$ gives the same result as in generic dimension:
\begin{subequations} \label{check-iso-ideal_modified}
\begin{align}
[\cJ_{ab}\,,\, \cI^\l{}_{\!\!\!cdef} ] &= 4\, \h_{b[c} \cI^\l{}_{\!\!\!def]a} - 4\, \h_{a[c} \cI^\l{}_{\!\!\!def]b} \, , \\[5pt]
[\cP_a\,,\, \cI^\l{}_{\!\!\!bcde} ] &= -\, 4\, \h_{a[b} \cI_{cde]} \,.
\end{align}
\end{subequations}
%

\subsection{Relativistic conformal algebra in the Carrollian basis ($D=5$)} \label{sec:carrollian-commutators}

We now display the schematic structure of the first commutators of the algebras $\mfk{hs}_5[\l]$ in the flat/Carrollian basis \eqref{Z^(s,t)}. We focus on this case because it allows one to evaluate the commutation relations on the faithful finite-dimensional representations that can be established for $\l = \frac{M}{2}+1$ with $M \in \mathbb{N}$. Computing the commutators for different values of $M$, one can then check that the structure stabilises, up to some $\l$ dependent factors in the structure constants.

\toclesslab\subsubsection{Finite-dimensional matrix representations of $\hs_5[\frac{M}{2}+1]$}{app:finite-dim_irreps_5D}

We wish to find a systematic way to obtain matrix representations of the algebra $\so(2,4) \simeq \sl(4,\mathbb R)$ bigger than the fundamental in order to realise higher-spin generators via matrix multiplication. To this end, we use a technique that applies to any $\sl(N,\mathbb R)$ algebra and that amounts to realise the generators as differential operators acting on a space of monomials of a given order. We shall then check a posteriori that, for $N=4$, the representations obtained in this way satisfy the relations $\cI_{AB} \sim 0$ and $\cI^\l_{ABCD} \sim 0$ that define the algebras $\mfk{hs}_5[\l]$ (see section~\ref{sec:5D_rel}).

If an element $a$ of $\sl(N,\mathbb R)$ is represented in the fundamental by a traceless matrix $A \in GL(N,\mathbb R)$, then we will denote by $\pi_A$ the following differential operator
\be \label{pi_A}
\pi_A = \sum_{i,j = 1}^N A^{ij} z_i \partial_j \,,
\ee
where we introduced $z_i$ and $\partial_i = \delta_{ij} \frac{\partial}{\partial z_j}$ as canonical differential operators acting on the algebra of real polynomials $\mathbb R [z_1, \ldots, z_N]$. This amounts to making the identification $\mathfrak{gl}(N,\mathbb R) \simeq End\left(\mathbb R^1 [z_1, \ldots, z_N]\right)$ where $\mathbb R^M [z_1, \ldots, z_N]$ is the space of monomials of degree $M$. A basis for those $\pi_A$ is given by the traceless set of operators 
\be \label{Lij}
L^i{}_j \equiv z^i\,\pr_j - \frac{1}{N}\,\d^i{}_j\,z^k\,\pr_k \, .
\ee
It is easy to see that $\pi$ is an algebra automorphism preserving the Lie bracket
\be
\left[\pi_A , \pi_B\right](v) \equiv \pi_A (\pi_B(v)) - \pi_A(\pi_B(v)) = \pi_{[A,B]}(v)
\ee
for any element $v$ of $\mathbb R^1 [x_1, \ldots, x_N]$, seen this time as a vector space. Conversely, given an operator $\pi_A$, one can reconstruct the associated matrix $A \in GL(N,\mathbb R)$ by observing that
\be
\pi_A(z_i) = \sum_{j=1}^N A^{ji} z_j \,,
\ee
and solve the associated linear system (well defined, since $A$ has non-zero determinant). 

The idea now is to replace the original vector space by a bigger one by acting with the operators \eqref{pi_A} on monomials of degree $M \geq 1$ (we could even produce a class of infinite-dimensional representations considering non-integer values of $M$). Consider the vectors
\be
v^M_{i_1\,\cdots\,i_N} = z_1^{i_1} \, \cdots \, z_N^{i_N} \,, \qquad 0 \leq i_1,\, \ldots,\, i_N \leq M \,,\quad i_1 + \cdots + i_N = M \,.
\ee
There are $\binom{N+M-1}{M}$ independent instances of them (the number of different monomials of degree $M$ in $N$ variables). The action of $\pi_A$ on those $v^M$'s defines another matrix, this time in $GL\left(\binom{N+M-1}{M},\mathbb R\right)$. One can then use symmetrised products of the generators of $\sl(N,\mathbb R)$ in this representation to build a finite-dimensional ``higher-spin'' algebra in the sense of \cite{Joung:2014qya}.

In particular, we can generate a matrix representation for the finite-dimensional truncations of $\hs_5[\l]$ with $\l$ half-integer containing generators of all spins from $1$ to $s = 2\,\l-1$ included. In the previous setup, this corresponds to set $N = 4$ and $M = s-1$. First of all, one can take all symmetrised products of the $\sl(4, \mathbb{R})$ generators in this representation and check that they span the fundamental representation of the  $\mfk{gl}\left(\binom{s+2}{s-1},\mathbb{R}\right)$ algebra. Its dimension then matches the number of independent components of the sum of all rectangular two-row $\mfk{o}(6)$ Young diagrams of length smaller or equal to $s-1$:\footnote{A similar counting works also for $N = 2$, which is the other case in which higher-spin algebras can be obtained as cosets of a certain $\cU(\sl(N,\mathbb R))$. Indeed, $\sum_{s'=1}^s (2s'-1) = s^2 = \binom{s}{s-1}^2$, thus matching the spectrum of $\mfk{hs}[\l]$.}
\be
\sum_{s'=1}^{s} \textrm{dim}_{\mfk{o}(6)}[\{s'-1,s'-1\}] = \sum_{s'=1}^{s} \frac{(s')^2(s'+1)^2(2s'+1)}{12}\, = \binom{s+2}{s-1}^2 \, .
\ee

We can also check that the eigenvalue of the quadratic Casimir matches that in \eqref{C2_5D}:
\be \label{C2_fin-dim}
C_2\left(\sl_N\right)\,v^M \equiv \frac{N^2}{2}\,\k^{ab}\,T_a\,T_b\,v^M = \frac{N}{4}\,L^i{}_j\,L^j{}_i\,v^M = \frac{N-1}{4}\,M\,(M+N)\,v^M \,.
\ee
For $N = 2$ eq.~\eqref{C2_fin-dim} gives $C_2 \sim \frac{\l^2-1}{4}\,id$ with $\l = M+1$ as in eq.~\eqref{C2_hs[lambda]} and for $N = 4$ it gives $C_2 \sim 3(\l^2-1)\,id$ with $\l = \frac{M}{2}+1$ as in eq.~\eqref{C2_5D}. In general, imposing $\l = \frac{2M}{N}+1$ we get $C_2 \sim \frac{N^2(N-1)}{16}\,(\l^2-1)\,id$.

Furthermore, following the notation of \cite{Joung:2014qya}, we checked that the ideal
\be
{I^\l}^{ac}_{bd} \equiv L^{[a}{}_b \otimes L^{c]}{}_d + \d^{[a}{}_{(b}\,L^{c]}{}_{d)} + \l\,\d^{[a}{}_{[b}\,L^{c]}{}_{d]} + \frac14\,(\l^2-1)\,\d^{[a}{}_{[b}\,\d^{c]}{}_{d]} \,,
\ee
corresponding to the rewriting of eqs.~\eqref{5D-ideal_1}, \eqref{5D-ideal_2} and \eqref{C2_5D} in the basis \eqref{Lij}, is factorised in the representations introduced above (that is ${I^\l}^{ac}_{bd}\,v^M = 0$), provided that one makes again the identification $N(1-\l) = -2M$.

\tocless\subsubsection{Some commutations relations of the $\hs_5[\l]$ algebra}

\noindent  
In the Carrollian basis \eqref{Poincare} the adjoint action of $\cJ_{ab}$ on a generator $\cZ^{(s,t)}_{a_1\cdots a_s|b_1 \cdots b_{s-t-1}}$ of the algebras $\mfk{hs}_D$ or $\mfk{hs}_5[\l]$ takes the form
\begin{align}
\left[\cJ_{ab}, \cZ^{(s,t)}_{a'(s-1)|b'(s-1-t)}\right] &= (s-1)\,\h_{aa'}\,\cZ^{(s,t)}_{ba'(s-2)|b'(s-1-t)} + (s-t-1)\,\h_{ab'}\,\cZ^{(s,t)}_{a'(s-1)|bb'(s-2-t)} \nn \\
&\hspace{-50pt} - (s-1)\,\h_{ba'}\,\cZ^{(s,t)}_{aa'(s-2)|b'(s-1-t)} - (s-t-1)\,\h_{bb'}\,\cZ^{(s,t)}_{a'(s-1)|ab'(s-2-t)} \,, \label{[J,Z]_app}
\end{align}
where we use a notation in which repeated indices denote a symmetrisation (again with the convention that dividing by the number of terms used in the symmetrisation is understood), while, e.g., $a(n)$ is a shorthand for a set of $n$ symmetrised indices. The adjoint action of $\cP_a$, in general, gives instead contributions involving both  $\cZ^{(s,t+1)}_{a_1\cdots a_s|b_1 \cdots b_{s-t-2}}$ and $\cZ^{(s,t-1)}_{a_1\cdots a_s|b_1 \cdots b_{s-t}}$:
\be \label{[P,Z]_app}
\begin{split}
\left[\cP_a , \cZ^{(s,t)}_{a'(s-1)|b'(s-1-t)}\right] &= c_1\,\h_{aa'}\,\cZ^{(s,t+1)}_{b'a'(s-2)|b'(s-2-t)} + c_2\,\h_{ab'}\,\cZ^{(s,t+1)}_{a'(s-1)|b'(s-2-t)} \\
& + c_3\,\h_{a'a'}\,\cZ^{(s,t+1)}_{b'aa'(s-3)|b'(s-2-t)} + c_4\,\h_{a'b'}\,\cZ^{(s,t+1)}_{aa'(s-2)|b'(s-2-t)} \\
& + c_5\,\h_{b'b'}\,\cZ^{(s,t+1)}_{a'(s-1)|ab'(s-3-t)} + c_6\,\cZ^{(s,t-1)}_{a'(s-1)|ab'(s-1-t)} \,,
\end{split}
\ee
where the constants $c_k$ are certain (possibly zero) functions of the spin $s$ and the translation number $t$. In the flat/Carrollian contraction $\cZ^{(s,t)} \to \e^{-(t \text{ mod } 2)} \cZ^{(s,t)}$, so that the commutators with $t$ odd vanish while those with $t$ even remain untouched (cf.\ \eqref{[P,3]}). 

The structure of commutators involving generators with $s > 2$ only are in general more complicated. For this reason, we limit ourself to present the schematic structure of the first few of them for the case of $\mfk{hs}_5[\l]$, that we obtained using the matrix representation of section~\ref{app:finite-dim_irreps_5D}. Commutators between spin-three fields take the form
\begin{subequations} \label{comm_5D_1}
\begin{align}
[\cK, \cK] &\propto \h\,\cZ^{(4,0)} + \h^2\,\cZ^{(4,2)} + i\l\,\ve\,\cM + \h^3\,\cJ \,,\\
[\cK, \cM] &\propto \h\,\cZ^{(4,1)} + \h^2\,\cZ^{(4,3)} + i\l\,\ve\,\cK + i\l\,\ve\,\cS + \h^3\,\cP \,,\\
[\cK, \cS] &\propto \h\,\cZ^{(4,2)} + i\l\,\ve\,\cM \,,\\
[\cM, \cM] &\propto \cZ^{(4,0)} + \h\,\cZ^{(4,2)} + i\l\,\ve\,\cM + \h^2\,\cJ \,,\\
[\cM, \cS] &\propto \cZ^{(4,1)} + \h\,\cZ^{(4,3)} + i\l\,\ve\,\cK + \h^2\,\cP \,,\\
[\cS, \cS] &\propto \cZ^{(4,2)} + \h\,\cJ \,,
\end{align}
\end{subequations}
where we showed all contributions that have non-vanishing structure constants without displaying their precise expression. For instance, a power of $\h$ represents in general a sum of several terms involving permutations of the indices in which the Minkowski metric appears. In this schematic form, eqs.~\eqref{[J,Z]_app} and \eqref{[P,Z]_app} would read
\be
[\cJ, \cZ^{(s,t)} ] \propto  \h\, \cZ^{(s,t)} \,,
\qquad
[\cP , \cZ^{(s,t)} ] \propto \h\, \cZ^{(s,t+1)} + \cZ^{(s,t-1)} \,.
\ee
Commutators of generators that have a spin strictly greater than two (e.g.\ \eqref{comm_5D_1} and \eqref{comm_5D_2}) also involve the symbol $\ve$, indicating that a dualisation was performed by using the Levi-Civita tensor in five dimensions $\ve_{abcde}$. The $i\l$ in front of those terms (that do not respect the parity rules on $s$ spelled out in section~\ref{sec:carrollian_inonu-wigner}) indicates that the relations \eqref{ideal_5D_carrollian} have also been used. These terms are present only when $D=5$ and they disappear when $\l = 0$: in this case one recovers the general structure of the commutators of $\mfk{hs}_D$.
In the contraction giving rise to the algebra $\ihs_5$, all $i\l\,\ve$ terms drop since $\l$ is set to zero and the $[\cM,\cM]$ commutator Abelianises. In the contraction obtained by rescaling the parameter $\l \to \e\,\l$ that lead to the algebra $\widetilde{\ihs}_5$, we retain terms involving a dualisation in the $[\cK,\cK]$ and $[\cK,\cS]$ commutators while they drop in the $[\cK,\cM]$, $[\cM,\cM]$ and $[\cM,\cS]$ commutators. 

The commutators between spin-three and spin-four fields are
\begin{subequations} \label{comm_5D_2}
\begin{align}
[\cK , \cZ^{(4,0)}] &\propto \h\,\cZ^{(5,0)} + \h^2\,\cZ^{(5,2)} + i\l\,\ve\,\cZ^{(4,1)} + \h^3\,\cK \,,\\
[\cK , \cZ^{(4,1)}] &\propto \h\,\cZ^{(5,1)} + \h^2\,\cZ^{(5,3)} + i\l\,\ve\,\cZ^{(4,0)} + i\l\,\ve\,\cZ^{(4,2)} + \h^3\,\cM \,,\\
[\cM , \cZ^{(4,0)}] &\propto \h\,\cZ^{(5,1)} + \h^2\,\cZ^{(5,3)} + i\l\,\ve\,\cZ^{(4,0)} + i\l\,\ve\,\cZ^{(4,2)} + \h^3\,\cM \,,\\
[\cK , \cZ^{(4,2)}] &\propto \h\,\cZ^{(5,2)} + \h^2\,\cZ^{(5,4)} + i\l\,\ve\,\cZ^{(4,1)} + i\l\,\ve\,\cZ^{(4,3)} + \h^2\,\cK + \h^3\,\cS \,,\\
[\cM , \cZ^{(4,1)}] &\propto \cZ^{(5,0)} + \h\,\cZ^{(5,2)} + \h^2\,\cZ^{(5,4)} + i\l\,\ve\,\cZ^{(4,1)} + i\l\,\ve\,\cZ^{(4,3)} + \h^2\,\cK + \h^3\,\cS \,,\\
[\cS , \cZ^{(4,0)}] &\propto \h\,\cZ^{(5,2)} + i\l\,\ve\,\cZ^{(4,1)} \,,\\
[\cK , \cZ^{(4,3)}] &\propto \h\,\cZ^{(5,3)} + i\l\,\ve\,\cZ^{(4,2)} + \h^2\,\cM \,,\\
[\cM , \cZ^{(4,2)}] &\propto \cZ^{(5,1)} + \h\,\cZ^{(5,3)} + i\l\,\ve\,\cZ^{(4,0)} + i\l\,\ve\,\cZ^{(4,2)} + \h^2\,\cM \,,\\
[\cS , \cZ^{(4,1)}] &\propto \cZ^{(5,1)} + \h\,\cZ^{(5,3)} + i\l\,\ve\,\cZ^{(4,0)} + i\l\,\ve\,\cZ^{(4,2)} + \h^2\,\cM \,,\\
[\cM , \cZ^{(4,3)}] &\propto \cZ^{(5,2)} + \h\,\cZ^{(5,4)} + i\l\,\ve\,\cZ^{(4,1)} + \h\,\cK + \h^2\,\cS \,,\\
[\cS , \cZ^{(4,2)}] &\propto \cZ^{(5,2)} + \h\,\cZ^{(5,4)} + i\l\,\ve\,\cZ^{(4,1)} + \h\,\cK + \h^2\,\cS \,,\\
[\cS , \cZ^{(4,3)}] &\propto \cZ^{(5,3)} + \h\,\cM \,.
\end{align}
\end{subequations}
In analogy with our previous discussion, in the limit in which $\l \to \e\,\l$, only the commutators $[\cK,\cZ^{(4,0)}]$, $[\cK,\cZ^{(4,2)}]$, $[\cS,\cZ^{(4,0)}]$ and $[\cS,\cZ^{(4,2)}]$ retain terms involving a dualisation.\\

\section{More about Galilean-conformal higher-spin algebras}

\subsection{Galilean-conformal ideal in any dimension} \label{sec:galilean-ideal}

We recall the relations \eqref{gal_D_singleton_ideal} obtained via the limit of the annihilator of the singleton:
\begin{subequations} \label{galilean-ideal}
\begin{alignat}{5}
\label{galilean-ideal:1} \g^{mn} \{T_{i,m}, T_{j,n}\} &\sim 0 \,, & \quad 
\cI_{ijk,m} \equiv \{ J_{[ij}, T_{k],m} \} &\sim 0 \,,\\[5pt]
\label{galilean-ideal:2} \d^{ij} \{T_{i,m}, T_{j,n}\} &\sim 0 \,, & \quad
\cI_i \equiv \g^{mn} \,\{ \Lb_m, T_{i,n} \} &\sim 0 \,,\\
\label{galilean-ideal:3} J^2 - \Lb^2 + \frac{(D-3)(D-5)}{4} \, id &\sim 0  \,, & \quad 
\cI_{ij,mn} \equiv \{ T_{[i,|m|}, T_{j],n} \} &\sim 0 \,,\\
\label{galilean-ideal:4} \cI_{i,m} \equiv \{{J_i}^j, T_{j,m} \} - (m-n) \g^{kn}  \{ \Lb_k, T_{i,m+n} \} &\sim 0 \,, & \quad
\cI_{ijkl} \equiv J_{[ij} J_{kl]} &\sim 0 \,.
\end{alignat}
\end{subequations}
We now check explicitly that they form an ideal. There are many commutation relations to be checked and the outcomes can be summarised as follows:
\begin{itemize}
\item the commutator of any expression in the first or in the second column of \eqref{galilean-ideal} with $J$, $\Lb$ or $T$ gives a result belonging to the same group;
\item the commutator of any expression with either $J$ or $\Lb$ gives back either the same expression or identically zero;
\item there is a natural hierarchy associated with the number $t$ of Galilean translations $T_{i,m}$: acting with $T_{j,n}$ on an expression with a given $t$, we end up with a relation containing $t+1$ translations, or identically zero if the original expression contains only translations;
\item expressions with $t=0$ cannot be obtained by the adjoint action of any $\gca_{D-1}$ generator on another relation of the ideal;
\item in contrast, expressions with $t=2$ can be recovered by the adjoint action of $T$ on other expressions of the ideal.
\end{itemize}
More precisely, referring at this stage only to the elements in the first column of eq.~\eqref{galilean-ideal}, the Lie bracket with $T$ of the combination \eqref{galilean-ideal:3} gives back exactly \eqref{galilean-ideal:4}:
\be
[T_{a,p}, J^2 - \Lb^2] = \{{J_a}^j, T_{j,p}\} - \g^{kn}(p-n)\{\Lb_k, T_{a,p+n}\} = \cI_{a,p} \,.
\ee
Commuting again with $T$ gives back \eqref{galilean-ideal:1} and \eqref{galilean-ideal:2}:
\be \label{comm-T-ideal}
[T_{a,p}, \cI_{i,m}] = \g_{pm} \g^{kn} \{ T_{a,k},T_{i,n} \} - \d_{ai} \d^{jk} \{ T_{j,p}, T_{k,m} \} \,.
\ee
We can indeed compute the double trace of the previous expressions. This gives 
\be
0 \sim \g_{pm} \d_{ai} \left( \frac{2}{D-2} - \frac23 \right) T^2 \sim 0 \,,
\ee
and therefore we must identify $T^2 \sim 0$ if $D \neq 5$. We can then compute a trace in each set of indices to conclude that both expressions in the first column of \eqref{galilean-ideal:1} and \eqref{galilean-ideal:2} have to be identified to zero in the quotient of the ideal. When $D = 5$, $T^2$ is set by computing the product $0 \sim \d^{ik}\,\g^{mp}\,\cI_{ij,mn} T_{k,p} \sim \frac43\,T^2\,T_{j,n}$ and one can extract from \eqref{comm-T-ideal} the conditions \eqref{gal_5D_ideal_modified:1} and \eqref{gal_5D_ideal_modified:2}.

Now moving to the elements in the second column of \eqref{galilean-ideal}, the Lie bracket of \eqref{galilean-ideal:4} (the latter exists only if $D-2 \geq 4$) with $T$ gives \eqref{galilean-ideal:1}
\be
[T_{a,p}, \cI_{ijkl}] = [T_{a,p}, J_{[ij} J_{kl]}] = 4 \d_{a[i} \{ J_{jk}, T_{l],p} \} = 4 \d_{a[i} \cI_{jkl],p} \,,
\ee
and the Lie bracket of \eqref{galilean-ideal:1} or \eqref{galilean-ideal:2} with $T$ gives back a certain combination of the elements in eq.~\eqref{galilean-ideal:3}:
\begin{align}
[T_{a,p}, \cI_{ijk,m}] &= [T_{a,p}, \{ J_{[ij}, T_{k],m} \}] = 2 \d_{a[i} \{T_{j,|p|}, T_{k],m}\} = 2\, \d_{a[i} \cI_{jk],pm} \,,\\[5pt]
[T_{a,p}, \cI_i] &= [T_{a,p}, \g^{mn} \{ \Lb_m , T_{i,n} \} ] = (p-m)\, \g^{mn} \{ T_{a,p+m}, T_{i,n} \} \nn \\
&\quad = (p-m)\, \g^{mn} \{ T_{[a,|p+m|}, T_{i],n} \} = (p-m)\, \g^{mn} \cI_{ai,(p+m)\,n} \,.
\end{align}
Finally, the Lie bracket with $T$ of the combinations entering eqs.~\eqref{galilean-ideal:1}, \eqref{galilean-ideal:2} and \eqref{galilean-ideal:3} clearly vanishes identically since in the Galilean conformal algebra 
\be
[T_{i,m}, T_{j,n} ] = 0 \,.
\ee

The neat separation between the two columns in the ideal \eqref{galilean-ideal} originates from their interpretation as the limit of the independent components of the relations $\cI_{AB} \sim 0$ and $\cI_{ABCD} \sim 0$ in the annihilator of the scalar singleton. Notice that in the non-relativistic limit it is not possible to obtain $J^2 - \Lb^2$, $\cI_i$ and $\cI_{ijkl}$ by acting on another expression of the ideal with $[T,\cdot]$. Still, the eigenvalue of $J^2 - \Lb^2$ is fixed by eq.~\eqref{fix-central-Galilei}, that mixes the two groups of elements of the ideal by considering products with $\gca_{D-1}$ generators. On the contrary, in the relativistic theory one can recover any expression from any other expression in the same group by acting with $[T,\cdot]$ multiple times.

For what concerns the ideal $\tilde \cI_\mfk g$ in $D=5$ defined in eq.~\eqref{gal_5D_ideal_modified}, we introduce
\begin{subequations}
\begin{align}
\cI^\l{}_k \equiv \g^{mn} \left\{ L_m, T_{n,k} \right\} - 2 \l \Lb_k &\sim 0 \,,\\
\bar\cI^\l{}_k \equiv \g^{mn} \left\{ \Lb_m, T_{k,n} \right\} - 2 \l L_k &\sim 0 \,,\\
\cI^\l{}_{mn,kl} \equiv \left\{ T_{m,k}, T_{n,l} \right\} - \left\{ T_{n,k}, T_{m,l} \right\} - 2\,\l\,(m-n)\,(k-l)\,T_{m+n,k+l} &\sim 0 \,.
\end{align}
\end{subequations}
The proof that the modified set of relations still span an ideal follows the same lines:
\be
\cI^\l{}_{mn,kl} = (m-n)[T_{m+n,k}, \cI^\l{}_l] = (k-l)[T_{m,k+l}, \bar\cI^\l{}_n ] \,.
\ee
In this case, however,
\be
0 \sim \g^{mp}\,\g^{kq}\,\cI^\l{}_{mn,kl}\,T_{p,q} \sim \frac43\,\left(T^2 - 3\,\l^2\,id\right)\,T_{n,l} \,,
\ee
which is solved by \eqref{gal_5D_ideal_modified:5}. Still, as we anticipated, one can consider the traceless projections in the indices $a,i$ and $p,m$ in eq.~\eqref{comm-T-ideal} to derive the relations \eqref{gal_5D_ideal_modified:1} and \eqref{gal_5D_ideal_modified:2}.

\subsection{Relativistic conformal algebra in the Galilean basis ($D=5$)} \label{sec:galilean-5D-commutators}

We now present explicitly the commutation relations of the spin-three generators of $\mfk{hs}_5[\l]$ in the Galilean basis (see Table~\ref{table3}) among themselves and with spin-two generators. To this end, we recall that in section~\ref{sec:galilean-5D} we labelled the generators as
\be
\galfiveD{q}{\qb}{s}{t}_{m,n} \,,
\ee
where $s \geq 2$ is the spin, $0 \leq t \leq s-1$ the translation number, while $(\mathbf q, \mathbf \qb) \in \mathbb N^2$ labels the two $\sl(2,\mathbb R)$ total spins. The indices $|m| \leq \mathbf q$ and $|n| \leq \mathbf \qb$ are instead tensor indices that account for the axial quantum numbers distinguishing states within the two $\sl(2,\mathbb R)$ irreps. In analogy with Appendix~\ref{sec:carrollian-commutators}, we focus on the case of $D=5$ because it allows one to compute commutators using matrix representations, while providing hints on the general structure of the algebras $\mfk{hs}_D$ in the Galilean basis.

\tocless\subsubsection{Commutators involving generators with spin 2 and 3}

In section~\ref{sec:galilean-5D} we introduced higher-spin generators by defining their highest-weight components and by acting on them with $L_m , \Lb_n$. As a result, their commutation relations with the generators of the two $\sl(2,\mathbb R)$ subalgebras read
\begin{subequations}
\begin{align} \label{gal_5D_commutators:charge}
\left[ L_m , \galfiveD{q}{\qb}{s}{t}_{n,l} \right] &= (\mathbf q\,m-n)\, \galfiveD{q}{\qb}{s}{t}_{m+n,l} \,, \\
\left[ \Lb_k , \galfiveD{q}{\qb}{s}{t}_{n,l} \right] &= (\mathbf \qb\,k-l)\, \galfiveD{q}{\qb}{s}{t}_{n,k+l} \,.
\end{align}
\end{subequations}

Introducing the spin-three generator $\Th_{m,n} \equiv L_m \Lb_n - \frac{\l}{3} T_{m,n}$ (for the reasons explained in section \ref{sec:galilean-5D}), the commutation relations of spin-three generators with the remaining generators $T_{m,n}$ of $\mfk{so}(2,D-1)$ are given, in the notation of Table~\ref{table3}, by
\begin{subequations} \label{gal_comm_2-3_explicit}
\begin{align}
[T_{m,k} , W_{n,l} ] &= \pol{1}{2}{2}(m,n) \, \pol{1}{0}{1}(k,l) \, \t_{m+n,k+l} + \frac23\, \pol{1}{2}{1}(m,n) \, \pol{1}{0}{1}(k,l) \, \Tt_{m+n,k+l} \,, 
\\
[T_{m,k} , \Wb_{n,l} ] &= \pol{1}{0}{1}(m,n) \, \pol{1}{2}{2}(k,l)\, \tb_{m+n,k+l} +  \frac23\, \pol{1}{0}{1}(m,n) \pol{1}{2}{1}(k,l)\, \Tt_{m+n,k+l} \,, 
\\[3pt]
[T_{m,k} , A_{n,l} ] &= 2\,\pol{1}{0}{1}(m,n)\,\pol{1}{0}{1}(k,l) \Tt_{m+n,k+l} \,, 
\\[5pt]
[T_{m,k} , \Th_{n,l} ] &= \pol{1}{1}{2}(m,n) \, \pol{1}{1}{1}(k,l) \, \t_{m+n,k+l} + \pol{1}{1}{1}(m,n) \, \pol{1}{1}{2}(k,l) \, \tb_{m+n,k+l} \nn \\
&\quad + \pol{1}{1}{1}(m,n) \, \pol{1}{1}{1}(k,l) \, \Tt_{m+n,k+l} 
\\
[T_{m,k} , \t_{n,l}] &= \pol{1}{2}{2}(m,n) \, \pol{1}{1}{2}(k,l) \, U_{m+n,k+l} - \frac83\, \pol{1}{2}{2}(m,n)\, \pol{1}{1}{0}(k,l)\, W_{m+n,k+l} \nn \\
&\quad - \pol{1}{2}{1}(m,n)\,\pol{1}{1}{1}(k,l)\, \Th_{m+n,k+l} \,, 
\\
[T_{m,k} , \tb_{n,l}] &= \pol{1}{1}{2}(m,n) \, \pol{1}{2}{2}(k,l) \, U_{m+n,k+l} - \frac83\,\pol{1}{1}{0}(m,n)\,\pol{1}{2}{2}(k,l)\, \Wb_{m+n,k+l} \nn \\
&\quad - \pol{1}{1}{1}(m,n)\, \pol{1}{2}{1}(k,l) \, \Th_{m+n,k+l} \,, 
\\
[ T_{m,k} , \Tt_{n,l} ] &= 2\,\pol{1}{1}{2}(m,n) \, \pol{1}{1}{2}(k,l)\, U_{m+n,k+l} + \frac83\,\pol{1}{1}{2}(m,n) \, \pol{1}{1}{0}(k,l)\, W_{m+n,k+l} \nn \\
&\quad + \frac83\,\pol{1}{1}{0}(m,n) \, \pol{1}{1}{2}(k,l)\, \Wb_{m+n,k+l} + \frac32\,\pol{1}{1}{1}(m,n)\,\pol{1}{1}{1}(k,l)\, \Th_{m+n,k+l} \nn \\
& \qquad + \frac{160}{9} \, \pol{1}{1}{0}(m,n) \,\pol{1}{1}{0}(k,l)\, A_{m+n,k+l} \,, 
\\
[T_{m,k} , U_{n,l}] &= - \frac13 \, \pol{1}{2}{2}(m,n)\,  \pol{1}{2}{1}(k,l) \, \t_{m+n,k+l} - \frac13 \, \pol{1}{2}{1}(m,n) \, \pol{1}{2}{2}(k,l) \,  \tb_{m+n,k+l} \nn \\
& \quad + \frac49\, \pol{1}{2}{1}(m,n) \,  \pol{1}{2}{1}(k,l) \, \Tt_{m+n,k+l} \,,
\end{align}
\end{subequations}
with the structure polynomials $\text{P}^{ab}_c(m,n)$ defined in eq.~\eqref{Pabc}. Using these objects, we can also present all commutation relations obtained so far in a compact form:
\begin{subequations} \label{gal_comm_2-3}
\begin{align}
\left[ L_m , \galfiveD{q}{\qb}{3}{t}_{n,l} \right] = \left[ \galfiveD{1}{0}{2}{0}_{m,0}\! , \galfiveD{q}{\qb}{3}{t}_{n,l} \right] &= \pol{1}{\mathbf q}{\mathbf q}(m,n) \, \pol{0}{\mathbf \qb}{\mathbf \qb}(0,l) \, \galfiveD{q}{\qb}{3}{t}_{m+n,l} \,,\\[5pt]
\left[ \Lb_k , \galfiveD{q}{\qb}{3}{t}_{n,l} \right] = \left[ \galfiveD{0}{1}{2}{0}_{0,k}\! , \galfiveD{q}{\qb}{3}{t}_{n,l} \right] &= \pol{0}{\mathbf q}{\mathbf q}(0,n) \, \pol{1}{\mathbf \qb}{\mathbf \qb}(k,l) \, \galfiveD{q}{\qb}{3}{t}_{n,k+l} \,,\\[5pt]
\left[ T_{m,k} , \galfiveD{q}{\qb}{3}{t}_{n,l} \right] = \left[ \galfiveD{1}{1}{2}{1}_{m,k}\! , \galfiveD{q}{\qb}{3}{t}_{n,l} \right] &\propto \!\!\sum_{t' = t \pm 1} \sum_{\mathbf q',\mathbf \qb'} \pol{1}{\mathbf q}{\mathbf q'}(m,n) \, \pol{1}{\mathbf \qb}{\mathbf \qb'}(k,l) \, \galfiveD{q'}{\qb'}{3}{t'}_{m+n,k+l} \,, \label{[T,q]}
\end{align}
\end{subequations}
where in the last line we omitted the overall coefficients depending on the labels $t'$, $\mathbf q'$ and $\mathbf \qb'$ that enter \eqref{gal_comm_2-3_explicit}.

\toclesslab\subsubsection{Commutators involving only generators with spin 3}{gal_5D_commutators_spin3_spin3}

\noindent The previous structure generalises to commutation relations involving generators with arbitrary spin as follows:
\be \label{comm-gal-gen}
\left[ \galfiveD{q_1}{\qb_1}{s_1}{t_1}_{m,k} , \galfiveD{q_2}{\qb_2}{s_2}{t_2}_{n,l} \right] \propto \sum_{s_3,t_3} \sum_{\mathbf q_3,\mathbf \qb_3} \pol{\mathbf q_1}{\mathbf q_2}{\mathbf q_3}(m,n) \, \pol{\mathbf \qb_1}{\mathbf \qb_2}{\mathbf \qb_3}(k,l) \, \galfiveD{q_3}{\qb_3}{s_3}{t_3}_{m+n,k+l} \,,
\ee
where $\propto$ means that we are omitting the portion of the structure constants that do not depend on the axial quantum numbers as in \eqref{[T,q]} (that might also vanish in some cases).

In the following, we present the missing information for all commutators of spin-three generators, discarding both the indices $m,n,k,l$ on both sides of \eqref{comm-gal-gen} and the structure polynomials. The omitted information can be recovered by remembering the charges of each object. As an example, eq.~\eqref{example} below is a shorthand for
\be
\begin{split}
& [\Th_{m,k},\Th_{n,l}] = \pol{1}{1}{2}(m,n)\,\pol{1}{1}{1}(k,l)\,Y_{m+n,k+l} + \pol{1}{1}{1}(m,n)\,\pol{1}{1}{2}(k,l)\,\Yb_{m+n,k+l} \\
& \quad - \frac83 \left(\pol{1}{1}{1}(m,n)\,\pol{1}{1}{0}(k,l)\,B_{m+n,k+l} +\pol{1}{1}{0}(m,n)\,\pol{1}{1}{1}(k,l)\,\Bb_{m+n,k+l}\right) \\
& \quad - \frac{\l}{6} \left(\pol{1}{1}{2}(m,n)\,\pol{1}{1}{1}(k,l)\,\t_{m+n,k+l} + \pol{1}{1}{1}(m,n)\,\pol{1}{1}{2}(k,l)\,\tb_{m+n,k+l} \right) \\
& \quad - \frac{8(4\l^2-9)}{63} \left(\pol{1}{1}{1}(m,n)\,\pol{1}{1}{0}(k,l)\,L_{m+n,k+l} + \pol{1}{1}{0}(m,n)\,\pol{1}{1}{1}(k,l)\,\Lb_{m+n,k+l} \right) .
\end{split}
\ee
The commutation relations are grouped in the following according to the number of ``Galilean translations'' $T_{m,n}$ they contain. We also resort again to the notation introduced in Table~\ref{table3} to label generators so that, e.g., $X_{m,n} \simeq \galfiveD{3}{0}{4}{0}_{m,n}$. Moreover, we only display non-vanishing commutators, so that we omit $[W,\Wb] = [W,A] = [\Wb,A] = 0$ from \eqref{RR-gal}.\\

\paragraph{Rotation with rotation: $[\galfiveD{q}{\qb}{3}{0} , \galfiveD{q}{\qb}{3}{0}]$}

\begin{subequations} \label{RR-gal}
\begingroup
\allowdisplaybreaks
\begin{align}
[W,W] &= X - \frac{8}{15}\, B - \frac{4\l^2-9}{63}\, L \, , \\[3pt]
[\Wb,\Wb] &= \Xb - \frac{8}{15}\, \Bb - \frac{4\l^2-9}{63}\, \Lb \, , \\[3pt]
[W,\Th] &= Y + \frac{\l}{6} \left(\t + \frac43\, \Tt\right) , \\[3pt]
[\Wb,\Th] &= \Yb + \frac{\l}{6} \left(\tb + \frac43\, \Tt\right) , \\[3pt]
[\Th,A] &= -\frac{2\l}{3}\, \Tt \, , \\[3pt]
[\Th,\Th] &= Y + \Yb - \frac83\, B - \frac83\, \Bb -\frac{\l}{6} \left(\t + \tb \right) - 8\,\frac{4\l^2-9}{63} \left(L + \Lb \right) . \label{example}
\end{align}
\endgroup
\end{subequations}

\paragraph{Rotation with mixed: $[\galfiveD{q}{\qb}{3}{0} , \galfiveD{q}{\qb}{3}{1}]$}

\begin{subequations}
\begingroup
\allowdisplaybreaks
\begin{align}
[W,\t] &= \s - \frac29\, \pi - \frac15\, \pit - \frac{\l}{6}\, \Th - 2\,\frac{4\l^2-9}{63}\, T \, , \\[3pt]
[W,\tb] &= \sh - \frac23\, \pib + \frac{\l}{2}\, U \, , \\[3pt]
[W,\Tt] &= \pi - \frac23\, \pit + \frac{\l}{3}\, \Th - 2\,\frac{4\l^2-9}{63}\, T \, , \\
[\Wb,\t] &= \sh - \frac23\, \pi + \frac{\l}{2}\, U \, , \\[3pt]
[\Wb,\tb] &= \sb - \frac29\, \pib - \frac15\, \pit - \frac{\l}{6}\, \Th - 2\,\frac{4\l^2-9}{63}\, T \, , \\[3pt]
[\Wb,\Tt] &= \pib - \frac23\, \pit + \frac{\l}{3}\, \Th - 2\,\frac{4\l^2-9}{63}\, T \, , \\[3pt]
[\Th,\t] &= \s + \sh - \frac23\, \pi - \frac15\, \pit + \frac{\l}{6}\left(U + \Th - \frac83\, W\right) - 2\,\frac{4\l^2-9}{63}\, T \, , \\[3pt]
[\Th,\tb] &= \sb + \sh - \frac23\, \pib - \frac15\, \pit + \frac{\l}{6}\left(U + \Th - \frac83\, \Wb \right) - 2\, \frac{4\l^2-9}{63}\, T \, , \\[3pt]
[\Th,\Tt] &= \pi - \pit + \pib - \frac{2\l}{3}\left(U + \frac43\, W + \frac43\, \Wb + \frac{80}{9}\, A \right) - \frac{4\l^2-9}{21}\, T \, , \\[3pt]
[A,\t] &= -2\, \pi \, , \\[3pt]
[A,\tb] &= -2\, \pib \, , \\[3pt]
[A,\Tt] &= -2\, \pit + \l\,\Th - 2\, \frac{4\l^2-9}{21}\, T \, .
\end{align}
\endgroup
\end{subequations}

\paragraph{Rotation with translation: $[\galfiveD{q}{\qb}{3}{0} , \galfiveD{q}{\qb}{3}{2}]$}

\begin{subequations}
\begin{align}
[W,U] &= \r - \frac29\, \rt + \frac{1}{15}\, \Yb - \frac{\l}{6}\, \tb \, , \\[3pt]
[\Wb,U] &= \rb - \frac29\, \rt + \frac{1}{15}\, Y - \frac{\l}{6}\, \t \, , \\[3pt]
[\Th,U] &= \r + \rb - \frac13\, \rt + \frac{1}{15} \left(\Yb + Y\right) - \frac{\l}{36}\left(2\, \tb + 2\, \t + \frac13\, \Tt \right) , \\[3pt]
[A,U] &= -2 \,  \rt \, .
\end{align}
\end{subequations}

\paragraph{Mixed with mixed: $[\galfiveD{q}{\qb}{3}{1} , \galfiveD{q}{\qb}{3}{1}]$}

\begin{subequations}
\begingroup
\allowdisplaybreaks
\begin{align}
[\t,\t] &= \frac32 \r - \frac{1}{15}\, \Yb - \frac13\, Y - 4\, X + \frac{8}{45}\, B + \frac83\, \Bb - \frac{\l}{6}\left(-\frac13\, \t + \tb\right) \nn \\
& + \frac{4\l^2-9}{21}\left(\frac83\, L + \frac83\,\Lb \right) , 
\\[3pt]
[\t,\tb] &= \rb + \r - \frac35\, Y - \frac35\, \Yb - \frac{\l}{6}\left(\t + \tb\right) , 
\\[3pt]
[\tb,\tb] &= \frac32\, \rb - \frac{1}{15}\, Y - \frac13\, \Yb - 4\, \Xb + \frac{8}{45}\, \Bb + \frac83\, B - \frac{\l}{6}\left(- \frac13\, \tb + \t\right) \nn \\
& + \frac{4\l^2-9}{21}\left(\frac83\,\Lb + \frac83\,L \right) , 
\\[3pt]
[\t,\Tt] &= 2 \r + \frac13\, \rt - Y + \frac45\, \Yb + \frac83\, X + \frac{112}{15}\, B + \frac{\l}{3}\, \Tt \, , 
\\[3pt]
[\tb,\Tt] &= 2 \rb + \frac13\, \rt - \Yb + \frac45\, Y + \frac83\, \Xb + \frac{112}{15}\, \Bb + \frac{\l}{3}\, \Tt \, , 
\\[3pt]
[\Tt,\Tt] &= Y + \Yb + \frac{40}{3} \left(B + \Bb\right) + \frac{\l}{2}(\t + \tb) + 4\,\frac{4\l^2-9}{21}\left(L + \Lb\right) .
\end{align}
\endgroup
\end{subequations}

\paragraph{Mixed with translation: $[\galfiveD{q}{\qb}{3}{1} , \galfiveD{q}{\qb}{3}{2}]$}

\begin{subequations}
\begingroup
\allowdisplaybreaks
\begin{align}
[\t,U] &= \frac12 V - \frac{7}{30}\, \s - \frac19\, \sh - \frac{1}{15}\, \sb + \frac{4}{27} \pi - \frac{3}{25}\, \pit + \frac{\l}{18} \left(U + \Th + 8\,\Wb\right) + 2\,\frac{4\l^2-9}{189}\, T \, , \\
[\tb,U] &= \frac12\, V - \frac{7}{30}\, \sb - \frac19\, \sh - \frac{1}{15}\, \s + \frac{4}{27}\, \pib - \frac{3}{25}\, \pit + \frac{\l}{18} \left(U + \Th + 8\, W\right) + 2\,\frac{4\l^2-9}{189}\, T \, , \\
[\Tt,U] &= -2\, V - \frac25\left(\s + \sb\right) - \frac13\, \sh + \frac13\left(\pi + \pib\right) - \frac{152}{225}\, \pit - \frac{\l}{3}\, \Th + 2\frac{4\l^2-9}{63}\, T \, .
\end{align}
\endgroup
\end{subequations}

\paragraph{Translation with translation: $[\galfiveD{q}{\qb}{3}{2} , \galfiveD{q}{\qb}{3}{2}]$}

\be
\begin{split}
[U,U] &= - \frac{1}{18} \left(\r + \rb\right) + \frac{2}{9} \left(X + \Xb\right) + \frac{7}{135} \left(Y + \Yb\right) + \frac{104}{135}\left(B + \Bb \right) \\
& - \frac{\l}{54}\left(\t + \tb\right) - 8\frac{4\l^2-9}{189}\left(L + \Lb\right) .
\end{split}
\ee

\paragraph{Structure Polynomials}

The structure polynomials entering eqs.~\eqref{gal_comm_2-3_explicit} read
\begin{subequations}
\begin{align}
\pol{a}{b}{c}(m,n) &= 0 \quad \text{for } c > a+b \quad \textrm{and} \quad \pol{a}{b}{c}(m,n) = 1 \quad \text{for } c = a+b \,,\\
\pol{a}{b}{c}(m,n) &= b\, m - a\,n \quad \textrm{for } c = a+b-1 \,,\\
\pol{1}{2}{1}(m,n) &= \frac14 \left(6\,m^2 - 3\,m\,n + n^2 - 4\right) ,\\
\pol{2}{2}{1}(m,n) &= \frac14\, (m-n)\left(2\,m^2 - m\,n + 2\,n^2 - 8\right) ,\\
\pol{2}{2}{2}(m,n) &= \frac34 \left(2\,m^2 - 3\,m\,n + 2\,n^2 - 4\right) ,\\
\pol{1}{1}{0}(m,n) &= \frac14 \left(m^2 - m\,n + n^2 - 1\right) = -\frac14\, \g_{mn} \quad \text{(the Killing metric of } \sl(2,\mathbb R)) \,,\\
\pol{2}{2}{0}(m,n) &= \frac{1}{16} \left(m^4 - m^3\,n + m^2\,n^2 - m\,n^3 + n^4 - 5\,m^2 + 5\,m\,n - 5\,n^2 + 4\right) ,
\end{align}
\end{subequations}
and we conjecture that, in general, they take the form
\be \label{Pabc}
\text{P}^{ab}_c(m,n) = \frac{2^{c-a-b}}{(a+b-c)!}\sum_{k=0}^{a+b-c} (-1)^k \binom{a+b-c}{k} [a+m]_{a+b-c-k} [a-m]_k [b+n]_k [b-n]_{a+b-c-k} \,,
\ee
inspired by the structure constants of $\mfk{hs}[\l]$ (these objects just encode the action of the two $\sl(2,\mathbb{R})$ subalgebras $L_m$ and $\Lb_m$ on objects with a certain conformal weight).

\section{Killing tensors in flat space and the Schouten bracket} \label{sec:properties-Killing}

\paragraph{Schouten bracket of Killing tensors}

The vector space of Killing tensors endowed with the Schouten bracket naturally forms an algebra because the Schouten bracket of two Killing tensors is again a Killing tensor. For $K_1$ and $K_2$, two traceless Killing tensors with rank $p$ and $q$ we have
\be \label{schouten_app}
[K_1,K_2]^{\m(p+q-1)} \equiv k[p,q] \left( p\, K_1^{\l \m(p-1)} \pr_\l K_2^{\m(q)} - q\, K_2^{\l \m(q-1)} \pr_\l K_1^{\m(p)} \right) ,
\ee
with $k[p,q] = \frac{(p+q-1)!}{p!q!}$, and
\be \label{killing-to-killing}
\begin{split}
\pr^\m [K_1,K_2]^{\m(p+q-1)} &= k[p,q] \left( p\, \pr^\m K_1^{\l \m(p-1)} \pr_\l K_2^{\m(q)} - q\, \pr^\m K_2^{\l \m(q-1)} \pr_\l K_1^{\m(p)} \right) \\
&= k[p,q] \left( \left(- \pr^\l K_1^{\m(p)}\right) \pr_\l K_2^{\m(q)} - \left(- \pr^\l K_2^{\m(p)}\right) \pr_\l K_1^{\m(q)} \right) = 0
\end{split}
\ee
by means of the Killing equation.

\subparagraph{Proposition (3D-like subalgebra)}

The algebra $\widetilde{\mfk{Sch}}_D$ ---~obtained by identifying $\h^{\m\n} \sim id$ within the Schouten bracket algebra of Minkowski Killing tensors in any dimensions~--- contains a subalgebra spanned by only two classes of generators per value of the spin (and only the identity for $s = 1$) and containing a Poincar\'e subalgebra. It is composed of the traceless projections of the tensors
\be
\cP_{i(s-1)}^{\m(s-1)} \equiv (\cP_i)^\m \cdots (\cP_i)^\m \,,\quad \cJ_{i(s-1)|j}^{\m(s-1)} \equiv (\cP_i)^\m \cdots (\cP_i)^\m\, (\cJ_{ij})^\m \,,
\ee
with homogeneity degree $0$ and $1$ respectively. We have the following commutation relations under the Schouten bracket
\begin{subequations}
\begin{align}
\left[\cJ_{i(p)|j}, \cJ_{k(q)|l}\right]^{\m(p+q-1)} &= f_{i(p)|j,k(q)|l}^{m(p+q-1)|n} \cJ_{m(p+q-1)|n}^{\m(p+q-1)} \nn \\
& + \h^{\m\m} {f'}_{i(p)|j,k(q)|l}^{m(p+q-3)|n} \cJ_{m(p+q-3)|n}^{\m(p+q-3)} + \ldots \,,\\[5pt]
\left[\cJ_{i(q)|j}, \cP_{k(p)}\right]^{\m(p+q-1)} &= g_{i(p)|j,k(q)}^{m(p+q-1)} \cP_{m(p+q-1)}^{\m(p+q-1)} + \h^{\m\m} {g'}_{i(p)|j,k(q)}^{m(p+q-3)} \cP_{m(p+q-3)}^{\m(p+q-3)} + \ldots\,,\\[5pt]
\left[\cP_{i(p)}, \cP_{k(q)}\right]^{\m(p+q-1)} &= 0 \,,
\end{align}
\end{subequations}
where $f, f', \ldots$ and $g, g', \ldots$ are structure constants build out of the tensor $\h_{ij}$.

\subparagraph{Proof}

The $\cP$'s span the basis of traceless constant tensors so $[\cP,\cP] = 0$ because the Schouten bracket involves one derivative. Also, $[\cJ,\cP]$ is a constant (traceful) tensor which can be naturally decomposed on the basis of traceless constant tensors which is given by the $\cP$'s. Therefore, it remains to prove that the Schouten bracket of two $\cJ$'s can be decomposed into a sum of $\cJ$'s. To this end, we first show that $[\cJ_1,\cJ_2]$ is divergence-free:
\begin{align}
& \pr_\m [\cJ_1, \cJ_2]^{\m(p+q-1)} = k[p,q] \pr_\m \left(p\,\cJ_1^{\l\m(p-1)}\pr_\l\,\cJ_2^{\m(q)} - q\,\cJ_2^{\l\m(q-1)}\pr_\l\,\cJ_2^{\m(p)}\right)  \nn \\
&\qquad = k[p,q] \left(\binom{p}{2}\,\pr_\n\,\cJ_1^{\l\n\m(p-2)}\pr_\l\,\cJ_2^{\m(q)} + p\,q\,\pr_\n\,\cJ_1^{\l\m(p-1)}\pr_\l\,\cJ_2^{\n\m(q-1)} \right. \nn \\
&\qquad\quad \left. - \binom{q}{2}\,\pr_\n\,\cJ_2^{\l\n\m(q-2)}\pr_\l\,\cJ_1^{\m(p)} - p\,q\,\pr_\n\,\cJ_2^{\l\m(q-1)} \pr_\l\,\cJ_1^{\n\m(p-1)} \right) = 0 \,,
\end{align}
where we used the Leibniz rule and the fact that $\pr_\n \pr_\l \cJ^{\m(p)} = 0$ because $\cJ$ has homogeneity 1 in $x$ and also that the $\cJ$'s are divergence-free (any traceless Killing tensor is also divergenceless). Note that in general, the Schouten bracket of two divergence-free Killing tensors is not a divergence-free Killing tensor, but it is true in this particular case because of homogeneity. By taking successive traces of this relation, we immediately know that
\be
\pr_\m [\cJ_1, \cJ_2]^{\m(p+q-1-2n)} = 0 \,, \quad \text{for } 0 \leq n \leq \left\lfloor \frac{p+q-2}{2} \right\rfloor \,,
\ee
where we denoted the $n$-th trace of a tensor by the same object with $2n$ indices less. We can now prove by simple recursion that all the successive traces of $[\cJ_1, \cJ_2]^{\m(p+q-1)}$ are Killing tensors by simply noticing that if $[\cJ_1,\cJ_2]^{\m(p+q-1-2n)}$ verifies the Killing equation then
\begin{align}
0 &= \h_{\m\m} \left(\pr^\m [\cJ_1,\cJ_2]^{\m(p+q-1-2n)}\right) \nn \\
&= \binom{p+q-1-2n}{1} \pr_\m [\cJ_1,\cJ_2]^{\m(p+q-1-2n)} + \binom{p+q-1-2n}{2} \pr^\m [\cJ_1,\cJ_2]^{\m(p+q-1-2n-2)} \nn \\
&= \binom{p+q-1-2n}{2} \pr^\m [\cJ_1,\cJ_2]^{\m(p+q-1-2n-2)} \,,
\end{align}
thanks to the property proved above. This constitutes the property that we want to prove for $n \to n+1$ and the case $n = 0$ has been proved in eq. \eqref{killing-to-killing}.
Therefore, we can decompose $[\cJ_1,\cJ_2]^{\m(p+q-1)}$ into a sum of traceless tensors, all of which satisfy the Killing equation. We know that the only traceless tensors verifying the Killing equation with homogeneity 1 are the $\cJ$'s which concludes the proof.

\paragraph{Useful properties of Minkowski Killing tensors}

We will now prove a certain number of properties of Minkowski Killing tensors.

\subparagraph{Proposition (solutions to the Killing equation)} For a given rank $p$, there are three classes of solutions to the Killing equation: solutions that are traceless (we have already characterised them using eq.~\eqref{sol-killing}: there are $\{p,p\}$ of them in $\mfk{o}(D+1)$ and we also know that their homogeneity in $x$ goes from $p$ to $0$) that represent the global symmetries of massless fields; solutions that are ``pure traces'' (all the solutions to the Killing equation of rank $p-2$, $p-4$ etc. form perfectly valid solutions of rank $p$ once symmetrised with the proper number of $\h$'s) that are to be identified with lower spins; and finally solutions that don't fit in either categories (i.e.\ those that are neither traceless nor have a zero traceless projection) and from counting we identify them with the global isometries of ``partially-massless-like'' fields in flat space (there are $\{p,p-2\}$, $\{p,p-4\}$ etc. of them in $\mfk{o}(D+1)$).

\subparagraph{Proposition (higher-order Killing equation)} if $K^{\m(p)}$ verifies the Killing equation then its traces $K^{\m(p-2n)}$ verify an order $2n+1$ Killing equation for $n \leq \left\lfloor \frac{p}{2} \right\rfloor$.

\subparagraph{Proof} the case $n=0$ is trivially satisfied so let us assume by recursion that the case $n < \left\lfloor \frac{p}{2} \right\rfloor$ is true. Then, taking two gradients and a trace of the equation $\pr^{\m(2n+1)}\,K^{\m(p-2n)} = 0$ we have that
\begin{align}
0 &= \binom{2n+3}{2}\,\Box\,\pr^{\m(2n+1)}\,K^{\m(p-2n)} + \binom{2n+3}{1}\,\binom{p-2n}{1}\,\pr^{\m(2n+2)}\,\pr\,\cdot\,K^{\m(p-1-2n)} \nn \\
&\quad + \binom{p-2n}{2}\,\pr^{\m(2n+3)}\,K^{\m(p-2n-2)} \,.
\end{align}
The first term vanishes by hypothesis and the second term can be rewritten as
\be
\pr_\a\,\pr^{\m(2n+2)}\,K^{\a \m(p-1-2n)} \\
= \frac{1}{p-2\,n} \pr_\a\,\pr^{\m(2n+1)}\,\left(2\,\pr^{(\m}\,K^{\a \m(p-1-2n))} - \pr^\a\,K^{\m(p-2n)}\right) ,
\ee
where the two terms vanish separately as a consequence of the hypothesis. Therefore, we are left with $\pr^{\m(2n+3)}\,K^{\m(p-2n-2)} = 0$ which concludes the proof.


\bibliographystyle{JHEP}

\begin{thebibliography}{200}
\expandafter\ifx\csname url\endcsname\relax
  \def\url#1{\texttt{#1}}\fi
\expandafter\ifx\csname urlprefix\endcsname\relax\def\urlprefix{URL }\fi
\expandafter\ifx\csname href\endcsname\relax
  \def\href#1#2{#2} \def\path#1{#1}\fi

\bibitem{Fradkin:1986ka}
E.S.~Fradkin and M.A.~Vasiliev, \emph{{Candidate to the Role of Higher Spin
  Symmetry}}, \href{https://doi.org/10.1016/S0003-4916(87)80025-8}{\emph{Annals
  Phys.} {\bfseries 177} (1987) 63}.

\bibitem{Vasiliev:1990en}
M.A.~Vasiliev, \emph{{Consistent equation for interacting gauge fields of all
  spins in (3+1)-dimensions}},
  \href{https://doi.org/10.1016/0370-2693(90)91400-6}{\emph{Phys. Lett. B}
  {\bfseries 243} (1990) 378}.

\bibitem{Vasiliev:1986qx}
M.A.~Vasiliev, \emph{{Extended Higher Spin Superalgebras and Their Realizations
  in Terms of Quantum Operators}},
  \href{https://doi.org/10.1002/prop.2190360104}{\emph{Fortsch. Phys.}
  {\bfseries 36} (1988) 33}.

\bibitem{Konstein:1989ij}
S.E.~Konstein and M.A.~Vasiliev, \emph{{Extended Higher Spin Superalgebras and
  Their Massless Representations}},
  \href{https://doi.org/10.1016/0550-3213(90)90216-Z}{\emph{Nucl. Phys. B}
  {\bfseries 331} (1990) 475}.

\bibitem{Sezgin:2001zs}
E.~Sezgin and P.~Sundell, \emph{{Doubletons and 5-D higher spin gauge theory}},
  \href{https://doi.org/10.1088/1126-6708/2001/09/036}{\emph{JHEP} {\bfseries
  09} (2001) 036} [\href{https://arxiv.org/abs/hep-th/0105001}{{\ttfamily
  hep-th/0105001}}].

\bibitem{Vasiliev:2001wa}
M.A.~Vasiliev, \emph{{Cubic interactions of bosonic higher spin gauge fields in
  AdS$_5$}}, \href{https://doi.org/10.1016/S0550-3213(01)00433-3}{\emph{Nucl.
  Phys. B} {\bfseries 616} (2001) 106}
  [\href{https://arxiv.org/abs/hep-th/0106200}{{\ttfamily hep-th/0106200}}].

\bibitem{Sezgin:2001ij}
E.~Sezgin and P.~Sundell, \emph{{7-D bosonic higher spin theory: Symmetry
  algebra and linearized constraints}},
  \href{https://doi.org/10.1016/S0550-3213(02)00299-7}{\emph{Nucl. Phys. B}
  {\bfseries 634} (2002) 120}
  [\href{https://arxiv.org/abs/hep-th/0112100}{{\ttfamily hep-th/0112100}}].

\bibitem{Eastwood:2002su}
M.G.~Eastwood, \emph{{Higher symmetries of the Laplacian}},
  \href{https://doi.org/10.4007/annals.2005.161.1645}{\emph{Annals Math.}
  {\bfseries 161} (2005) 1645}
  [\href{https://arxiv.org/abs/hep-th/0206233}{{\ttfamily hep-th/0206233}}].

\bibitem{Vasiliev:2003ev}
M.A.~Vasiliev, \emph{{Nonlinear equations for symmetric massless higher spin
  fields in (A)dS(d)}},
  \href{https://doi.org/10.1016/S0370-2693(03)00872-4}{\emph{Phys. Lett. B}
  {\bfseries 567} (2003) 139}
  [\href{https://arxiv.org/abs/hep-th/0304049}{{\ttfamily hep-th/0304049}}].

\bibitem{Vasiliev:2004qz}
M.A.~Vasiliev, \emph{{Higher spin gauge theories in various dimensions}},
  \href{https://doi.org/10.1002/prop.200410167}{\emph{Fortsch. Phys.}
  {\bfseries 52} (2004) 702}
  [\href{https://arxiv.org/abs/hep-th/0401177}{{\ttfamily hep-th/0401177}}].

\bibitem{Vasiliev:2004cm}
M.A.~Vasiliev, \emph{{Higher spin superalgebras in any dimension and their
  representations}},
  \href{https://doi.org/10.1088/1126-6708/2004/12/046}{\emph{JHEP} {\bfseries
  12} (2004) 046} [\href{https://arxiv.org/abs/hep-th/0404124}{{\ttfamily
  hep-th/0404124}}].

\bibitem{Boulanger:2011se}
N.~Boulanger and E.D.~Skvortsov, \emph{{Higher-spin algebras and cubic
  interactions for simple mixed-symmetry fields in AdS spacetime}},
  \href{https://doi.org/10.1007/JHEP09(2011)063}{\emph{JHEP} {\bfseries 09}
  (2011) 063} [\href{https://arxiv.org/abs/1107.5028}{{\ttfamily 1107.5028}}].

\bibitem{Joung:2015jza}
E.~Joung and K.~Mkrtchyan, \emph{{Partially-massless higher-spin algebras and
  their finite-dimensional truncations}},
  \href{https://doi.org/10.1007/JHEP01(2016)003}{\emph{JHEP} {\bfseries 01}
  (2016) 003} [\href{https://arxiv.org/abs/1508.07332}{{\ttfamily
  1508.07332}}].

\bibitem{Alkalaev:2014nsa}
K.B.~Alkalaev, M.~Grigoriev and E.D.~Skvortsov, \emph{{Uniformizing higher-spin
  equations}}, \href{https://doi.org/10.1088/1751-8113/48/1/015401}{\emph{J.
  Phys. A} {\bfseries 48} (2015) 015401}
  [\href{https://arxiv.org/abs/1409.6507}{{\ttfamily 1409.6507}}].

\bibitem{Brust:2016zns}
C.~Brust and K.~Hinterbichler, \emph{{Partially Massless Higher-Spin Theory}},
  \href{https://doi.org/10.1007/JHEP02(2017)086}{\emph{JHEP} {\bfseries 02}
  (2017) 086} [\href{https://arxiv.org/abs/1610.08510}{{\ttfamily
  1610.08510}}].

\bibitem{Blencowe:1988gj}
M.P.~Blencowe, \emph{{A Consistent Interacting Massless Higher Spin Field
  Theory in $D$ = (2+1)}},
  \href{https://doi.org/10.1088/0264-9381/6/4/005}{\emph{Class. Quant. Grav.}
  {\bfseries 6} (1989) 443}.

\bibitem{Bergshoeff:1989ns}
E.~Bergshoeff, M.P.~Blencowe and K.S.~Stelle, \emph{{Area Preserving
  Diffeomorphisms and Higher Spin Algebra}},
  \href{https://doi.org/10.1007/BF02108779}{\emph{Commun. Math. Phys.}
  {\bfseries 128} (1990) 213}.

\bibitem{Campoleoni:2010zq}
A.~Campoleoni, S.~Fredenhagen, S.~Pfenninger and S.~Theisen, \emph{{Asymptotic
  symmetries of three-dimensional gravity coupled to higher-spin fields}},
  \href{https://doi.org/10.1007/JHEP11(2010)007}{\emph{JHEP} {\bfseries 11}
  (2010) 007} [\href{https://arxiv.org/abs/1008.4744}{{\ttfamily 1008.4744}}].

\bibitem{Grigoriev:2020lzu}
M.~Grigoriev, K.~Mkrtchyan and E.~Skvortsov, \emph{{Matter-free higher spin
  gravities in 3D: Partially-massless fields and general structure}},
  \href{https://doi.org/10.1103/PhysRevD.102.066003}{\emph{Phys. Rev. D}
  {\bfseries 102} (2020) 066003}
  [\href{https://arxiv.org/abs/2005.05931}{{\ttfamily 2005.05931}}].

\bibitem{Prokushkin:1998bq}
S.F.~Prokushkin and M.A.~Vasiliev, \emph{{Higher spin gauge interactions for
  massive matter fields in 3-D AdS space-time}},
  \href{https://doi.org/10.1016/S0550-3213(98)00839-6}{\emph{Nucl. Phys. B}
  {\bfseries 545} (1999) 385}
  [\href{https://arxiv.org/abs/hep-th/9806236}{{\ttfamily hep-th/9806236}}].

\bibitem{Henneaux:1985ey}
M.~Henneaux, \emph{{Asymptotically anti-de Sitter universes in $d = 3, 4$ and
  higher dimensions}},  in \emph{{Proceedings of the 4th Marcel Grossmann
  Meeting on General Relativity}}, R.~Ruffini, ed., pp.~959--966, Elsevier
  Science Publishers, 1986.

\bibitem{Campoleoni:2016uwr}
A.~Campoleoni, M.~Henneaux, S.~H\"ortner and A.~Leonard, \emph{{Higher-spin
  charges in Hamiltonian form. I. Bose fields}},
  \href{https://doi.org/10.1007/JHEP10(2016)146}{\emph{JHEP} {\bfseries 10}
  (2016) 146} [\href{https://arxiv.org/abs/1608.04663}{{\ttfamily
  1608.04663}}].

\bibitem{Campoleoni:2017vds}
A.~Campoleoni, M.~Henneaux, S.~H\"ortner and A.~Leonard, \emph{{Higher-spin
  charges in Hamiltonian form. II. Fermi fields}},
  \href{https://doi.org/10.1007/JHEP02(2017)058}{\emph{JHEP} {\bfseries 02}
  (2017) 058} [\href{https://arxiv.org/abs/1701.05526}{{\ttfamily
  1701.05526}}].

\bibitem{Brown:1986nw}
J.D.~Brown and M.~Henneaux, \emph{{Central Charges in the Canonical Realization
  of Asymptotic Symmetries: An Example from Three-Dimensional Gravity}},
  \href{https://doi.org/10.1007/BF01211590}{\emph{Commun. Math. Phys.}
  {\bfseries 104} (1986) 207}.

\bibitem{Henneaux:2010xg}
M.~Henneaux and S.-J.~Rey, \emph{{Nonlinear $W_{\infty}$ as Asymptotic
  Symmetry of Three-Dimensional Higher Spin Anti-de Sitter Gravity}},
  \href{https://doi.org/10.1007/JHEP12(2010)007}{\emph{JHEP} {\bfseries 12}
  (2010) 007} [\href{https://arxiv.org/abs/1008.4579}{{\ttfamily 1008.4579}}].

\bibitem{Gaberdiel:2011wb}
M.R.~Gaberdiel and T.~Hartman, \emph{{Symmetries of Holographic Minimal
  Models}}, \href{https://doi.org/10.1007/JHEP05(2011)031}{\emph{JHEP}
  {\bfseries 05} (2011) 031} [\href{https://arxiv.org/abs/1101.2910}{{\ttfamily
  1101.2910}}].

\bibitem{Campoleoni:2011hg}
A.~Campoleoni, S.~Fredenhagen and S.~Pfenninger, \emph{{Asymptotic W-symmetries
  in three-dimensional higher-spin gauge theories}},
  \href{https://doi.org/10.1007/JHEP09(2011)113}{\emph{JHEP} {\bfseries 09}
  (2011) 113} [\href{https://arxiv.org/abs/1107.0290}{{\ttfamily 1107.0290}}].

\bibitem{Campoleoni:2017mbt}
A.~Campoleoni, D.~Francia and C.~Heissenberg, \emph{{On higher-spin
  supertranslations and superrotations}},
  \href{https://doi.org/10.1007/JHEP05(2017)120}{\emph{JHEP} {\bfseries 05}
  (2017) 120} [\href{https://arxiv.org/abs/1703.01351}{{\ttfamily
  1703.01351}}].

\bibitem{Campoleoni:2020ejn}
A.~Campoleoni, D.~Francia and C.~Heissenberg, \emph{{On asymptotic symmetries
  in higher dimensions for any spin}},
  \href{https://doi.org/10.1007/JHEP12(2020)129}{\emph{JHEP} {\bfseries 12}
  (2020) 129} [\href{https://arxiv.org/abs/2011.04420}{{\ttfamily
  2011.04420}}].

\bibitem{Mikhailov:2002bp}
A.~Mikhailov, \emph{{Notes on higher spin symmetries}},
  \href{https://arxiv.org/abs/hep-th/0201019}{{\ttfamily hep-th/0201019}}.

\bibitem{Bekaert:2007mi}
X.~Bekaert, \emph{{Higher spin algebras as higher symmetries}}, {\emph{Ann. U.
  Craiova Phys.} {\bfseries 16} (2006) 58}
  [\href{https://arxiv.org/abs/0704.0898}{{\ttfamily 0704.0898}}].

\bibitem{Iazeolla:2008ix}
C.~Iazeolla and P.~Sundell, \emph{{A Fiber Approach to Harmonic Analysis of
  Unfolded Higher-Spin Field Equations}},
  \href{https://doi.org/10.1088/1126-6708/2008/10/022}{\emph{JHEP} {\bfseries
  10} (2008) 022} [\href{https://arxiv.org/abs/0806.1942}{{\ttfamily
  0806.1942}}].

\bibitem{Bekaert:2008sa}
X.~Bekaert, \emph{{Comments on higher-spin symmetries}},
  \href{https://doi.org/10.1142/S0219887809003527}{\emph{Int. J. Geom. Meth.
  Mod. Phys.} {\bfseries 6} (2009) 285}
  [\href{https://arxiv.org/abs/0807.4223}{{\ttfamily 0807.4223}}].

\bibitem{Joung:2014qya}
E.~Joung and K.~Mkrtchyan, \emph{{Notes on higher-spin algebras: minimal
  representations and structure constants}},
  \href{https://doi.org/10.1007/JHEP05(2014)103}{\emph{JHEP} {\bfseries 05}
  (2014) 103} [\href{https://arxiv.org/abs/1401.7977}{{\ttfamily 1401.7977}}].

\bibitem{Feigin:1988}
B.L.~Feigin, \emph{{The Lie algebras $\mathfrak{gl}_\lambda$ and cohomologies
  of Lie algebras of differential operators}},
  \href{https://doi.org/10.1070/RM1988v043n02ABEH001720}{\emph{Russ. Math.
  Surv.} {\bfseries 43} (1988) 169}.

\bibitem{Bordemann:1989zi}
M.~Bordemann, J.~Hoppe and P.~Schaller, \emph{{Infinite dimensional matrix
  algebras}}, \href{https://doi.org/10.1016/0370-2693(89)91687-0}{\emph{Phys.
  Lett. B} {\bfseries 232} (1989) 199}.

\bibitem{Fradkin:1990ir}
E.S.~Fradkin and V.Y.~Linetsky, \emph{{Infinite dimensional generalizations of
  simple Lie algebras}},
  \href{https://doi.org/10.1142/S0217732390002249}{\emph{Mod. Phys. Lett. A}
  {\bfseries 5} (1990) 1967}.

\bibitem{Boulanger:2013zza}
N.~Boulanger, D.~Ponomarev, E.D.~Skvortsov and M.~Taronna, \emph{{On the
  uniqueness of higher-spin symmetries in AdS and CFT}},
  \href{https://doi.org/10.1142/S0217751X13501625}{\emph{Int. J. Mod. Phys. A}
  {\bfseries 28} (2013) 1350162}
  [\href{https://arxiv.org/abs/1305.5180}{{\ttfamily 1305.5180}}].

\bibitem{Maldacena:2011jn}
J.~Maldacena and A.~Zhiboedov, \emph{{Constraining Conformal Field Theories
  with A Higher Spin Symmetry}},
  \href{https://doi.org/10.1088/1751-8113/46/21/214011}{\emph{J. Phys. A}
  {\bfseries 46} (2013) 214011}
  [\href{https://arxiv.org/abs/1112.1016}{{\ttfamily 1112.1016}}].

\bibitem{Stanev:2013qra}
Y.S.~Stanev, \emph{{Constraining conformal field theory with higher spin
  symmetry in four dimensions}},
  \href{https://doi.org/10.1016/j.nuclphysb.2013.09.002}{\emph{Nucl. Phys. B}
  {\bfseries 876} (2013) 651}
  [\href{https://arxiv.org/abs/1307.5209}{{\ttfamily 1307.5209}}].

\bibitem{Alba:2013yda}
V.~Alba and K.~Diab, \emph{{Constraining conformal field theories with a higher
  spin symmetry in d=4}},  \href{https://arxiv.org/abs/1307.8092}{{\ttfamily
  1307.8092}}.

\bibitem{Alba:2015upa}
V.~Alba and K.~Diab, \emph{{Constraining conformal field theories with a higher
  spin symmetry in $d > 3$ dimensions}},
  \href{https://doi.org/10.1007/JHEP03(2016)044}{\emph{JHEP} {\bfseries 03}
  (2016) 044} [\href{https://arxiv.org/abs/1510.02535}{{\ttfamily
  1510.02535}}].

\bibitem{Bengtsson:1986bz}
A.K.H.~Bengtsson and I.~Bengtsson, \emph{{Massless higher-spin fields
  revisited}}, \href{https://doi.org/10.1088/0264-9381/3/5/022}{\emph{Class.
  Quant. Grav.} {\bfseries 3} (1986) 927}.

\bibitem{Bekaert:2006us}
X.~Bekaert, N.~Boulanger, S.~Cnockaert and S.~Leclercq, \emph{{On killing
  tensors and cubic vertices in higher-spin gauge theories}},
  \href{https://doi.org/10.1002/prop.200510274}{\emph{Fortsch. Phys.}
  {\bfseries 54} (2006) 282}
  [\href{https://arxiv.org/abs/hep-th/0602092}{{\ttfamily hep-th/0602092}}].

\bibitem{Joung:2013nma}
E.~Joung and M.~Taronna, \emph{{Cubic-interaction-induced deformations of
  higher-spin symmetries}},
  \href{https://doi.org/10.1007/JHEP03(2014)103}{\emph{JHEP} {\bfseries 03}
  (2014) 103} [\href{https://arxiv.org/abs/1311.0242}{{\ttfamily 1311.0242}}].

\bibitem{Sleight:2016xqq}
C.~Sleight and M.~Taronna, \emph{{Higher-Spin Algebras, Holography and Flat
  Space}}, \href{https://doi.org/10.1007/JHEP02(2017)095}{\emph{JHEP}
  {\bfseries 02} (2017) 095}
  [\href{https://arxiv.org/abs/1609.00991}{{\ttfamily 1609.00991}}].

\bibitem{Ponomarev:2017nrr}
D.~Ponomarev, \emph{{Chiral Higher Spin Theories and Self-Duality}},
  \href{https://doi.org/10.1007/JHEP12(2017)141}{\emph{JHEP} {\bfseries 12}
  (2017) 141} [\href{https://arxiv.org/abs/1710.00270}{{\ttfamily
  1710.00270}}].

\bibitem{Campoleoni:2011tn}
A.~Campoleoni, \emph{{Higher Spins in D = 2 + 1}},
  \href{https://doi.org/10.1142/9789814522519_0020}{\emph{Subnucl. Ser.}
  {\bfseries 49} (2013) 385} [\href{https://arxiv.org/abs/1110.5841}{{\ttfamily
  1110.5841}}].

\bibitem{Afshar:2013vka}
H.~Afshar, A.~Bagchi, R.~Fareghbal, D.~Grumiller and J.~Rosseel, \emph{{Spin-3
  Gravity in Three-Dimensional Flat Space}},
  \href{https://doi.org/10.1103/PhysRevLett.111.121603}{\emph{Phys. Rev. Lett.}
  {\bfseries 111} (2013) 121603}
  [\href{https://arxiv.org/abs/1307.4768}{{\ttfamily 1307.4768}}].

\bibitem{Gonzalez:2013oaa}
H.A.~Gonzalez, J.~Matulich, M.~Pino and R.~Troncoso, \emph{{Asymptotically flat
  spacetimes in three-dimensional higher spin gravity}},
  \href{https://doi.org/10.1007/JHEP09(2013)016}{\emph{JHEP} {\bfseries 09}
  (2013) 016} [\href{https://arxiv.org/abs/1307.5651}{{\ttfamily 1307.5651}}].

\bibitem{Ammon:2017vwt}
M.~Ammon, D.~Grumiller, S.~Prohazka, M.~Riegler and R.~Wutte,
  \emph{{Higher-Spin Flat Space Cosmologies with Soft Hair}},
  \href{https://doi.org/10.1007/JHEP05(2017)031}{\emph{JHEP} {\bfseries 05}
  (2017) 031} [\href{https://arxiv.org/abs/1703.02594}{{\ttfamily
  1703.02594}}].

\bibitem{Ammon:2020fxs}
M.~Ammon, M.~Pannier and M.~Riegler, \emph{{Scalar Fields in 3D Asymptotically
  Flat Higher-Spin Gravity}},
  \href{https://doi.org/10.1088/1751-8121/abdbc6}{\emph{J. Phys. A} {\bfseries
  54} (2021) 105401} [\href{https://arxiv.org/abs/2009.14210}{{\ttfamily
  2009.14210}}].

\bibitem{Fuentealba:2015jma}
O.~Fuentealba, J.~Matulich and R.~Troncoso, \emph{{Extension of the Poincar\'e
  group with half-integer spin generators: hypergravity and beyond}},
  \href{https://doi.org/10.1007/JHEP09(2015)003}{\emph{JHEP} {\bfseries 09}
  (2015) 003} [\href{https://arxiv.org/abs/1505.06173}{{\ttfamily
  1505.06173}}].

\bibitem{Fuentealba:2019bgb}
O.~Fuentealba, J.~Matulich and R.~Troncoso, \emph{{Hypergravity in five
  dimensions}}, \href{https://doi.org/10.1103/PhysRevD.101.124002}{\emph{Phys.
  Rev. D} {\bfseries 101} (2020) 124002}
  [\href{https://arxiv.org/abs/1910.03179}{{\ttfamily 1910.03179}}].

\bibitem{Weinberg:1964ew}
S.~Weinberg, \emph{{Photons and Gravitons in $S$-Matrix Theory: Derivation of
  Charge Conservation and Equality of Gravitational and Inertial Mass}},
  \href{https://doi.org/10.1103/PhysRev.135.B1049}{\emph{Phys. Rev.} {\bfseries
  135} (1964) B1049}.

\bibitem{Bekaert:2010hp}
X.~Bekaert, N.~Boulanger and S.~Leclercq, \emph{{Strong obstruction of the
  Berends-Burgers-van Dam spin-3 vertex}},
  \href{https://doi.org/10.1088/1751-8113/43/18/185401}{\emph{J. Phys. A}
  {\bfseries 43} (2010) 185401}
  [\href{https://arxiv.org/abs/1002.0289}{{\ttfamily 1002.0289}}].

\bibitem{Bekaert:2010hw}
X.~Bekaert, N.~Boulanger and P.~Sundell, \emph{{How higher-spin gravity
  surpasses the spin two barrier: no-go theorems versus yes-go examples}},
  \href{https://doi.org/10.1103/RevModPhys.84.987}{\emph{Rev. Mod. Phys.}
  {\bfseries 84} (2012) 987} [\href{https://arxiv.org/abs/1007.0435}{{\ttfamily
  1007.0435}}].

\bibitem{Porrati:2012rd}
M.~Porrati, \emph{{Old and New No Go Theorems on Interacting Massless Particles
  in Flat Space}},  in \emph{{17th International Seminar on High Energy
  Physics}}, 9, 2012 [\href{https://arxiv.org/abs/1209.4876}{{\ttfamily
  1209.4876}}].

\bibitem{Taronna:2017wbx}
M.~Taronna, \emph{{On the Non-Local Obstruction to Interacting Higher Spins in
  Flat Space}}, \href{https://doi.org/10.1007/JHEP05(2017)026}{\emph{JHEP}
  {\bfseries 05} (2017) 026}
  [\href{https://arxiv.org/abs/1701.05772}{{\ttfamily 1701.05772}}].

\bibitem{Sagnotti:2013bha}
A.~Sagnotti, \emph{{Notes on Strings and Higher Spins}},
  \href{https://doi.org/10.1088/1751-8113/46/21/214006}{\emph{J. Phys. A}
  {\bfseries 46} (2013) 214006}
  [\href{https://arxiv.org/abs/1112.4285}{{\ttfamily 1112.4285}}].

\bibitem{Bengtsson:1983pd}
A.K.H.~Bengtsson, I.~Bengtsson and L.~Brink, \emph{{Cubic Interaction Terms for
  Arbitrary Spin}},
  \href{https://doi.org/10.1016/0550-3213(83)90140-2}{\emph{Nucl. Phys. B}
  {\bfseries 227} (1983) 31}.

\bibitem{Bengtsson:1986kh}
A.K.H.~Bengtsson, I.~Bengtsson and N.~Linden, \emph{{Interacting Higher Spin
  Gauge Fields on the Light Front}},
  \href{https://doi.org/10.1088/0264-9381/4/5/028}{\emph{Class. Quant. Grav.}
  {\bfseries 4} (1987) 1333}.

\bibitem{Fradkin:1991iy}
E.S.~Fradkin and R.R.~Metsaev, \emph{{A Cubic interaction of totally symmetric
  massless representations of the Lorentz group in arbitrary dimensions}},
  \href{https://doi.org/10.1088/0264-9381/8/4/004}{\emph{Class. Quant. Grav.}
  {\bfseries 8} (1991) L89}.

\bibitem{Metsaev:1993ap}
R.R.~Metsaev, \emph{{Generating function for cubic interaction vertices of
  higher spin fields in any dimension}},
  \href{https://doi.org/10.1142/S0217732393003706}{\emph{Mod. Phys. Lett. A}
  {\bfseries 8} (1993) 2413}.

\bibitem{Metsaev:2005ar}
R.R.~Metsaev, \emph{{Cubic interaction vertices of massive and massless higher
  spin fields}},
  \href{https://doi.org/10.1016/j.nuclphysb.2006.10.002}{\emph{Nucl. Phys. B}
  {\bfseries 759} (2006) 147}
  [\href{https://arxiv.org/abs/hep-th/0512342}{{\ttfamily hep-th/0512342}}].

\bibitem{Metsaev:2007rn}
R.R.~Metsaev, \emph{{Cubic interaction vertices for fermionic and bosonic
  arbitrary spin fields}},
  \href{https://doi.org/10.1016/j.nuclphysb.2012.01.022}{\emph{Nucl. Phys. B}
  {\bfseries 859} (2012) 13} [\href{https://arxiv.org/abs/0712.3526}{{\ttfamily
  0712.3526}}].

\bibitem{Manvelyan:2010jr}
R.~Manvelyan, K.~Mkrtchyan and W.~Ruhl, \emph{{General trilinear interaction
  for arbitrary even higher spin gauge fields}},
  \href{https://doi.org/10.1016/j.nuclphysb.2010.04.019}{\emph{Nucl. Phys. B}
  {\bfseries 836} (2010) 204}
  [\href{https://arxiv.org/abs/1003.2877}{{\ttfamily 1003.2877}}].

\bibitem{Sagnotti:2010at}
A.~Sagnotti and M.~Taronna, \emph{{String Lessons for Higher-Spin
  Interactions}},
  \href{https://doi.org/10.1016/j.nuclphysb.2010.08.019}{\emph{Nucl. Phys. B}
  {\bfseries 842} (2011) 299}
  [\href{https://arxiv.org/abs/1006.5242}{{\ttfamily 1006.5242}}].

\bibitem{Francia:2016weg}
D.~Francia, G.L.~Monaco and K.~Mkrtchyan, \emph{{Cubic interactions of
  Maxwell-like higher spins}},
  \href{https://doi.org/10.1007/JHEP04(2017)068}{\emph{JHEP} {\bfseries 04}
  (2017) 068} [\href{https://arxiv.org/abs/1611.00292}{{\ttfamily
  1611.00292}}].

\bibitem{Ponomarev:2016lrm}
D.~Ponomarev and E.D.~Skvortsov, \emph{{Light-Front Higher-Spin Theories in
  Flat Space}}, \href{https://doi.org/10.1088/1751-8121/aa56e7}{\emph{J. Phys.
  A} {\bfseries 50} (2017) 095401}
  [\href{https://arxiv.org/abs/1609.04655}{{\ttfamily 1609.04655}}].

\bibitem{Skvortsov:2018jea}
E.D.~Skvortsov, T.~Tran and M.~Tsulaia, \emph{{Quantum Chiral Higher Spin
  Gravity}}, \href{https://doi.org/10.1103/PhysRevLett.121.031601}{\emph{Phys.
  Rev. Lett.} {\bfseries 121} (2018) 031601}
  [\href{https://arxiv.org/abs/1805.00048}{{\ttfamily 1805.00048}}].

\bibitem{Bonora:2018ggh}
L.~Bonora, M.~Cvitan, P.~Dominis~Prester, S.~Giaccari and T.~Stemberga,
  \emph{{HS in flat spacetime. YM-like models}},
  \href{https://arxiv.org/abs/1812.05030}{{\ttfamily 1812.05030}}.

\bibitem{Bonora:2021pcj}
L.~Bonora and S.~Giaccari, \emph{{HS Yang-Mills-like models: a review}},
  {\emph{Ann. U. Craiova Phys.} {\bfseries 30} (2020) 1}
  [\href{https://arxiv.org/abs/2103.10105}{{\ttfamily 2103.10105}}].

\bibitem{Krasnov:2021nsq}
K.~Krasnov, E.~Skvortsov and T.~Tran, \emph{{Actions for Self-dual Higher Spin
  Gravities}}, \href{https://doi.org/10.1007/JHEP08(2021)076}{JHEP \textbf{08} (2021) 076} [\href{https://arxiv.org/abs/2105.12782}{{\ttfamily
  2105.12782}}].

\bibitem{Fradkin:1985am}
E.S.~Fradkin and A.A.~Tseytlin, \emph{Conformal supergravity},
  \href{https://doi.org/10.1016/0370-1573(85)90138-3}{\emph{Phys. Rept.}
  {\bfseries 119} (1985) 233}.

\bibitem{Segal:2002gd}
A.Y.~Segal, \emph{{Conformal higher spin theory}},
  \href{https://doi.org/10.1016/S0550-3213(03)00368-7}{\emph{Nucl. Phys. B}
  {\bfseries 664} (2003) 59}
  [\href{https://arxiv.org/abs/hep-th/0207212}{{\ttfamily hep-th/0207212}}].

\bibitem{talk_quarks}
X.~Bekaert and B.~Oblak, ``{Higher-spin BMS algebras}.'' {Talk given at the
  Quarks Online Workshop-2021 ``Integrability, Holography, Higher-Spin Gravity
  and Strings''}, 1/6/2021.

\bibitem{Grumiller:2019fmp}
D.~Grumiller, A.~P\'erez, M.M.~Sheikh-Jabbari, R.~Troncoso and C.~Zwikel,
  \emph{{Spacetime structure near generic horizons and soft hair}},
  \href{https://doi.org/10.1103/PhysRevLett.124.041601}{\emph{Phys. Rev. Lett.}
  {\bfseries 124} (2020) 041601}
  [\href{https://arxiv.org/abs/1908.09833}{{\ttfamily 1908.09833}}].

\bibitem{Duval:2014uva}
C.~Duval, G.W.~Gibbons and P.A.~Horvathy, \emph{{Conformal Carroll groups and
  BMS symmetry}},
  \href{https://doi.org/10.1088/0264-9381/31/9/092001}{\emph{Class. Quant.
  Grav.} {\bfseries 31} (2014) 092001}
  [\href{https://arxiv.org/abs/1402.5894}{{\ttfamily 1402.5894}}].

\bibitem{Basile:2018eac}
T.~Basile, X.~Bekaert and E.~Joung, \emph{{Conformal Higher-Spin Gravity:
  Linearized Spectrum = Symmetry Algebra}},
  \href{https://doi.org/10.1007/JHEP11(2018)167}{\emph{JHEP} {\bfseries 11}
  (2018) 167} [\href{https://arxiv.org/abs/1808.07728}{{\ttfamily
  1808.07728}}].

\bibitem{Grigoriev:2019xmp}
M.~Grigoriev, I.~Lovrekovic and E.~Skvortsov, \emph{{New Conformal Higher Spin
  Gravities in $3d$}},
  \href{https://doi.org/10.1007/JHEP01(2020)059}{\emph{JHEP} {\bfseries 01}
  (2020) 059} [\href{https://arxiv.org/abs/1909.13305}{{\ttfamily
  1909.13305}}].

\bibitem{Bagchi:2009my}
A.~Bagchi and R.~Gopakumar, \emph{{Galilean Conformal Algebras and AdS/CFT}},
  \href{https://doi.org/10.1088/1126-6708/2009/07/037}{\emph{JHEP} {\bfseries
  07} (2009) 037} [\href{https://arxiv.org/abs/0902.1385}{{\ttfamily
  0902.1385}}].

\bibitem{Bagchi:2009pe}
A.~Bagchi, R.~Gopakumar, I.~Mandal and A.~Miwa, \emph{{GCA in 2d}},
  \href{https://doi.org/10.1007/JHEP08(2010)004}{\emph{JHEP} {\bfseries 08}
  (2010) 004} [\href{https://arxiv.org/abs/0912.1090}{{\ttfamily 0912.1090}}].

\bibitem{Bagchi:2010zz}
A.~Bagchi, \emph{{Correspondence between Asymptotically Flat Spacetimes and
  Nonrelativistic Conformal Field Theories}},
  \href{https://doi.org/10.1103/PhysRevLett.105.171601}{\emph{Phys. Rev. Lett.}
  {\bfseries 105} (2010) 171601}
  [\href{https://arxiv.org/abs/1006.3354}{{\ttfamily 1006.3354}}].

\bibitem{Fronsdal:1978rb}
C.~Fronsdal, \emph{{Massless Fields with Integer Spin}},
  \href{https://doi.org/10.1103/PhysRevD.18.3624}{\emph{Phys. Rev. D}
  {\bfseries 18} (1978) 3624}.

\bibitem{Fronsdal:1978vb}
C.~Fronsdal, \emph{{Singletons and Massless, Integral Spin Fields on de Sitter
  Space (Elementary Particles in a Curved Space. 7.}},
  \href{https://doi.org/10.1103/PhysRevD.20.848}{\emph{Phys. Rev. D} {\bfseries
  20} (1979) 848}.

\bibitem{Bekaert:2005ka}
X.~Bekaert and N.~Boulanger, \emph{{Gauge invariants and Killing tensors in
  higher-spin gauge theories}},
  \href{https://doi.org/10.1016/j.nuclphysb.2005.06.009}{\emph{Nucl. Phys. B}
  {\bfseries 722} (2005) 225}
  [\href{https://arxiv.org/abs/hep-th/0505068}{{\ttfamily hep-th/0505068}}].

\bibitem{Thompson:1986}
G.~Thompson, \emph{{Killing tensors in spaces of constant curvature}},
  \href{https://doi.org/10.1063/1.527288}{\emph{J. Math. Phys.} {\bfseries 27}
  (1986) 2693}.

\bibitem{Barnich:2005bn}
G.~Barnich, N.~Bouatta and M.~Grigoriev, \emph{{Surface charges and dynamical
  Killing tensors for higher spin gauge fields in constant curvature spaces}},
  \href{https://doi.org/10.1088/1126-6708/2005/10/010}{\emph{JHEP} {\bfseries
  10} (2005) 010} [\href{https://arxiv.org/abs/hep-th/0507138}{{\ttfamily
  hep-th/0507138}}].

\bibitem{McLenaghan:2004}
R.G.~McLenaghan, R.~Milson and R.G.~Smirnov, \emph{{Killing tensors as
  irreducible representations of the general linear group}},
  \href{https://doi.org/10.1016/j.crma.2004.07.017}{\emph{C.R. Acad. Sci.
  Paris, Ser. I} {\bfseries 339} (2004) 621-624}.

\bibitem{Vasiliev:1986td}
M.A.~Vasiliev, \emph{{Free Massless Fields of Arbitrary Spin in the De Sitter
  Space and Initial Data for a Higher Spin Superalgebra}},
  \href{https://doi.org/10.1002/prop.2190351103}{\emph{Fortsch. Phys.}
  {\bfseries 35} (1987) 741}.

\bibitem{Karapetyan:2021wdc}
M.~Karapetyan, R.~Manvelyan and G.~Poghosyan, \emph{{On special quartic
  interaction of higher spin gauge fields with scalars and gauge symmetry
  commutator in the linear approximation}},
  \href{https://doi.org/10.1016/j.nuclphysb.2021.115512}{\emph{Nucl. Phys. B}
  {\bfseries 971} (2021) 115512}
  [\href{https://arxiv.org/abs/2104.09139}{{\ttfamily 2104.09139}}].

\bibitem{Bengtsson:1986ys}
A.K.H.~Bengtsson, \emph{{A Unified Action for Higher Spin Gauge Bosons From
  Covariant String Theory}},
  \href{https://doi.org/10.1016/0370-2693(86)90100-0}{\emph{Phys. Lett. B}
  {\bfseries 182} (1986) 321}.

\bibitem{Henneaux:1988}
M.~Henneaux and C.~Teitelboim, \emph{First and second quantized point particles
  of any spin},  in \emph{Quantum Mechanics of Fundamental Systems 2},
  C.~Teitelboim and J.~Zanelli, eds., pp.~113--152, Plenum Press, New York
  (1988), \href{https://doi.org/10.1007/978-1-4613-0797-6}{DOI}.

\bibitem{Francia:2002pt}
D.~Francia and A.~Sagnotti, \emph{{On the geometry of higher spin gauge
  fields}}, \href{https://doi.org/10.1088/0264-9381/20/12/313}{\emph{Class.
  Quant. Grav.} {\bfseries 20} (2003) S473}
  [\href{https://arxiv.org/abs/hep-th/0212185}{{\ttfamily hep-th/0212185}}].

\bibitem{Sagnotti:2003qa}
A.~Sagnotti and M.~Tsulaia, \emph{{On higher spins and the tensionless limit of
  string theory}},
  \href{https://doi.org/10.1016/j.nuclphysb.2004.01.024}{\emph{Nucl. Phys. B}
  {\bfseries 682} (2004) 83}
  [\href{https://arxiv.org/abs/hep-th/0311257}{{\ttfamily hep-th/0311257}}].

\bibitem{Campoleoni:2012th}
A.~Campoleoni and D.~Francia, \emph{{Maxwell-like Lagrangians for higher
  spins}}, \href{https://doi.org/10.1007/JHEP03(2013)168}{\emph{JHEP}
  {\bfseries 03} (2013) 168} [\href{https://arxiv.org/abs/1206.5877}{{\ttfamily
  1206.5877}}].

\bibitem{Vasiliev:1999ba}
M.A.~Vasiliev, \emph{{Higher spin gauge theories: Star product and AdS space}},
   \href{https://arxiv.org/abs/hep-th/9910096}{{\ttfamily hep-th/9910096}}.
   
\bibitem{Alkalaev:2014qpa}
K.~B.~Alkalaev, \emph{{Global and local properties of AdS$_{2}$ higher spin gravity}},
\href{https://doi.org/10.1007/JHEP10(2014)122}{\emph{JHEP} {\bfseries 10} (2014) 122}
[\href{https://arxiv.org/abs/1404.5330}{{\ttfamily 1404.5330}}].

\bibitem{Gaberdiel:2012uj}
M.R.~Gaberdiel and R.~Gopakumar, \emph{{Minimal Model Holography}},
  \href{https://doi.org/10.1088/1751-8113/46/21/214002}{\emph{J. Phys. A}
  {\bfseries 46} (2013) 214002}
  [\href{https://arxiv.org/abs/1207.6697}{{\ttfamily 1207.6697}}].

\bibitem{Fernando:2009fq}
S.~Fernando and M.~Gunaydin, \emph{{Minimal unitary representation of SU(2,2)
  and its deformations as massless conformal fields and their supersymmetric
  extensions}}, \href{https://doi.org/10.1063/1.3447773}{\emph{J. Math. Phys.}
  {\bfseries 51} (2010) 082301}
  [\href{https://arxiv.org/abs/0908.3624}{{\ttfamily 0908.3624}}].

\bibitem{Manvelyan:2013oua}
R.~Manvelyan, K.~Mkrtchyan, R.~Mkrtchyan and S.~Theisen, \emph{{On Higher Spin
  Symmetries in $AdS_{5}$}},
  \href{https://doi.org/10.1007/JHEP10(2013)185}{\emph{JHEP} {\bfseries 10}
  (2013) 185} [\href{https://arxiv.org/abs/1304.7988}{{\ttfamily 1304.7988}}].

\bibitem{Deser:1983mm}
S.~Deser and R.I.~Nepomechie, \emph{{Gauge Invariance Versus Masslessness in De
  Sitter Space}},
  \href{https://doi.org/10.1016/0003-4916(84)90156-8}{\emph{Annals Phys.}
  {\bfseries 154} (1984) 396}.

\bibitem{Higuchi:1986wu}
A.~Higuchi, \emph{{Symmetric Tensor Spherical Harmonics on the $N$ Sphere and
  Their Application to the De Sitter Group SO($N$,1)}},
  \href{https://doi.org/10.1063/1.527513}{\emph{J. Math. Phys.} {\bfseries 28}
  (1987) 1553}.

\bibitem{Deser:2001us}
S.~Deser and A.~Waldron, \emph{{Partial masslessness of higher spins in
  (A)dS}}, \href{https://doi.org/10.1016/S0550-3213(01)00212-7}{\emph{Nucl.
  Phys. B} {\bfseries 607} (2001) 577}
  [\href{https://arxiv.org/abs/hep-th/0103198}{{\ttfamily hep-th/0103198}}].

\bibitem{Zinoviev:2001dt}
Y.M.~Zinoviev, \emph{{On massive high spin particles in AdS}},
  \href{https://arxiv.org/abs/hep-th/0108192}{{\ttfamily hep-th/0108192}}.

\bibitem{Skvortsov:2006at}
E.D.~Skvortsov and M.A.~Vasiliev, \emph{{Geometric formulation for partially
  massless fields}},
  \href{https://doi.org/10.1016/j.nuclphysb.2006.06.019}{\emph{Nucl. Phys. B}
  {\bfseries 756} (2006) 117}
  [\href{https://arxiv.org/abs/hep-th/0601095}{{\ttfamily hep-th/0601095}}].

\bibitem{AIF_2014__64_4_1581_0}
J.-P.~Michel, \emph{Higher symmetries of the {Laplacian} via~quantization},
  \href{https://doi.org/10.5802/aif.2891}{\emph{Annales de l'Institut Fourier}
  {\bfseries 64} (2014) 1581}.

\bibitem{Bekaert:2013zya}
X.~Bekaert and M.~Grigoriev, \emph{{Higher order singletons, partially massless
  fields and their boundary values in the ambient approach}},
  \href{https://doi.org/10.1016/j.nuclphysb.2013.08.015}{\emph{Nucl. Phys. B}
  {\bfseries 876} (2013) 667}
  [\href{https://arxiv.org/abs/1305.0162}{{\ttfamily 1305.0162}}].

\bibitem{gover2012higher}
A.R.~Gover and J.~Silhan, \emph{{Higher symmetries of the conformal powers of
  the Laplacian on conformally flat manifolds}},
  \href{https://doi.org/10.1063/1.3692324}{\emph{J. Math. Phys.} {\bfseries 53}
  (2012) 032301} [\href{https://arxiv.org/abs/0911.5265}{{\ttfamily
  0911.5265}}].

\bibitem{Basile:2014wua}
T.~Basile, X.~Bekaert and N.~Boulanger, \emph{{Flato-Fronsdal theorem for
  higher-order singletons}},
  \href{https://doi.org/10.1007/JHEP11(2014)131}{\emph{JHEP} {\bfseries 11}
  (2014) 131} [\href{https://arxiv.org/abs/1410.7668}{{\ttfamily 1410.7668}}].

\bibitem{Pope:1989sr}
C.N.~Pope, L.J.~Romans and X.~Shen, \emph{{$W$(infinity) and the Racah-wigner
  Algebra}}, \href{https://doi.org/10.1016/0550-3213(90)90539-P}{\emph{Nucl.
  Phys. B} {\bfseries 339} (1990) 191}.

\bibitem{Bagchi:2012cy}
A.~Bagchi and R.~Fareghbal, \emph{{BMS/GCA Redux: Towards Flatspace Holography
  from Non-Relativistic Symmetries}},
  \href{https://doi.org/10.1007/JHEP10(2012)092}{\emph{JHEP} {\bfseries 10}
  (2012) 092} [\href{https://arxiv.org/abs/1203.5795}{{\ttfamily 1203.5795}}].

\bibitem{Campoleoni:2016vsh}
A.~Campoleoni, H.A.~Gonzalez, B.~Oblak and M.~Riegler, \emph{{BMS Modules in
  Three Dimensions}},
  \href{https://doi.org/10.1142/S0217751X16500688}{\emph{Int. J. Mod. Phys. A}
  {\bfseries 31} (2016) 1650068}
  [\href{https://arxiv.org/abs/1603.03812}{{\ttfamily 1603.03812}}].

\bibitem{Gaberdiel:2012ku}
M.R.~Gaberdiel and R.~Gopakumar, \emph{{Triality in Minimal Model Holography}},
  \href{https://doi.org/10.1007/JHEP07(2012)127}{\emph{JHEP} {\bfseries 07}
  (2012) 127} [\href{https://arxiv.org/abs/1205.2472}{{\ttfamily 1205.2472}}].

\bibitem{AIHPA_1984__40_1_35_0}
S.M.~Paneitz, \emph{All linear representations of the poincar\'e group up to
  dimension 8}, {\emph{Annales de l'I.H.P. Physique th\'eorique} {\bfseries 40}
  (1984) 35}.

\bibitem{Campoleoni:2014tfa}
A.~Campoleoni and M.~Henneaux, \emph{{Asymptotic symmetries of
  three-dimensional higher-spin gravity: the metric approach}},
  \href{https://doi.org/10.1007/JHEP03(2015)143}{\emph{JHEP} {\bfseries 03}
  (2015) 143} [\href{https://arxiv.org/abs/1412.6774}{{\ttfamily 1412.6774}}].

\bibitem{Schouten:1940}
J.A.~Schouten, \emph{{Uber Differentialkomitanten zweier kontravarianter
  Grossen}}, {\emph{Proc. Kon. Ned. Akad. Wet. Amst.} {\bfseries 43} (1940)
  449}.

\bibitem{Nijenhuis:1955a}
A.~Nijenhuis, \emph{{Jacobi-type identities for bilinear differential
  concomitants of certain tensor fields, I}}, {\emph{Indag. Math.} {\bfseries
  17} (1955) 390}.

\bibitem{Nijenhuis:1955b}
A.~Nijenhuis, \emph{{Jacobi-type identities for bilinear differential
  concomitants of certain tensor fields, II}}, {\emph{Indag. Math.} {\bfseries
  17} (1955) 398}.

\bibitem{Dubois-Violette:1994tlf}
M.~Dubois-Violette and P.W.~Michor, \emph{{A Common generalization of the
  Frohlicher-Nijenhuis bracket and the Schouten bracket for symmetric
  multivector fields}},
  \href{https://doi.org/10.1016/0019-3577(95)98200-U}{\emph{Indag. Math., N.
  S.} {\bfseries 6} (1995) 51}
  [\href{https://arxiv.org/abs/alg-geom/9401006}{{\ttfamily
  alg-geom/9401006}}].

\bibitem{Ponomarev:2021xdq}
D.~Ponomarev, \emph{{3d conformal fields with manifest sl(2, \ensuremath{\mathbb{C}})}},
\href{https://doi.org/10.1007/JHEP06(2021)055}{\emph{JHEP} {\bfseries 06} (2021) 055}
[\href{https://arxiv.org/abs/2104.02770}{{\ttfamily 2104.02770}}].

\bibitem{Flato:1978qz}
M.~Flato and C.~Fronsdal, \emph{{One Massless Particle Equals Two Dirac
  Singletons: Elementary Particles in a Curved Space. 6.}},
  \href{https://doi.org/10.1007/BF00400170}{\emph{Lett. Math. Phys.} {\bfseries
  2} (1978) 421}.

\bibitem{Kuzenko:2020ayk}
S.M.~Kuzenko and A.E.~Pindur, \emph{{Massless particles in five and higher
  dimensions}},
  \href{https://doi.org/10.1016/j.physletb.2020.136020}{\emph{Phys. Lett. B}
  {\bfseries 812} (2021) 136020}
  [\href{https://arxiv.org/abs/2010.07124}{{\ttfamily 2010.07124}}].

\bibitem{Poincare_Casimirs}
L.~Barannik and W.~Fushchich, \emph{{Casimir operators of the generalised
  Poincar\'e and Galilei groups}},  in \emph{{Group Theoretical Methods in
  Physics: Proceedings of the Third Yurmala Seminar}}, M.A.~Markov, V.I.~Manko
  and V.V.~Dodonov, eds., pp.~275--282, VNU Science Press, 1986.

\bibitem{Bekaert:2006py}
X.~Bekaert and N.~Boulanger, \emph{{The unitary representations of the
  Poincar\'e group in any spacetime dimension}},
  \href{https://doi.org/10.21468/SciPostPhysLectNotes.30}{\emph{SciPost Phys.
  Lect. Notes} {\bfseries 30} (2021) 1}
  [\href{https://arxiv.org/abs/hep-th/0611263}{{\ttfamily hep-th/0611263}}].

\bibitem{Berends:1984rq}
F.A.~Berends, G.J.H.~Burgers and H.~van Dam, \emph{{On the Theoretical Problems
  in Constructing Interactions Involving Higher Spin Massless Particles}},
  \href{https://doi.org/10.1016/0550-3213(85)90074-4}{\emph{Nucl. Phys. B}
  {\bfseries 260} (1985) 295}.
  
\bibitem{Bekaert:2021sfc}
X.~Bekaert,
\emph{{Notes on Higher-Spin Diffeomorphisms}},
\href{https://doi.org/10.3390/universe7120508}{Universe \textbf{7} (2021) no.12, 508}
[\href{https://arxiv.org/abs/2108.09263}{{\ttfamily 2108.09263}}].

\bibitem{Gomis:2005pg}
J.~Gomis, J.~Gomis and K.~Kamimura, \emph{{Non-relativistic superstrings: A New
  soluble sector of AdS(5) x S**5}},
  \href{https://doi.org/10.1088/1126-6708/2005/12/024}{\emph{JHEP} {\bfseries
  12} (2005) 024} [\href{https://arxiv.org/abs/hep-th/0507036}{{\ttfamily
  hep-th/0507036}}].

\bibitem{Joung:2012qy}
E.~Joung and K.~Mkrtchyan, \emph{{A note on higher-derivative actions for free
  higher-spin fields}},
  \href{https://doi.org/10.1007/JHEP11(2012)153}{\emph{JHEP} {\bfseries 11}
  (2012) 153} [\href{https://arxiv.org/abs/1209.4864}{{\ttfamily 1209.4864}}].

\bibitem{Francia:2012rg}
D.~Francia, \emph{{Generalised connections and higher-spin equations}},
  \href{https://doi.org/10.1088/0264-9381/29/24/245003}{\emph{Class. Quant.
  Grav.} {\bfseries 29} (2012) 245003}
  [\href{https://arxiv.org/abs/1209.4885}{{\ttfamily 1209.4885}}].

\bibitem{Brink:2000ag}
L.~Brink, R.R.~Metsaev and M.A.~Vasiliev, \emph{{How massless are massless
  fields in AdS(d)}},
  \href{https://doi.org/10.1016/S0550-3213(00)00402-8}{\emph{Nucl. Phys. B}
  {\bfseries 586} (2000) 183}
  [\href{https://arxiv.org/abs/hep-th/0005136}{{\ttfamily hep-th/0005136}}].

\bibitem{Drew:1980yk}
M.S.~Drew and J.D.~Gegenberg, \emph{Conformally covariant massless spin-2 field
  equations}, \href{https://doi.org/10.1007/BF02776555}{\emph{Nuovo Cim. A}
  {\bfseries 60} (1980) 41}.

\bibitem{deWit:1979sib}
B.~de~Wit and D.Z.~Freedman, \emph{{Systematics of Higher Spin Gauge Fields}},
  \href{https://doi.org/10.1103/PhysRevD.21.358}{\emph{Phys. Rev. D} {\bfseries
  21} (1980) 358}.

\bibitem{Francia:2002aa}
D.~Francia and A.~Sagnotti, \emph{{Free geometric equations for higher spins}},
  \href{https://doi.org/10.1016/S0370-2693(02)02449-8}{\emph{Phys. Lett. B}
  {\bfseries 543} (2002) 303}
  [\href{https://arxiv.org/abs/hep-th/0207002}{{\ttfamily hep-th/0207002}}].

\bibitem{Andringa:2010it}
R.~Andringa, E.~Bergshoeff, S.~Panda and M.~de~Roo, \emph{{Newtonian Gravity
  and the Bargmann Algebra}},
  \href{https://doi.org/10.1088/0264-9381/28/10/105011}{\emph{Class. Quant.
  Grav.} {\bfseries 28} (2011) 105011}
  [\href{https://arxiv.org/abs/1011.1145}{{\ttfamily 1011.1145}}].

\bibitem{Hartong:2015xda}
J.~Hartong, \emph{{Gauging the Carroll Algebra and Ultra-Relativistic
  Gravity}}, \href{https://doi.org/10.1007/JHEP08(2015)069}{\emph{JHEP}
  {\bfseries 08} (2015) 069}
  [\href{https://arxiv.org/abs/1505.05011}{{\ttfamily 1505.05011}}].

\bibitem{Bergshoeff:2017btm}
E.~Bergshoeff, J.~Gomis, B.~Rollier, J.~Rosseel and T.~ter Veldhuis,
  \emph{{Carroll versus Galilei Gravity}},
  \href{https://doi.org/10.1007/JHEP03(2017)165}{\emph{JHEP} {\bfseries 03}
  (2017) 165} [\href{https://arxiv.org/abs/1701.06156}{{\ttfamily
  1701.06156}}].

\bibitem{Boulanger:2003vs}
N.~Boulanger, S.~Cnockaert and M.~Henneaux, \emph{{A note on spin s duality}},
  \href{https://doi.org/10.1088/1126-6708/2003/06/060}{\emph{JHEP} {\bfseries
  06} (2003) 060} [\href{https://arxiv.org/abs/hep-th/0306023}{{\ttfamily
  hep-th/0306023}}].

\bibitem{Henneaux:2021yzg}
M.~Henneaux and P.~Salgado-Rebolledo, \emph{{Carroll contractions of
  Lorentz-invariant theories}},
\href{https://doi.org/10.1007/JHEP11(2021)180}{JHEP \textbf{11} (2021) 180}
  [\href{https://arxiv.org/abs/2109.06708}{{\ttfamily 2109.06708}}].

\bibitem{Bagchi:2017yvj}
A.~Bagchi, J.~Chakrabortty and A.~Mehra, \emph{{Galilean Field Theories and
  Conformal Structure}},
  \href{https://doi.org/10.1007/JHEP04(2018)144}{\emph{JHEP} {\bfseries 04}
  (2018) 144} [\href{https://arxiv.org/abs/1712.05631}{{\ttfamily
  1712.05631}}].

\bibitem{Bagchi:2019xfx}
A.~Bagchi, A.~Mehra and P.~Nandi, \emph{{Field Theories with Conformal
  Carrollian Symmetry}},
  \href{https://doi.org/10.1007/JHEP05(2019)108}{\emph{JHEP} {\bfseries 05}
  (2019) 108} [\href{https://arxiv.org/abs/1901.10147}{{\ttfamily
  1901.10147}}].

\bibitem{Ciambelli:2018wre}
L.~Ciambelli, C.~Marteau, A.C.~Petkou, P.M.~Petropoulos and K.~Siampos,
  \emph{{Flat holography and Carrollian fluids}},
  \href{https://doi.org/10.1007/JHEP07(2018)165}{\emph{JHEP} {\bfseries 07}
  (2018) 165} [\href{https://arxiv.org/abs/1802.06809}{{\ttfamily
  1802.06809}}].

\bibitem{Bacry:1968zf}
H.~Bacry and J.~Levy-Leblond, \emph{{Possible kinematics}},
  \href{https://doi.org/10.1063/1.1664490}{\emph{J. Math. Phys.} {\bfseries 9}
  (1968) 1605}.

\bibitem{Bacry:1986pm}
H.~Bacry and J.~Nuyts, \emph{{Classification of Ten-dimensional Kinematical
  Groups With Space Isotropy}},
  \href{https://doi.org/10.1063/1.527306}{\emph{J. Math. Phys.} {\bfseries 27}
  (1986) 2455}.

\bibitem{Figueroa-OFarrill:2018ilb}
J.~Figueroa-O'Farrill and S.~Prohazka, \emph{{Spatially isotropic homogeneous
  spacetimes}}, \href{https://doi.org/10.1007/JHEP01(2019)229}{\emph{JHEP}
  {\bfseries 01} (2019) 229}
  [\href{https://arxiv.org/abs/1809.01224}{{\ttfamily 1809.01224}}].

\bibitem{Bergshoeff:2016soe}
E.~Bergshoeff, D.~Grumiller, S.~Prohazka and J.~Rosseel,
  \emph{{Three-dimensional Spin-3 Theories Based on General Kinematical
  Algebras}}, \href{https://doi.org/10.1007/JHEP01(2017)114}{\emph{JHEP}
  {\bfseries 01} (2017) 114}
  [\href{https://arxiv.org/abs/1612.02277}{{\ttfamily 1612.02277}}].

\bibitem{Gary:2012ms}
M.~Gary, D.~Grumiller and R.~Rashkov, \emph{{Towards non-AdS holography in
  3-dimensional higher spin gravity}},
  \href{https://doi.org/10.1007/JHEP03(2012)022}{\emph{JHEP} {\bfseries 03}
  (2012) 022} [\href{https://arxiv.org/abs/1201.0013}{{\ttfamily 1201.0013}}].

\bibitem{Afshar:2012nk}
H.~Afshar, M.~Gary, D.~Grumiller, R.~Rashkov and M.~Riegler, \emph{{Non-AdS
  holography in 3-dimensional higher spin gravity - General recipe and
  example}}, \href{https://doi.org/10.1007/JHEP11(2012)099}{\emph{JHEP}
  {\bfseries 11} (2012) 099} [\href{https://arxiv.org/abs/1209.2860}{{\ttfamily
  1209.2860}}].

\bibitem{Gary:2014mca}
M.~Gary, D.~Grumiller, S.~Prohazka and S.-J.~Rey, \emph{{Lifshitz Holography
  with Isotropic Scale Invariance}},
  \href{https://doi.org/10.1007/JHEP08(2014)001}{\emph{JHEP} {\bfseries 08}
  (2014) 001} [\href{https://arxiv.org/abs/1406.1468}{{\ttfamily 1406.1468}}].

\bibitem{Prohazka:2017lqb}
S.~Prohazka and M.~Riegler, \emph{{Higher Spins Without (Anti-)de Sitter}},
  \href{https://doi.org/10.3390/universe4010020}{\emph{Universe} {\bfseries 4}
  (2018) 20} [\href{https://arxiv.org/abs/1710.11105}{{\ttfamily 1710.11105}}].

\bibitem{Chernyavsky:2019hyp}
D.~Chernyavsky and D.~Sorokin, \emph{{Three-dimensional (higher-spin) gravities
  with extended Schr\"odinger and $l$-conformal Galilean symmetries}},
  \href{https://doi.org/10.1007/JHEP07(2019)156}{\emph{JHEP} {\bfseries 07}
  (2019) 156} [\href{https://arxiv.org/abs/1905.13154}{{\ttfamily
  1905.13154}}].

\bibitem{Korybut:2014jza}
A.V.~Korybut, \emph{{Covariant structure constants for a deformed oscillator
  algebra}}, \href{https://doi.org/10.1134/S0040577917100014}{\emph{Theor.
  Math. Phys.} {\bfseries 193} (2017) 1409}
  [\href{https://arxiv.org/abs/1409.8634}{{\ttfamily 1409.8634}}].

\bibitem{Basile:2016goq}
T.~Basile, N.~Boulanger and F.~Buisseret, \emph{{Structure constants of
  shs$[\lambda]$ : the deformed-oscillator point of view}},
  \href{https://doi.org/10.1088/1751-8121/aa9af6}{\emph{J. Phys. A} {\bfseries
  51} (2018) 025201} [\href{https://arxiv.org/abs/1604.04510}{{\ttfamily
  1604.04510}}].

\end{thebibliography}

\end{document}